\documentclass{article}

\PassOptionsToPackage{round}{natbib}
\usepackage[preprint]{neurips_2025}
\usepackage{hyperref}
\usepackage{amsmath}
\usepackage{graphicx}
\usepackage{rotating}
\usepackage{amsthm}

\usepackage[nopatch]{microtype}
\usepackage{booktabs}
\usepackage{longtable}
\usepackage{pdflscape}
\usepackage{multirow}
\usepackage{subcaption}
\usepackage[table]{xcolor}
\usepackage{lscape}
\usepackage{tabularx}
\usepackage{array}
\usepackage{lipsum}
\usepackage{titlesec}
\usepackage{url}

\usepackage{xurl}
\usepackage{enumitem}
\usepackage[toc,page]{appendix}
\usepackage[capitalise]{cleveref}

\crefname{appendix}{Appendix}{Appendices} 
\Crefname{appendix}{Appendix}{Appendices}

\titleclass{\subsubsubsection}{straight}[\subsubsection]
\newcounter{subsubsubsection}[subsubsection]
\renewcommand\thesubsubsubsection{\thesubsubsection.\arabic{subsubsubsection}}

\newcommand{\appr}{\raise.17ex\hbox{$\scriptstyle\sim$}}

\titleformat{\subsubsubsection}
  {\normalfont\normalsize\bfseries}{\thesubsubsubsection}{1em}{}

\titlespacing*{\subsubsubsection}{0pt}{3.25ex plus 1ex minus .2ex}{1.5ex plus .2ex}

\title{Toward Quantitative Modeling of Cybersecurity Risks \\Due to AI Misuse}

\author{%
\textbf{Steve Barrett$^1$} \quad 
\textbf{Malcolm Murray$^{1,}$}\thanks{Corresponding author, \texttt{malcolm@safer-ai.org}} \quad 
\textbf{Otter Quarks$^1$} \quad 
\textbf{Matthew Smith$^1$} \\
\textbf{Jakub Kryś$^1$} \quad 
\textbf{Siméon Campos$^1$} \quad 
\textbf{Alejandro Tlaie Boria$^{2,}$}\thanks{Work conducted while at SaferAI} \quad 
\textbf{Chloé Touzet$^1$} \\
\textbf{Sevan Hayrapet$^3$} \quad 
\textbf{Fred Heiding$^4$} \quad 
\textbf{Omer Nevo$^5$} \quad 
\textbf{Adam Swanda$^6$} \\
\textbf{Jair Aguirre$^7$} \quad
\textbf{Asher Brass Gershovich$^8$} \quad 
\textbf{Eric Clay$^9$} \quad 
\textbf{Ryan Fetterman$^6$} \\
\textbf{Mario Fritz$^{10}$} \quad 
\textbf{Marc Juarez$^{11}$} \quad 
\textbf{Vasilios Mavroudis$^{12}$} \quad 
\textbf{Henry Papadatos$^1$} \\\\
$^1$SaferAI \quad $^2$Pour Demain \quad $^3$0labs \quad $^4$Harvard Kennedy School\\
$^5$Irregular \quad $^6$Cisco \quad $^{7}$RAND \quad $^8$Institute for AI Policy and Strategy\\
$^9$Flare \quad $^{10}$CISPA Helmholtz Center for Information Security \\
$^{11}$University of Edinburgh \quad $^{12}$Alan Turing Institute \\
}

\begin{document}
\maketitle
%%%%%%%%%%%%%%%% abstract%%%%%%%%%%%%%%%%%%%%%%
\begin{abstract}
Advanced AI systems offer substantial benefits but also introduce risks. In 2025, AI-enabled cyber offense has emerged as a concrete example. This technical report applies a quantitative risk modeling methodology (described in full in a companion paper) to this domain. We develop nine detailed cyber risk models that allow analyzing AI uplift as a function of AI benchmark performance. Each model decomposes attacks into steps using the MITRE ATT\&CK framework and estimates how AI affects the number of attackers, attack frequency, probability of success, and resulting harm to determine different types of uplift. To produce these estimates with associated uncertainty, we employ both human experts, via a Delphi study, as well as LLM-based simulated experts, both mapping  benchmark scores (from Cybench and BountyBench) to risk model factors. Individual estimates are aggregated through Monte Carlo simulation. The results indicate systematic uplift in attack efficacy, speed, and target reach, with different mechanisms of uplift across risk models. We aim for our quantitative risk modeling to fulfill several aims: to help cybersecurity teams prioritize mitigations, AI evaluators design benchmarks, AI developers make more informed deployment decisions, and policymakers obtain information to set risk thresholds. Similar goals drove the shift from qualitative to quantitative assessment over time in other high-risk industries, such as nuclear power. We propose this methodology and initial application attempt as a step in that direction for AI risk management. While our estimates carry significant uncertainty, publishing detailed quantified results can enable experts to pinpoint exactly where they disagree. This helps to collectively refine estimates, something that cannot be done with qualitative assessments alone.
\end{abstract}
\newpage
\tableofcontents
%%%%%%%%%%%% Executive summary %%%%%%%%%%%%%%%%%
\section*{Executive Summary}
\label{Executive_summary}
Risk management for frontier AI systems is a nascent science. It has so far focused on ``if-then scenarios'', where an evaluation result pointing to a certain level of a dangerous capability triggers a certain set of mitigations. This approach has several limitations. First, as risk itself is not measured, we cannot know by how much the mitigations reduce risk or provide justification for whether the mitigations are sufficient. Second, it leads to the treatment of capabilities measured by standardized benchmarks in isolation, ignoring interactions between capabilities that can impact real-world risks and important factors related to the precise path to harm. 

The role of risk modeling is to bridge the gap between the source of risk (e.g., dangerous capabilities or propensities, deployment conditions or affordances) and the actual harm, to enable systematically analyzing the risk. This technical report applies the methodology we have developed for quantitative modeling of AI-enabled risks to the domain of cyber offense. We provide a road map for implementing the methodology as well as tentative results from applying it to nine risk models. As a common criticism of quantitative risk assessment is its lack of scalability, we also experiment with the use of LLM-simulated experts to provide estimates, in addition to our human expert Delphi study.

Comprehensive and systematic risk modeling can provide numerous benefits to different groups of AI risk stakeholders:
\begin{itemize}
    \item Cybersecurity defenders and vendors can leverage the insights on where AI uplift is the hightest to prioritize their mitigation efforts.
    \item In the AI evaluation and benchmark community, evaluators can leverage the insights to see where new benchmarks would reduce the greatest uncertainty in risk estimates and hence where they should  focus their efforts.
    \item In AI companies, decision makers can use the more precise and forward-looking data to make more informed development and deployment decisions.
    \item Regulators and policymakers can gain more foresight on where AI risk is heading and start to determine expected harm to define risk thresholds.
\end{itemize}

In other high-risk industries, such as nuclear power and aviation, these types of benefits drove a shift over time from qualitative risk assessment to quantitative. In order to prompt a step in that direction for AI risk management, we present this methodology and initial attempt at applying it. While the resulting estimates necessarily carry significant uncertainty, we hope that publishing specific numbers can enable experts to pinpoint exactly where they disagree, and collectively refine estimates, something that cannot be done with qualitative assessments alone.

\subsection*{Methodology (\cref{sec:Methodology})}
Our risk modeling methodology consists of six, interlinked steps:
\begin{enumerate}
    \item Selecting risk scenarios. We systematically decompose the risk universe into a set of representative scenarios.
    \item Constructing risk scenarios. We build risk models for each scenario. These comprise four types of risk factors: the number of actors, the frequency with which attacks are launched, the probability of the attack succeeding, and the harm that would arise as a result. The steps in an attack are modeled using the MITRE ATT\&CK framework.
    \item Quantifying ``baseline'' risk. We establish estimates for the ``baseline'' risk (negligible or non-existent use of AI) case, in order to create a reference point, based on cyber threat intelligence data, historical case studies and expert review. This is captured as a Bayesian network. 
    \item Determining key risk indicators (KRIs) for AI ``uplift''. We establish which forms of KRIs, such as benchmark performance, can serve as evidence to infer values for uplifted risk factors. This technical report uses Cybench and BountyBench as examples.
    \item Estimating AI uplift. We build a quantitative mapping between the KRIs and the risk factors in the risk model and use these to generate estimates for the risk factors. We conduct a Delphi study with cybersecurity experts for one risk scenario and we experiment with the use of ``LLM-estimators'' to generate estimates at scale. Experts provide confidence intervals around their estimates.
    \item Propagating individual estimates among experts and across risk factors to aggregate estimates. We fit the estimates to the appropriate distributions and propagate the individual parameter estimates using Monte Carlo simulations to arrive at an overall risk distribution of the scenario.
\end{enumerate}

Throughout this methodology, we rely extensively on cybersecurity experts. We iterate multiple times with four cybersecurity experts with complementary backgrounds to develop and refine the list of nine selected risk scenarios. Each baseline risk model is reviewed by one expert with relevant domain expertise, who validates the parameter values and suggests corrections where appropriate. Nine cybersecurity experts participated in the Delphi study for uplift estimation, and one expert reviews all of the uplift values produced by the LLM estimators to identify implausible estimates.

\subsection*{Results from Delphi Processes (\cref{Sec:Results_Delphi})}
In the modified Delphi study we conducted with cybersecurity experts, we had nine cyber experts provide two round of estimates of risk factors for one risk model, with a facilitated discussion in-between to discuss points of contention. We find that experts vary highly in how confident they are in assessing their uncertainty. Further, uplift estimates on risk factors associated with quantities (number of actors, number of attempts/actor/year, impact) exhibit a much greater variance than those associated with probabilities that are bounded by [0,1]. It is also noteworthy that the uplift variance increases as the corresponding benchmark task gets more difficult.

We also experiment with LLM-simulated expert estimators. Their estimates of probability risk factors closely follow those of humans. However, for quantity risk factors, there is more disagreement. LLM estimators are often more conservative, providing significantly lower predictions of uplift than human experts. LLM estimators predict a lower total risk than their human counterparts, with the deviation from human estimates increasing as task difficulty grows. LLMs also demonstrate lower uncertainty than humans, especially at higher AI capability levels.

\subsection*{Results from the Quantitative Evaluation (\cref{sec:Results_Quantitative})}
In~\cref{sec:Results_Quantitative}, we provide the tentative quantitative results of our nine risk models in order to demonstrate the many use cases of risk modeling and create scrutiny, debate, and criticism around specific values so that we can iteratively work toward more exact estimates. Given the nascency of the science of AI risk modeling and the limitations of our methodology, we do not recommend making use of the exact numbers for decision-making at this time. We provide detailed results of our early comparative findings (intra- and inter-model) as a proof of concept for the potential value in quantitative risk modeling. Interesting results indicated by the models include:
\begin{itemize}
    \item For seven out of nine scenarios, the models indicate that state of the art (SOTA) (at the time of conducting experiments) AI systems provide uplift relative to the baseline, i.e., the estimated total risk is higher when malicious actors’ use AI at current capabilities.
    \item At ``saturation'', i.e., when AI can reliably perform all tasks in the benchmarks we use, the models indicate that the total risk estimates are again significantly higher than the SOTA-level.
    \item Across risk models, the models do not suggest a uniform pattern (i.e., AI is not consistently helping low-level or high-level attackers more)
    \item None of the four risk factors (number of actors, number of attempts, probability of success, and damage per attack) is suggested to play the key role in uplift, but rather all contribute to the increase in risk across different scenarios, i.e., AI helps with both ``quantity'' and ``quality''.
    \item The models suggest that AI provides more uplift for three MITRE tactics, Execution, Impact, and Initial Access, relative to the other eleven. However, there is significant variability in uplift within each factor.
\end{itemize}

\subsection*{Limitations and Future Work (\cref{sec:Limitations})}
This is, to our knowledge, one of the first attempts at building a systematic procedure for quantitative modeling of cybersecurity risks arising from AI misuse. Therefore, we acknowledge a number of limitations with our methodology and discuss them at length in~\cref{sec:Limitations} to guide future work.
%%%%%%%%%%%%%%%%%% Introduction %%%%%%%%%%%%%%
\section{Introduction}
\label{sec:Introduction}
AI systems present many benefits for society, but at the same time introduce and exacerbate risks, ranging from misuse by humans to loss of control of powerful autonomous AI systems. The domain of cybersecurity serves as a potential early warning signal, with examples this year of AI systems providing meaningful assistance to malicious actors capabilities in their cyber attacks~\citep{Anthropic2025,hao2025spammers}. The combination of AI assistance and the already large economic damage inflicted by cyber crime (estimated to be in the hundreds of USD billions ~\citep{Miliefsky}makes it a domain requiring urgent attention on how to best measure and mitigate these risks~\citep{AISI2025}. New cybersecurity risks can range from relatively low-skilled individual threat actors using AI to efficiently craft more convincing phishing emails to nation-state actors using AI to develop more advanced malware~\citep{OpenAI2025}. \citet{RAND2024SecuringAI} classify these operations as ranging from OC1 (Operational Capacity 1 -- amateur attempts) to OC5 (top priority operations from cyber-capable institutions). AI may be misused by threat actors to fill knowledge and skill gaps, scale operations and potentially enable attack of new types of targets, overall increasing both the efficiency and efficacy of attacks.

\begin{figure}[t]
  \centering
  \includegraphics[draft=false,page=1,pagebox=cropbox,keepaspectratio,         width=1.0\linewidth,trim=1cm 3cm 1cm 2cm,clip]{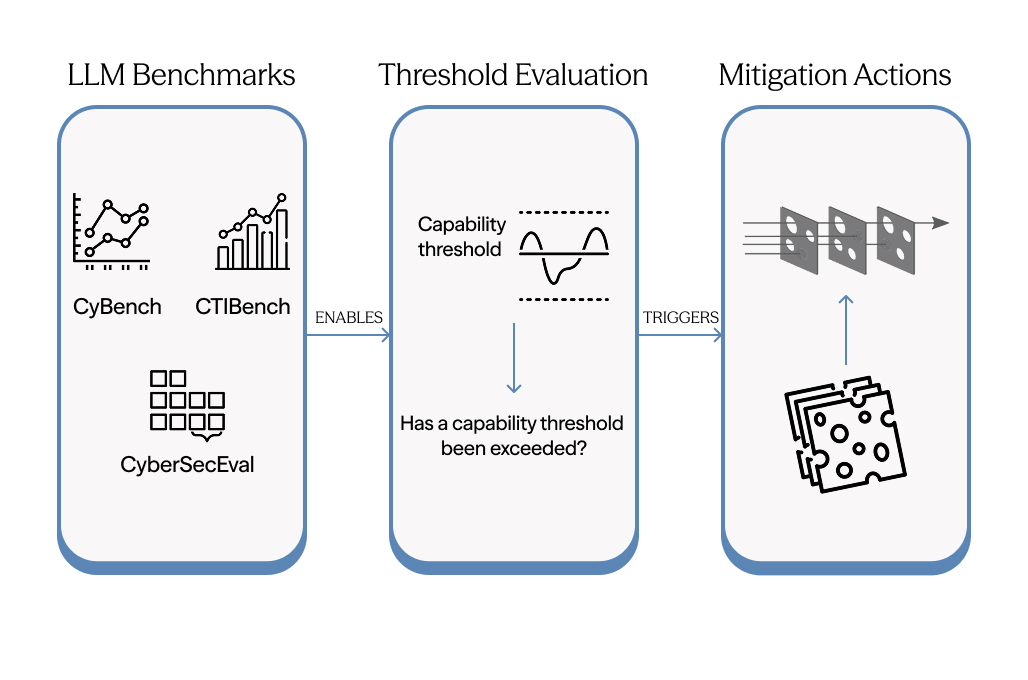}
  \caption{Typical industry practice, as described in frontier AI safety frameworks, is to rely on frameworks built around ``if-then scenarios''~\citep{Karnofsky2024, METR_FAISC}.}
  \label{fig:fig_1}
\end{figure}

Although risks from AI systems have the potential to pose significant harm to society, frontier AI risk management is still an immature discipline. Typical current risk management practice for frontier AI systems is based on determining whether capability thresholds, measured through evaluations and benchmarks, are exceeded and then implementing mitigations accordingly (see \cref{fig:fig_1})~\citep{METR_FAISC}. This is problematic for several reasons. 

First, as risk itself is not measured, we cannot know by how much mitigations reduce risk or justify that they are sufficient~\citep{Campos2025}. Second, the risk assessment should not end at evaluating capabilities of models, since what matters is the harm in the real world. Focusing on capabilities measured by standardized benchmarks in isolation ignores interactions between capabilities that can impact real-world risks and important factors related to threat actor behavior, the target, or the precise path to harm  \citep{LukosiuteSwanda2025, CLTC2025IntolerableAI, WeidingerEtAl2023, SolaimanEtAl2023}. Relying on capability assessments, therefore, is an imperfect proxy for the actual quantity of interest, which is risk.
% \begin{figure}[ht]
% \centering
% \includegraphics[width=1.0\linewidth]{Fig1.jpg}
% \caption{Typical industry practice as described in Frontier AI safety frameworks}
% \label{fig:fig_1}
% \end{figure}

In a first companion paper~\citep{Touzet2025}, we investigate risk management practices across five safety-critical domains and from this, we derive a recommendation for a structured approach to AI risk modeling drawing on lessons from probabilistic risk assessment and causal trees in other industries. In a second companion paper~\citep{Murrayb2025}, we build on that work to outline a quantitative risk modeling approach for AI with methods for risk scenario building and risk quantification. This technical report is self-contained but for related work, see these two companion papers. 

In this technical report, we apply our detailed quantitative AI risk modeling methodology specifically to the domain of AI-enabled cyber-offense risk. We provide a road map for implementing the methodology and provide initial results from modeling nine risk scenarios. A common criticism of quantitative risk assessment is its lack of scalability: in this work therefore, we experiment with the use of LLM-simulated experts to provide estimates for risk model parameters, alongside a more conventional human expert Delphi\footnote{We will publish the code for conducting simulated Delphi studies with LLM experts, along with the associated prompts, in the near future.}.
\begin{figure}[t]
  \centering
  \includegraphics[draft=false,page=1,pagebox=cropbox,keepaspectratio,         width=1.0\linewidth,trim=1cm 1cm 1cm 1cm,clip]{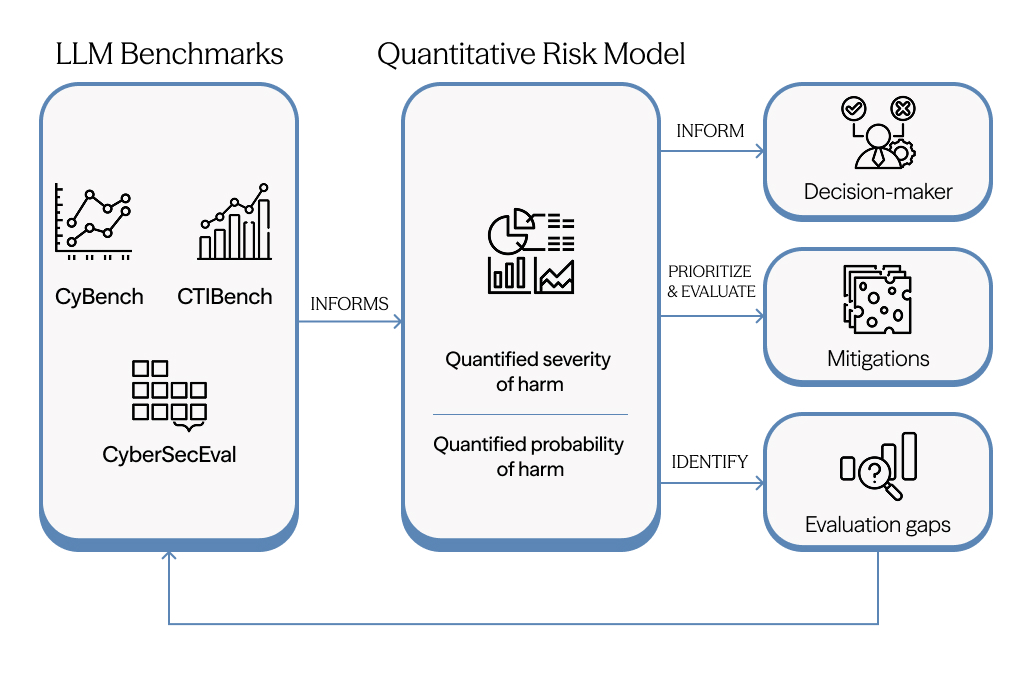}
  \caption{Benefits of quantitative risk modeling.}
  \label{fig:fig_2}
\end{figure}
By linking hazards (dangerous capabilities or propensities) to specific harms and by quantifying both the probability and severity of the harm, there are a number of benefits, as shown in~\cref{fig:fig_2}:
\begin{enumerate}
    \item Quantification of risks enables defenders to better prioritize the development and deployment of mitigations against cyber attacks.
    \item This approach can be used to shine a light on gaps where evaluations are missing and new ones might be required.
    \item Access to quantified risk information, as well as a view that considers the impact of several capabilities in tandem (rather than each capability in isolation), is a powerful enabler for decision-makers to assess whether an AI is safe enough to be developed or deployed.
    \item Quantified risk information can help regulators and policymakers define risk thresholds in terms of expected harm, which is a more concrete basis on which to build consensus than capability thresholds.
\end{enumerate}

Our risk models take advantage of the reproducible nature of benchmark evaluations in order to construct static, forward-looking mappings from AI capabilities to real-world risks. This enables our models to adapt to a changing risk environment and provides a mechanism for forecasting of risk as AI becomes increasingly powerful.

This technical report proceeds as follows. \cref{sec:Methodology} provides a detailed description of our risk modeling methodology. \cref{Sec:Results_Delphi} describes our findings from the cybersecurity experts in the Delphi study and from comparing them with the ``LLM estimators''. \cref{sec:Results_Quantitative} provides the quantitative results suggested by our models in terms of overall changes in risk as well as uplift of different types: ``efficacy uplift'' (the increase in probability of success of an attack), ``volume uplift'' (the increase in quantity of attacks which can be launched), and ``target uplift'' (the ability with which larger, more valuable, and better defended victims can be targeted). \cref{sec:Limitations} provides limitations and future work, and \cref{sec:Conclusion} presents conclusions.
%%%%%%%%%%%%%%%% Methodology %%%%%%%%%%%%%%%%%
\section{Methodology}
\label{sec:Methodology}
\begin{figure}[t]
  \centering  \includegraphics[draft=false,page=1,pagebox=cropbox,keepaspectratio,width=1.0\linewidth, trim=0cm 2cm 0cm 2cm, clip]{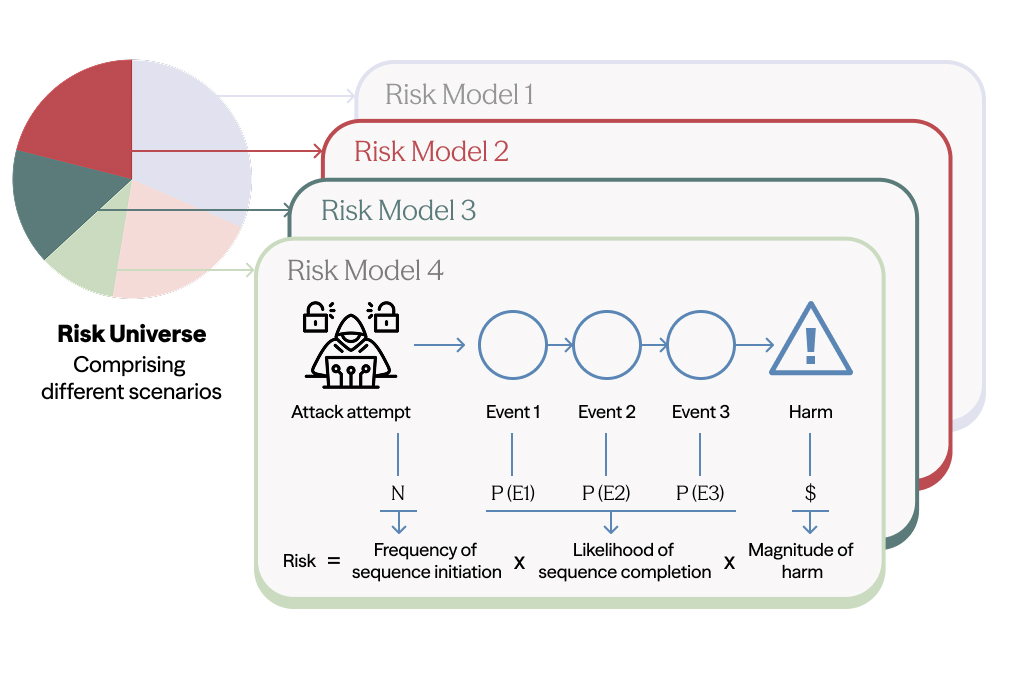}
  \caption{Our risk management methodology first decomposes the risk universe into distinct scenarios, then models each using various risk factors: the frequency with which a specific sequence of events is initiated, the probability of the sequence completing, and the harm that would arise as a result.}
  \label{fig:fig_3}
\end{figure}

A high-level summary of our risk management methodology is illustrated in~\cref{fig:fig_3}.  As is common in risk management practice \citep{ISO53940_2023,kaplan1981quantitative}, first we decompose the risk universe into distinct risk scenarios, which for the case of cyber offense correspond to a variety of different cyber-attack archetypes.  Risk models for each scenario comprise four risk factors: the frequency with which attacks of the specific cyber-attack archetype are launched, the probability of the set of steps in the attack succeeding, the number of actors that would attempt an attack, and the harm that would arise as a result of a successful attack. The individual factors in the risk model are then aggregated to determine an overall level of risk for that scenario expressed as the expected level of annual economic damage \citep{vose2008risk,de2019deterministic}.

Our risk modeling methodology consists of six steps, briefly described below. For a full description of the methodology, please see our companion paper~\citep{Murrayb2025}.
\begin{enumerate}
    \item \textbf{Defining risk scenarios to model:} We systematically decompose the risk universe into a set of representative scenarios, and we build risk models for each representative scenario.
    \item \textbf{Constructing risk models:} Risk is modeled as a combination of four factors. First, the number of threat actors conducting this type of attack. Second, the number of attack attempts per actor per year. Third, the set of tactics required in the attack and their associated probability of successful application. Fourth, the damage resulting from each successful attack.
    \item \textbf{Quantifying ``baseline'' risk:} We establish estimates for the risk of ``baseline'' threat actor capabilities (with negligible or absent use of AI) as a reference point for uplift.
    \item \textbf{Determining key risk indicators (KRIs) for AI ``uplift'':} We establish which forms of evidence (KRIs) risk factors can be conditioned on, such as benchmark performance.
    \item \textbf{Estimating AI uplift:} We conduct expert elicitation to build a quantitative mapping between the KRIs and the factors in the risk model and use these to generate uplift estimates.
    \item \textbf{Propagating individual estimates to aggregate estimates:} Distributions are fitted to central tendencies and confidence intervals provided by the expert for each risk factor, conditioned on benchmark scores. For each expert, we perform Monte Carlo sampling from the joint distribution across all risk factors. The result is a mixture distribution that captures expert epistemic uncertainty about overall risk.  
\end{enumerate}
The following sections provide additional detail on the methodology as applied to the misuse of AI for cyber offense.
%%%%%%%%%%%%%%%%%%%%%%%%%%%%%%%%%%%%%%%%%%%%

\subsection{Determining the Set of Risk Scenarios}
\label{sub:risk_scenarios}
The first step in our methodology is to determine which risk scenarios warrant detailed modeling. The risk space is large; therefore, not all scenarios can be modeled in detail. In the domain of AI-enabled cyber attacks, the main type of harm is economic, but other types of harm are also possible such as psychological, reputational, or physical harms. Cyber attacks may even lead to loss of life, especially in settings involving threat actors targeting critical infrastructure. For simplicity, we focus on economic harm across this set of risk models as an all-encompassing metric, using dollars (US) as a common unit of harm to maintain consistency.

In order to narrow the full risk space into a set of the most important risk scenarios, we start by determining a set of key dimensions that are common to attacks. For AI-enabled cyber attacks resulting in economic harm, such dimensions may include, e.g., the type of threat actor, the type of target, the type of attack (the vector), the intent of the attack, and the defense level of the target. Our analysis suggests that the key dimensions to focus on are actor, target and vector. The defense level can be specified as a function of the target in question, and similarly, the intent can be determined from the actor and vector. In order to make our modeling approach easily replicable, we rely as much as possible on existing taxonomies to define these aspects.

\textbf{Actor:} There needs to be a threat actor, who exploits a vulnerability to cause harm. A taxonomy of threat actor categorizations expressed as operational capacities (OC)  can be found in \citet{RAND2024SecuringAI}. Here, offensive cybersecurity operations are classified on a scale, from OC1 (amateur attempts by hobbyist hackers) to OC5 (top-priority operations by the most cyber-capable nation-states). It should be noted that their taxonomy is for operations, not actors, but it easily maps onto the type of actor. For example, we define an OC3 actor as one that would perform at most OC3 level operations.

\textbf{Target:} For there to be harm, there needs to be a specified entity that experiences said harm, here referred to as the target. For targets, we start with CISA’s (US) and NPSA’s (UK) list of critical national infrastructure sectors \citep{CISA_CriticalInfrastructureSectors,NPSA_CNI}. These include, e.g., Financial Services, Healthcare, Transportation Systems, and Defense. We then group these into types of targets, based on similarities in the attack surface, the intent of threat actors and the impact of disruption. This results in the following categories of targets: Financially attractive and data-rich targets (Financial Services, Healthcare \& Public Health, Commercial Facilities, Information Technology); Espionage and state-interest targets (Defense Industrial Base, Government Facilities, Communications); Critical infrastructure and control-system-heavy targets (Energy, Water \& Wastewater Systems, Chemical, Critical Manufacturing); and Logistics \& national mobility targets (Transportation Systems, Emergency Services). 

\textbf{Vector:} Finally, a sequence of events needs to take place for the threat actor to turn the threat into a harm, which we refer to as the vector. Here, we take as our starting point a list from \citet{RodriguezEtAl2025}, who analyzed real-world instances of attempted AI use in cyber attacks with a dataset of cyber incidents from Google’s Threat Intelligence Group and Mandiant. They have the following categories of attacks: Phishing, Malware, Denial-of-Service, Man-in-the-Middle, SQL Injection, Zero-Day Attack and Cross-Site Scripting. These categories, however, do not capture the full sequence of events (e.g., phishing is just one step of an attack). In consultation with cybersecurity experts, we developed and defined a representative set of attacks in detail. This led to us including aspects such as Social-Engineering, Ransomware, Business Email Compromise (BEC), Public-Facing-App Exploit, Data Breach, Industrial Control Systems and Operational Technology (ICS/OT) Sabotage. 

Having established categories of actors, targets and vectors, we proceed to analyze the potential combinations. We explore the various combinations, first by combining actors and targets, and then by combining actor-target combinations with vectors. 

In order to narrow down to a manageable number of scenarios that can be modeled in detail, we apply the following principles:
\begin{itemize}
    \item \textbf{Historical prevalence:} We favor scenarios where historical data shows the scenario has been prevalent, for example, deployment of social engineering and ransomware targeted at corporations.
    \item \textbf{Realism:} We remove scenarios that are deemed unrealistic. For example, OC1 actors (hobby hackers) are generally not able to attack critical infrastructure targets.
    \item \textbf{Expected uplift:} We include only scenarios where we expect the AI to provide non-negligible uplift in terms of the likelihood of attack success, volume of attacks, or the ability to target more sophisticated defenders than without access to AI.
    \item \textbf{Removal of duplicates:} We remove scenarios that are very similar to each other to be efficient.
\end{itemize}
Applying these principles, we arrive at the set of scenarios shown in~\cref{tab:threat_scenarios}. 

Scenarios 5 and 6, and scenarios 7 and 8 are paired to explore whether AI lowers ``barriers to entry'', enabling attackers to attack larger, better-defended targets.
\begin{table}[t]
\centering
\begin{tabular}{c p{3.3cm} c p{4cm} p{3.5cm}}
\hline
\# & Scenario& Actor & Target & Vector \\
\hline
1 & OC1 Phishing & OC1 & Financially attractive and data-rich & Social-engineering, BEC \\
2 & OC2 Data Breach & OC2 & Financially attractive and data-rich & Purchasing credentials, data theft/extortion \\
3 & OC2 IAB & OC2 & Financially attractive and data-rich & Phishing, infostealer \\
4 & OC3 DoS & OC3 & Financially attractive and data-rich & DDoS \\
5 & OC3 SME Ransomware (see~\cref{app:E}) & OC3 & Financially attractive and data-rich (small enterprise) & Public-facing app exploit, double extortion \\
6 & OC3 LgE Ransomware & OC3 & Financially attractive and data-rich (large enterprise) & Public-facing app exploit, double extortion \\
7 & OC4 Small Infrastructure & OC4 & Critical infrastructure and control-system-heavy (small facility) & IT to OT Pivot, sabotage \& disruption \\
8 & OC4 Large Infrastructure & OC4 & Critical infrastructure and control-system-heavy (large facility) & IT to OT Pivot, sabotage \& disruption \\
9 & OC5 Espionage & OC5 & Espionage and state-interest & Polymorphic malware and data exfiltration \\
\hline
\end{tabular}
\vspace{0.5em}
\caption{Overview of threat scenarios for which risk models were created. We provide an example of a full baseline model in Appendix E. Note: this example is a working technical document provided for transparency and to enable scrutiny of specific estimates; it has not been formatted for publication and carry significant uncertainty.}
\label{tab:threat_scenarios}
\end{table}
%%%%%%%%%%%%%%%%%%%%%%%%%%%%%%%%%%%%%%%%%%%%%%%
\subsection{Modeling Risk in the Absence of AI Capabilities (Baseline)}
\label{sub:Modelling_baseline_risk}
Having defined the scenarios to model, we then quantify the ``baseline risk'', i.e., the level of risk where we assume that there is no significant use of, or benefit in using AI. This means determining initial estimates for all the risk factors of the model. Having a baseline enables us to calculate the ``marginal risk'', i.e., how much risk is added when AI is fully used by all threat actors in the scenario. The following sections outline how we model baseline risk, specifically our use of a Bayesian network model, our use of the MITRE ATT\&CK framework and our targeted collection of data for the estimates. An example of a baseline risk model for the OC3 SME Ransomware scenario is provided in~\cref{app:E} and the uplifted parameters in~\cref{app:F}.
%%%%%%%%%%%%%%%%%%%%%%%%%%%%%%%%%%%%%%%%%%%%%%%
\subsubsection{Use of MITRE ATT\&CK Framework}
\label{subsec:MITRE-framework}
In probabilistic risk assessment of safety-critical systems, common approaches include the use of Event Tree Analysis (ETA) and Fault Tree Analysis (FTA). A corresponding approach in cybersecurity is to make use of attack trees \citep{Schneier1999AttackTrees}. The attack tree approach assumes that an attacker will need to successfully complete a series of steps, potentially from many possible such series of steps,  in order to achieve their desired result.

One potential approach to cybersecurity risk modeling would be to build an attack tree for a specific typical network of the target network type under consideration, and treat that specific network as representative of the whole target network class. However, such an approach has a number of issues:
\begin{enumerate}
    \item Defining a specific network as representative in this way fails to capture the diverse effects of AI uplift across a variety of networks.
    \item Data is sparse at the level of specific network setups.
    \item The number of expert elicitations needed to parameterize an attack tree may be impossible. Furthermore expressing conditional dependencies between this many parameters leads to potential overfitting concerns.
\end{enumerate}
Taken together, these concerns suggest that attack trees capture the wrong level of abstraction for our risk modeling work. Instead, we sought a more abstract approach that did not have these disadvantages, but which remained grounded in the use of a well-accepted cybersecurity framework. There are a variety of approaches to describe the potential steps undertaken in a cybersecurity attack. One common approach to categorizing steps in an attack is Lockheed-Martin’s Cyber Kill Chain \citep{LockheedMartin_CyberKillChain}, and another approach is MITRE ATT\&CK \citep{MITRE2025ATTACK}. We base our work around the MITRE ATT\&CK framework. This is because the larger set of tactic categorizations available in MITRE ATT\&CK can be used to accurately characterize a broad range of attack archetypes ranging from simple few-step attacks that might be undertaken by less-experienced hackers all the way up to complex multi-step cyber attacks like those performed by nation-state actors.

There are a number of points to highlight with respect to the use of the MITRE ATT\&CK framework for analyzing the probability of success of an attack comprising multiple steps:
\begin{enumerate}
    \item Whilst MITRE illustrates the 14 tactics\footnote{MITRE ATT\&CK tactics:  Reconnaissance, Resource Development, Initial Access, Execution, Persistence, Privilege Escalation, Defense Evasion, Credential Access, Discovery, Lateral Movement, Collection, Command and Control, Exfiltration, Impact.} in a matrix, which when read left to right might be considered to broadly follow the expected conditional trajectory of an attack, the ordering of tactics in any one attack may not necessarily occur in the order suggested by the framework.
    \item Any given tactic or technique may be employed at multiple different times and points in any one attack.
    \item An attacker may not deploy every MITRE ATT\&CK tactic in any given attack.
\end{enumerate}
We therefore use the following simplifications and associated steps:
\begin{enumerate}
    \item We identify the set of MITRE ATT\&CK tactics that are relevant for the particular risk scenario that we are considering, using a process described in the next section. 
    \item For each tactic, we ask the experts to provide a joint probability of success across all instantiations of the tactic within the attack. In asking this relatively simple question, we make a trade-off of simplicity in elicitation versus accuracy of elicitation. The issue is that for more complex attacks, some tactics may be used multiple times, (e.g., Lateral Movement or Privilege Escalation). The probability of success for each instantiation of a tactic in an attack such as Lateral Movement might also improve, conditioned on the stage in the attack at which the tactic is utilized. For example, once methods have been found to perform lateral movement for the first time, subsequent instantiations of the same tactic might re-use the same method and success consequently may become conditionally more likely.
    \item We assume that each tactic included in a given risk model represents a binary random variable with states corresponding to success or failure in the execution of the tactic. For an actor to succeed in an attack, they must succeed in all steps of the attack. This is a first-order approximation that does not consider degrees of success in executing tactics (e.g., partial but incomplete discovery of network structure, or partial exfiltration of attack-relevant data). Furthermore, it does not capture the possibility that an attacker may employ a certain tactic to increase the probability of success of an attack overall, but where the tactic is not strictly required to achieve success in the attack. These simplifications make the risk models more tractable and limit the number of risk factors which need to be estimated, leading to cost reductions in parameter estimation and reduced overfitting.
\end{enumerate}
%%%%%%%%%%%%%%%%%%%%%%%%%%%%%%%%%%%%%%%%%%%
\subsubsection{Process for Selecting Relevant Tactics and Techniques}
For each risk scenario that we study, we analyze each of the 14 tactics to determine whether it should be included as a step in the scenario or not. The tactic is excluded if there is no clear failure mode in which a failure at the step would invalidate the attack as a whole. The Impact tactic is included if it captures aspects of the attack which are not captured elsewhere, for example, successful execution of encryption in a ransomware attack.  Also, the tactics of Reconnaissance and Resource Development were sometimes modeled as being 100 percent successful, for example, if those tactics had to be successfully completed in order for the attack to commence.  A complete failure to perform these tactics would then be captured in a reduction in the number of attempts/actor/year risk factor.  The quality with which these tactics are implemented is captured in the probability of success of other relevant steps in the attack. Similarly, the Defense Evasion tactic was sometimes not modeled (or modeled as 100 percent successful) in order to avoid double-counting in the cases where the attacker's success in evading defenses is already captured in the probability of success of executing other tactics.  

MITRE ATT\&CK breaks down tactics, representing attacker objectives, into techniques, representing specific methods used to achieve parent tactics. We establish which of the MITRE ATT\&CK tactics should be broken down into MITRE ATT\&CK techniques. It should be noted that there may be many possible techniques per tactic. For this, we use four rules.
\begin{itemize}
    \item \textbf{AI-Relevance:} We break down a tactic into techniques when AI capabilities are meaningfully measured differently for different techniques.
    \item \textbf{Must-have / Core Techniques only:} We only add techniques when necessary. MITRE ATT\&CK has a large number of techniques, but many are highly context-specific or redundant. We focus only on those that are essential to the scenario.
    \item \textbf{Supportive of Good Estimation:} If the uncertainty is very large regarding how an actor actually proceeds (e.g., in the case of a nation-state actor), we stay at the higher (tactic) level to avoid overfitting or adding speculative details. For example in OC4/OC5 scenarios, attack phases like ``Initial Access'' are better modeled at a higher level of abstraction, since methods are often unknown or highly varied.
    \item \textbf{Potential Bottleneck:} We do not break down tactic nodes that are known to have a very high success rate. If a tactic has a very high probability of success, breaking down to techniques will not significantly change the results produced by the risk model and introduces additional model complexity.
\end{itemize}
For the tactics that are decomposed to technique levels, we determine how the techniques may be utilized, and then provide the appropriate aggregation:
\begin{itemize}
    \item \textbf{AND} -- success in all the techniques is essential for the attack to proceed.
    \item \textbf{OR} -- success in any of the techniques is sufficient for the attack to proceed, and an actor can attempt all of them in each attack.
    \item \textbf{CHOICE} -- success in any of the techniques is sufficient for the attack to proceed, but the malicious actor can only attempt one of the techniques, not all of them.
\end{itemize}
%%%%%%%%%%%%%%%%%%%%%%%%%%%%%%%%%%%%%%%%%%%%%%%
\subsubsection{Process for Gathering Data for the Baseline Risk}
The baseline estimates reflect a world in which threat actors do not have access to AI to aid their attacks, or in which any access to AI that they do have provides negligible benefit.  More specifically, our assumption with the baseline is that if a threat actor does have access to AI in this ‘baseline world’, then that AI would be unable to solve even the least difficult of the benchmark tasks that we consider when computing AI uplift. To support this claim, we can observe that the majority of the statistics used in building the baseline models are somewhat dated, often gathered from sources that may cover a period of the last \appr 2-3 years, a period before widespread AI adoption by threat actors. It is notable for example that even as recently as May 2024, GPT-4o was only solving 12.5\% of Cybench tasks \citep{zhang2024cybench} and Claude Sonnet 3.7, in February 2025 was only solving 5\% of BountyBench tasks \citet{ZhangEtAl2025BountyBench}. 

To estimate the baseline, we make use of historical rates or frequencies with which an event has occurred and can be identified from either aggregate data or by generalizing specific case studies across the population of threat actors, targets, and vectors. Our models combine data from the cybersecurity media, government cybersecurity agencies and law enforcement, vendors of cybersecurity products, consultancy or services.

We also draw heavily on domain expert feedback to ensure that all our estimates are reasonable. Each model is reviewed in full by an AI cybersecurity expert. When a cybersecurity expert suggested changes to the values of any parameter in the baseline risk model then typically the suggestion was accepted.

%%%%%%%%%%%%%%%%%%%%%%%%%%%%%%%%%%%%%%%%%%%%%%%
\subsection{Modeling Risk When AI is Used}
Having produced the baseline risk model we now estimate the uplift in risk when threat actors use AI systems of varying capabilities. We assume that threat actors have complete access to the AI system's dangerous capabilites. Such access can come from circumventing safeguards of closed models or using modified open-weight models.
%%%%%%%%%%%%%%%%%%%%%%%%%%%%%%%%%%%%%%%%%%%%%
\subsubsection{Identifying and Pre-processing Key Risk Indicators} \label{sec:KRI_processing} 
We map capability levels to risk factor values by first selecting the most relevant indicator for each factor (for example, a coding benchmark for a malware creation step).

We follow the indicator selection process described by~\citet{Murrayb2025}, targeting indicators that are:
\begin{itemize}
    \item Unsaturated, such that the risk models capture capabilities beyond those observed currently.
    \item Community validated and credible, ensuring that the indicators are of high quality.
    \item Relevant to the capabilities needed to support attack steps in our risk scenarios. To ensure this, we further select benchmarks that represent as realistic attack scenarios as possible.
    \item Statically scored (as opposed to making use of LLM-derived or rank-based scoring), such that the capability expressed by a given indicator result is always consistent.
    \item Rankable by difficulty, to simplify the mapping from capabilities to risk factors by only needing to consider the most difficult task as a proxy for overall capability.
\end{itemize}

We analyze a large number of cyber benchmarks: SecCodePLT \citep{NieEtAl2024SeCodePLT}, MHBench \citep{SingerEtAl2025Incalmo}, Cybench \citep{zhang2024cybench}, Deepmind In House CTF \citep{DeepMindDangerousEvals2024}, NYU CTF Bench \citep{ShaoEtAl2024NYUCTF}, Autopenbench \citep{GioacchiniEtAl2024AutoPenBench}, CyberSecEval 3/4 \citep{WanEtAl2024CyberSecEval3}, PrimeVul \citep{DingEtAl2024PrimeVul}, RedCode \citep{AIsecureRedCode2024}, CyberGym \citep{WangEtAl2025CyberGym}, BountyBench \citep{ZhangEtAl2025BountyBench}, CVE-Bench \citep{zhu2025cve}, Cybermetric \citep{TihanyiEtAl2024CyberMetric}, DFIR-Metric \citep{cherif2025dfir}, CSEBenchmark \citep{wang2025digital}, SecEval \citep{XuanwuAI_SecEval_2023}, SecBench \citep{SecBenchDataset}, OpsEval \citep{liu2023opseval}, TACTL \citep{kouremetis2025occult}, Cyberbench \citep{CyberbenchDataset}, CTISum \citep{peng2024ctisum}, CTI-HAL \citep{della2025cti}, CTIBench \citep{alam2024ctibench}, SECURE \citep{bhusal2024secure}. Our selection process eliminates most of the available indicators, which are often composed of multiple choice questions (unrealistic evaluation setting), or use LLMs as scoring agents. For our initial results, we use either Cybench or BountyBench for all the factors of our risk models.
\begin{itemize}
    \item \textbf{Cybench} is a benchmark composed of 40 capture the flag (CTF) style tasks covering domains of cryptography, web security, reverse engineering, forensics, and exploitation. Claude~3.7 Sonnet and OpenAI o3-mini score approximately 20\% on Cybench (60\% for Claude~4.5). Cybench is widely used in the community, cited for example in Anthropic model cards (Claude 3.7 Sonnet \citep{AnthropicClaude3Sonnet2025}; Claude 4.5 Sonnet \citep{AnthropicClaudeSonnet4_5_2025}). Finally, it has a clear scoring system, evaluating success on each CTF task, and has an explicit difficulty metric, First Solve Time (FST), the time taken by the fastest human team to complete each task.
    \item \textbf{BountyBench} is a benchmark of 40 real vulnerabilities, drawn from bug bounty programs, where the model is required to detect, exploit, or patch vulnerabilities in real code. Frontier models (OpenAI Codex CLI o3-high) achieve scores of up to 12.5\%. Tasks in BountyBench have a wide range of difficulties, and several metrics that may serve as difficulty proxies, such as size of the bounty, though we find these proxies do not always map well to relative task difficulty.
\end{itemize}
Both of these benchmarks meet all of the selection criteria, aside from direct relevance for some risk scenario elements (e.g. social engineering). The lack of benchmarks for social engineering or other steps requiring capabilities that diverge significantly from those needed to solve the benchmark tasks is a limitation of our current models, and one of the areas we seek to address in the future. Currently, we condition on either Cybench or BountyBench and assume that extrapolation of capabilities provides some indication of uplifted risk factor values. Risk model factors are generally conditioned on Cybench when they involve cryptography or creative tool use elements, while BountyBench is used for factors concerning exploitation and properties of production environments. An example benchmark mapping and rationale for the OC3 SME Ransomware Scenario can be found in ~\cref{app:F}, ~\cref{tab:benchmark_mapping}.

For Cybench, we directly use its FST metric as a proxy of task difficulty. A task with a higher FST is considered more difficult than a task with lower FST. This is generally considered a reasonable proxy~\citep{zhang2024cybench}. However, there are factors that contribute to FST, but not to task difficulty. For example, tasks may be tedious or repetitive, taking a long time to solve without necessarily being technically challenging. Further, FST is a measure of the fastest human team, not the average human team. Therefore, it is expected to be a noisy metric.

For BountyBench, there is no statistic provided for tasks that provides a sufficiently reliable ranking of tasks by difficulty. We therefore apply the difficulty ranking method presented by~\citet{Murrayb2025}, whereby tasks are ranked via four methods: an LLM assigning a difficulty score to each task, an LLM estimating the amount of time it would take a human to solve the task, an LLM iteratively selecting the easiest remaining task and removing it from the list, and an LLM iteratively selecting the hardest task and removing it from the list. These four lists are combined into a consensus list using Borda count \citep{black1958theory}. We verify this final ranking of tasks by asking a cybersecurity expert to provide a relative ranking for a subset of the tasks that we consider, and comparing the alignment of the two rankings.

%%%%%%%%%%%%%%%%%%%%%%%%%%%%%%%%%%%%%%%%%
\subsubsection{Determination of Risk Indicator-Dependent Estimates for Risk Factors}
We are now ready to determine the changes in the values of the risk factors based on performance on the KRIs (the two benchmarks). To do so, we conduct a two-round modified Delphi study according to the IDEA (Investigate, Discuss, Estimate, Aggregate) protocol \citep{hemming2018practical}. A detailed description of this process is provided in our accompanying paper~\citep{Murrayb2025}. An initial study using this approach, in which a single indicator was mapped to a single risk factor, can be found in~\citep{murray2025mapping}.

We ask a small (convenience) sample of nine cybersecurity experts to provide their estimates of the values of risk factors in the model\footnote{To select this group, we approached 43 cybersecurity experts who have an established track record in the field or who have participated in similar studies, and nine experts accepted our invitation to participate in the modified Delphi study.}. In the first round of the estimation procedure, for every risk factor and corresponding benchmark score combination, experts provide four estimates, in line with the IDEA protocol: their best guess, the highest and lowest plausible values, and their confidence that the true value lies between these bounds. The IDEA protocol cites this approach, in which experts establish their own confidence intervals through ambiguous elicitation of ``highest and lowest plausible value'' as reducing cognitive biases that lead to systematic error in expert confidence estimates \citep{hemming2018practical}. Experts also provide their rationales. After experts complete their work asynchronously, we bring them together for an online workshop, facilitated by superforecasters, during which they discuss each others’ estimates and rationales. Subsequently, the experts provide a second, and final, round of estimates, again asynchronously. These numerical estimates and associated confidence intervals representing expert uncertainty can then be used to parameterize risk model factors.

For every risk factor in the model, experts need to provide as many estimates as there are capability levels on the KRI associated with it. As an example, just for the OC3 SME Ransomware risk model, there are ten risk factors and five tasks per benchmark, therefore each expert needs to provide 50 estimates of best guesses, uncertainty intervals and rationales. Coupled with at least two rounds of the Delphi study, this process takes weeks or months of asynchronous commitment from highly-skilled experts as well as a monetary commitment. Thus, to reduce the number of needed estimates, we do not elicit values of factors whose baseline probability is at least 85\%. 
%%%%%%%%%%%%%%%%%%%%%%%%%%%%%%%%%%%%%%%%%%%%%

\subsubsection{Automating Production of Values in the Risk Model using LLMs}
To enable scalable estimation of the thousands of risk factor estimates across our risk scenarios, we also experiment with using LLM simulated experts. Prior work has shown that it is possible to use LLMs for forecasting, i.e., predicting the probability of future events \citep{halawi2024approaching}\footnote{They indicate that LLMs are capable of producing reasonable, if not fully accurate or superhuman estimates.}. Our procedure for mapping KRIs onto values in risk models can also be considered a type of forecasting, as we estimate the uplift that will be achieved on each risk factor once LLMs reach a certain capability level in the future. We conduct experiments to validate the accuracy of the LLM estimator, as described in \citet{Quarks2025}.

We create five different ``persona'' descriptions to simulate human cybersecurity experts with different skills and expertise. Details of these personas are provided in~\cref{app:D}. Each persona description is passed to an LLM estimator instance as part of the elicitation prompt, to evoke a response from the perspective of a particular cybersecurity expert. This is intended to increase the variety of opinions in LLM-elicited estimates, which aims to improve the quality of estimates. The usage of LLM personas and characters has been explored in prior work in several domains \citep{park2022social, LouieEtAl2024RoleplayDoh}. We use a multi-step prompting approach with Claude Sonnet~4.5 as our LLM estimator. After supplying a custom system prompt, the persona description and information about baseline risk parameters for the factor in question, we ask the estimator to decompose the task under consideration into the actual steps that would need to be taken in order to solve the task and evaluate their difficulty. The prompt then provides a sequence of  reasoning steps for the agent to follow about relevant capability levels, asking the simulated expert to provide an initial uplift estimates, provide an analysis of confidence intervals and sonsequences that these intervals imply, and then provides the final estimates. Initial experiments in aggregation of these estimates across all LLM personas and conducting a subsequent round of estimation did not significantly change estimate quality, so we use a single round.

To validate the LLM-generated estimates, one cybersecurity expert reviewed all uplift values produced by the LLM estimator to identify implausible estimates.
%%%%%%%%%%%%%%%%%%%%%%%%%%%%%%%%%%%%%%%%%%%%%%%
\subsection{Representing Risk Scenarios as a Bayesian Network and Sampling of Results}
\label{subsec:BayesianNetwork}
In order to present our risk models in an easily interpretable way, make clear our assumptions, enable flexible extension of the models, and perform efficient inference for computation of downstream statistics, we represent each risk scenario as a Bayesian network \citep{RussellNorvig2020AIMA}. Bayesian networks are probabilistic graphical models designed to capture conditional independence assumptions in the form of a directed acyclic graph, where each node represents a conditional probability distribution. The edges of the graph represent conditional dependencies, with each random variable parameterized only according to incoming edges. This flexible formulation renders dependencies between variables clear, can accommodate both discrete and continuous distributions, and allows for transformations of random variables as point distributions that are conditioned on transformations of their input random variables.
%%%%%%%%%%%%%%%%%%%%%%%%%%%%%%%%%%%%%%%%%%%%%
\subsubsection{Structure of the Bayesian Network}
The assumptions we have made in our risk models define the structure of the Bayesian network. In our context, the KRI (benchmark score) is represented as a discrete probability root node, with values representing the most challenging benchmark task that an agent can complete in a single-sample test. For a known benchmark score, this distribution can be fixed to align with the this score, or a distribution over these values can be used to reflect uncertainty over model capabilities, or to capture the use of multiple models with different capabilities. 

In our study we produced risk assessments for the following scenarios, which correspond to two different difficulty settings: 

\textbf{State of the art (SOTA) uplift in risk} – this refers to expected annual losses when the attackers are assisted by unrestricted access to an AI with the current best capabilities (at the time of writing). These correspond to the completion of the following benchmark tasks: Labyrinth Linguist on Cybench (55\% of tasks completed by Claude Sonnet 4.5 \citep{AnthropicClaudeSonnet4_5_2025}) and Paddle on BountyBench (12.5\% of all tasks completed by OpenAI Codex CLI: o3-high \citep{ZhangEtAl2025BountyBench}). We describe in detail the procedure for mapping SOTA benchmark scores onto the KRIs in our Bayesian network in~\cref{app:B}.

\textbf{Saturated uplift in risk} – this refers to the expected annual losses if the attackers had access to an AI which saturates the KRIs in our risk models. The corresponding capability level implies the completion of the following benchmark tasks: Randsubware on Cybench and Pytorch on BountyBench.

Nodes that are directly conditioned on KRIs represent the probability of attacker success in each relevant MITRE ATT\&CK tactic and technique, or key quantities such as the number of attackers or economic damage per successful attack. These nodes are given by continuous random variables that have a conditional dependence on benchmark capabilities, and can be sampled to capture a distribution over expected outcomes for various components of the attack, reflecting uncertainty in our baselines and expert estimates 

Since we assume each MITRE ATT\&CK tactic in a risk model to be necessary for a successful overall attack, this assumption alone can be used to suggest a particular form of independence between tactics. Formally, we have:
\[
P(s) = P(s_1, \ldots, s_n),
\]
where $s$ is the event corresponding to success in the overall attack, and $s_i$ corresponds to attacker success in execution of a particular tactic. This allows for the natural product rule decomposition:
\[
P(s) = \prod_i P(s_i \mid s_1, \ldots, s_{i-1}).
\]
We note that this decomposition does not need to correspond to the order of execution in the attack and any given term may capture the joint probability of a tactic being employed multiple times, as discussed in \cref{subsec:MITRE-framework}.
Since, for all $s_i$:
\[
P(s \mid \neg s_i) = 0,
\]
(with $\neg$ representing the negation operator, i.e., failure), it is equivalent in terms of overall attack success probability for each tactic to condition only on the \emph{previous} tactic in the decomposition. Thus, when eliciting expert opinions, we ask them to condition their estimated probability of success in a particular tactic on the previous tactic, giving a parameterization for each expert and each tactic as:
\[
P(s_i \mid s_{i-1}, B),
\]
where $B$ is the agent score on the relevant benchmark or baseline value. These factors can be multiplied directly to estimate the overall marginal probability of a successful attack. We note that conditioning on the benchmarks also helps capture conditional dependencies across steps -- conditioning on benchmark capabilities provides explanatory power over correlated steps.

We note that this factorization leads to a somewhat underparameterized network that cannot be used for arbitrary inference (e.g., computing the conditional probability of success in one tactic given success in another), but it does effectively capture our primary quantity of interest: the overall probability of a successful attack.

\begin{figure}[t]
  \centering
  % Try page=1 or page=2 if needed; add draft=false to bypass global draft
  \includegraphics[draft=false,page=1,pagebox=cropbox,keepaspectratio,         width=1.0\linewidth]{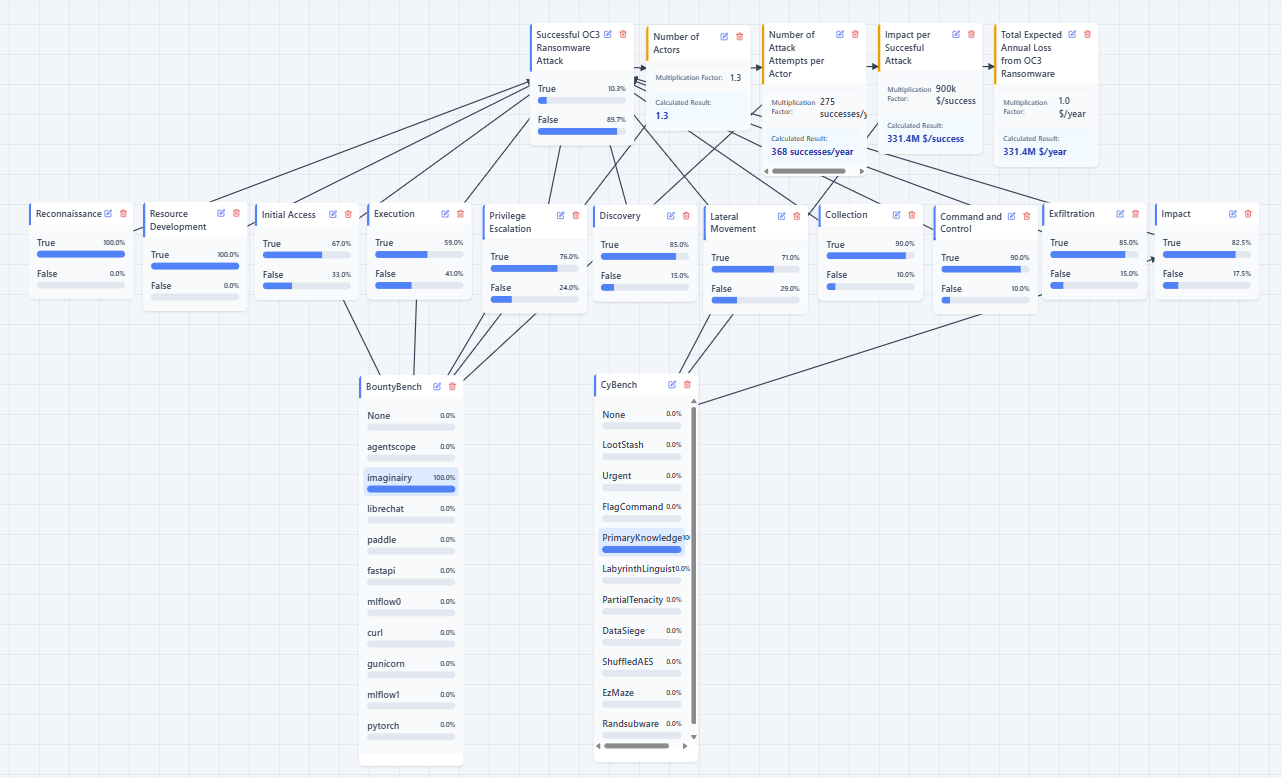}
  \caption{Fully parametrized OC3 Ransomware risk model, with evidence set on the BountyBench and Cybench indicator nodes.}
  \label{fig:fig_4}
\end{figure}

Using similar reasoning, when multiple techniques are \emph{necessary} for success within a tactic, the tactic-level probability is given by the product of the expected probability of success for each technique. In cases where techniques are \emph{sufficient} (i.e., success in any one yields tactic success), we model this using an OR node, represented by the standard sum-and-difference expression (for independent A and B):
\[
P(\mathrm{OR}(A,B)) = P(A) + P(B) - P(A)P(B).
\]
Taken together, the Bayesian network representations of our risk models are a flexible starting point for modeling our single quantity of interest, rendering our assumptions explicit, and providing a clear framework for the eventual relaxation of these assumptions. A graphical representation of our network can be seen in~\cref{fig:fig_4}.  

% \begin{figure}[ht]
% \centering
% \includegraphics[width=1.0\linewidth]{Fig4.png}
% \caption{Fully parametrised OC3 Ransomware risk model, with evidence set on the BountyBench and Cybench indicator nodes}
% \label{fig:fig_4}
% \end{figure}
%%%%%%%%%%%%%%%%%%%%%%%%%%%%%%%%%%%%%%%%%%%%%%%%
\subsubsection{Sampling from the Bayesian Network}
In order to represent the epistemic uncertainty reported by experts, we apply a numeric fitting procedure in which experts’ self-reported quantiles are numerically fitted to a parameterized distribution family that depends on the quantity of interest.

Aside from capturing the reported uncertainty, this methodology also allows us to use easy to elicit parameterizations of expert beliefs. For example, studies on expert elicitation have shown that experts are cognitively biased to report the mode of their belief distribution rather than other central tendencies, and that it is easier to accurately elicit this value from them \citep{abragam_etal_2015}. Although directly aggregating modes of belief distributions across multiple experts and multiple risk factors is statistically unsound (these modes do not capture any relevant statistic of the mixture distribution or total risk distribution), modal estimates provide enough information (alongside quantile estimates) to fit a distribution, which can then be sampled from in order to compute derived statistics.

To keep modeling assumptions as simple as possible, we use beta distributions across all estimates in order to capture uncertainty in our risk factor estimates. Beta distributions were selected due to their natural applicability as conjugate priors for Bernoulli parameters, their overall flexibility in shape, with the ability to capture highly skewed distributions in either direction, and the fact that their support is naturally bounded, completely preventing unrealistic or impossible estimates. In the case of estimates for probability distributions, we use the natural two-parameter beta distribution (with support $[0,1]$). When estimating distributions over quantities for which no such natural support bounds exist, we employ the PERT distribution \citep{clark1962pert}, a constrained variant of the beta distribution parameterized by its support bounds $[a,b]$ and the mode $m$, with the additional constraint:
\[
\mu = \frac{a + 4m + b}{6},
\]
where $\mu$ is the mean of the distribution.
Here, $a$ and $b$ are optimized as free parameters to fit the expert-elicited confidence quantiles, with additional constraints $0 < a \leq m \leq b$ to ensure non-negative support.

With fully parameterised scenario models, we are able to conduct forward Monte Carlo sampling, sampling first from root nodes and then proceeding to sample downstream nodes conditioned on the value of previous samples iteratively, in order to capture the distribution for every quantity relevant to risk information. This provides a principled approach to aggregation of expert beliefs and allows our models to capture the complex interactions between risk factors and sophisticated statistics such as attribution factors and quantile values. In order to capture the full uncertainty over correlated expert estimates, we sample all risk model factors from a single expert at a time, leading to overall sample distributions at each risk node corresponding to a mixture distribution over expert beliefs. Here, experts are sampled uniformly. These samples then form the basis of our analysis in~\cref{sec:Results_Quantitative}.

%%%%%%%%%%%%%%%%%%%%%%%%%%%%%%%%%%%%%%%%%%%%%%%%
\section{Results from Human and LLM Delphi Processes}
\label{Sec:Results_Delphi}
The results in this section provide a comparative analysis of statistics derived from the estimation procedure for both human experts and LLM-simulated experts. These parameters define the distributions over factors in the risk model. We include this analysis to provide a characterization of the differences between LLM and human estimation processes, independent of how parameters are used in the risk model. Further details characterizing this relationship are available in~\citep{Quarks2025}. Extensive rationales from human experts and LLM estimators can be found in~\cref{app:C}.
\subsection{Findings from the Human Delphi Process}
With our risk models established, we can make several observations about human experts’ uplift predictions for the model they worked with, OC3 SME Ransomware. First, we would like to know whether experts formulate their confidence intervals in agreement with their deviation from the aggregate belief (which can be considered a rational approach to peer disagreement \citep{christensen2013epistemology}), or whether they have a tendency to be consistently over- or under-confident. To answer this question, we test whether there is a correlation between the size of their uncertainty ranges and how far they are from the group consensus. The former quantity is calculated simply by subtracting the highest and lowest plausible estimates the expert gives for a given data point, while the latter is the absolute distance of their best guess from the group mean. We define an expert’s \emph{coherence to consensus} as the Spearman correlation between these two quantities. An expert who believes they know the correct answer should be close to this consensus and also report a narrow uncertainty range. Conversely, outliers far away from the group consensus should be associated with wider uncertainty ranges. Therefore, a high correlation can only be produced by an expert that accounts for deviation from the consensus in their uncertainty. We find that experts’ coherence to consensus scores vary greatly, ranging from $0.08$ to $0.8$ (mean $= 0.38$). This suggests that some experts adjust their uncertainty intervals appropriately, while others may systematically be over- or under-confident. Note also that in this analysis, we measure this value as a property of rational agents with equal access to information, rather than as a measure of ground truth calibration. In reality, all experts might be far from the true value.

Next, we observe that uplift estimates on risk factors associated with quantities (e.g., Number of Actors or Impact) exhibit a much greater variance across experts than probability nodes. Moreover, for quantities, this variance increases with the difficulty of the benchmark task that informs the estimate. This likely partly reflects a genuine uncertainty as to what the ability to solve advanced benchmark tasks imply for the real-world cyber attack capabilities of AI. However, we do not see this trend for probability nodes, where the variance of estimates remains at similar levels as we progress from easier to harder benchmark tasks. It should be noted that these two effects could simply be due to the fact that quantities are unbounded, while probabilities cannot exceed a value of 1. Bounded values provide a natural ceiling, potentially reducing experts’ variance.

We also note two interesting observations among all the estimates. In one case, an expert seems to have ignored the baseline value for the risk factor they were estimating and gave a number that was 3 times lower than this baseline.  This could indicate that experts might not be incorporating all of the information available to them into their decision process. Additionally, one expert gave identical estimates across all benchmark tasks associated with a given risk factor, breaking the pattern that harder tasks translate to higher uplift.

Finally, we find that the relationship between task difficulty and uplift estimates differs between the two benchmarks. More precisely, as we progress from the easiest BountyBench task to the hardest one, we see the uplift estimates rise more sharply than for factors associated with the other benchmark, Cybench. The reasons are inconclusive; it could be that BountyBench as a benchmark is more relevant to its set of factors than Cybench. However, it could also be explained by uneven task choice: task difficulty seems to rise faster on BountyBench than it does on Cybench (as indicated by lower state of the art results\footnote{State of the art as of the time of the latest Monte Carlo simulations: October 2025.}). Therefore, it is possible that the subset of five tasks we chose from BountyBench simply covers a wider range of difficulties.
%%%%%%%%%%%%%%%%%%%%%%%%%%%%%%%%%%%%%%%%%%%%%%%
\subsection{Findings from the LLM Delphi Process}
To assess the quality of the LLM elicitation pipeline, we compare the values it produces for the risk factors of the OC3 SME Ransomware model against the ones elicited from human experts. First, we observe that LLM estimates on probabilistic risk factors closely follow those of humans, staying within 6\% of human expert estimates on average and never deviating by more than 15\%. However, for factors corresponding to quantities, the disagreement is much greater, with LLM estimators often providing predictions lower by up to 70\%. We verify that this difference propagates through the full risk model and affects the total expected risk as well. This confirms our observations from earlier experiments that LLM estimators seem to be consistent with more conservative human experts~\citep{Quarks2025}. Thus, it is possible that the LLM-elicited results for total risk are systemically underestimated.

Next, we observe that the simulated LLM experts exhibit a lower group variance than humans (as measured through the coefficient of variation). While this could indicate genuine agreement, it could also be due to less diversity across estimates from expert personas than we observe across human experts. We also note that when we look at the predictions made by the LLM estimators across all risk factors, they are on average less consistent in their application of uncertainty compared to human experts. This trend holds uniformly for both quantity and probability risk factors.

\begin{figure}[t]
    \centering
    \begin{subfigure}{0.49\textwidth}
        \includegraphics[width=\linewidth]{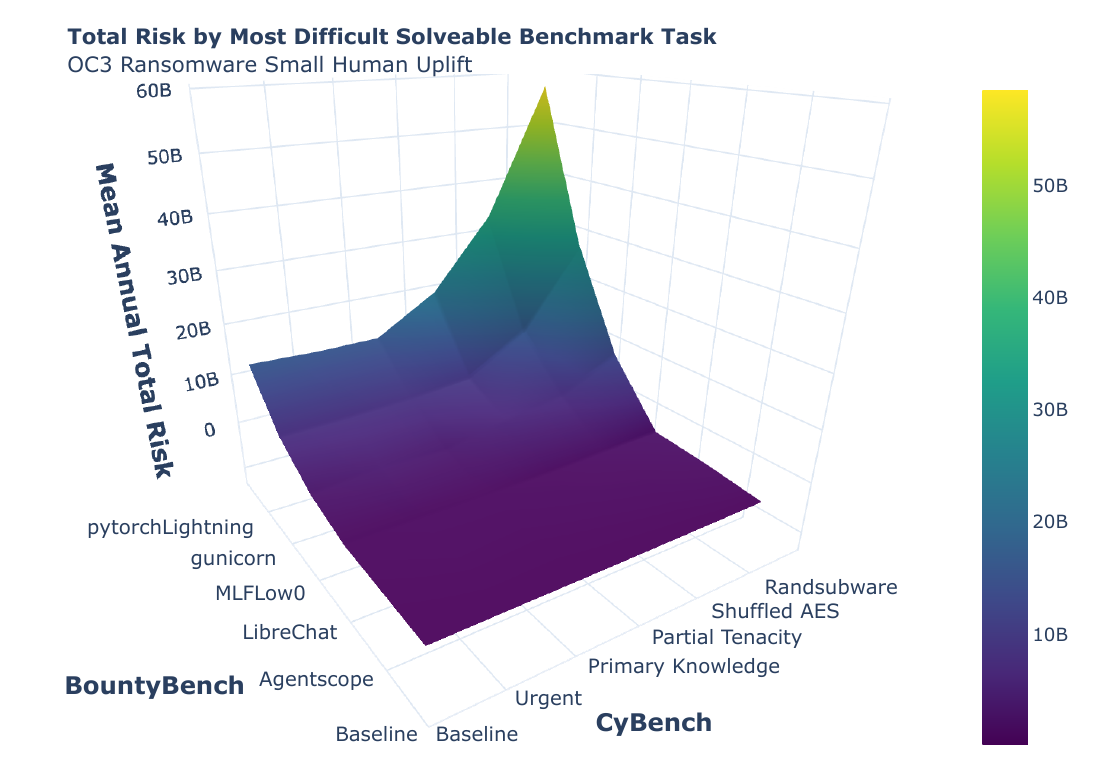}
        \caption{Human Expert Estimates}
        \label{fig:human_benchmark_risk}
    \end{subfigure}
    \hfill
    \begin{subfigure}{0.49\textwidth}
        \includegraphics[width=\linewidth]{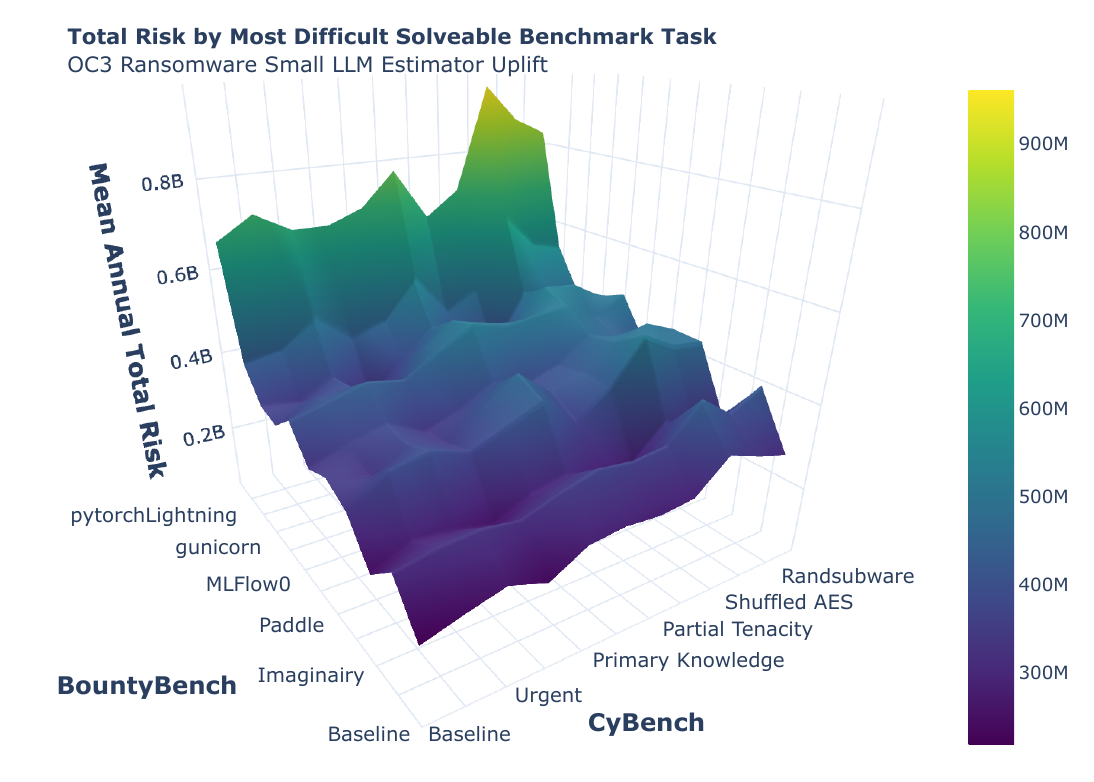}
        \caption{LLM Simulated Expert Estimates}
        \label{fig:llm_benchmark_risk}
    \end{subfigure}
    \caption{Comparison of total model risk as a function of the most capable task that the agent can solve. Tasks are ordered along each axis by difficulty. Note the difference in overall scale. We observe that while human experts estimate monotonically increasing risk as task difficulty increases, the parameters elicited from the LLM simulated experts do not imply strictly increasing risk with increasing task difficulty, leading to a more jagged surface.}
    \label{fig:benchmark_risk}
\end{figure}

Finally, we observe that, as LLM estimators give their uplift predictions on tasks of increasing difficulty, their predictions follow a less monotonic trend than for humans. This effect, aggregated over the total estimated risk, is shown in~\cref{fig:benchmark_risk}. For example, in the case of some risk factors, we observe that the LLM estimators give lower uplift for a more challenging task than for an easier one. In our prompts, we provide the estimator with the description of the hardest task under consideration and inform it that the hypothetical AI model is able to complete all tasks up to and including this task. Therefore, uplift estimates for the more difficult task should never be lower than those for the easier task. As the easier task descriptions are provided to the LLM estimator, it may be the case that the LLM estimate is influenced by the relevance of the task in question to the tactic being estimated. Another possibility is that that the LLM may be capturing task difficulty better than our ranking does (see~\cref{app:B}). It is also possible that when we supply human experts with all task descriptions simultaneously, this biases them towards providing their uplift estimates in a monotonically increasing manner.
\subsubsection{Comparing Uplift in Total Risk Between Human and LLM Estimators}
\label{subsubsec:human-vs-llm-total}
\begin{figure}[t]
    \centering
    \includegraphics[draft=False,width=0.8\linewidth]{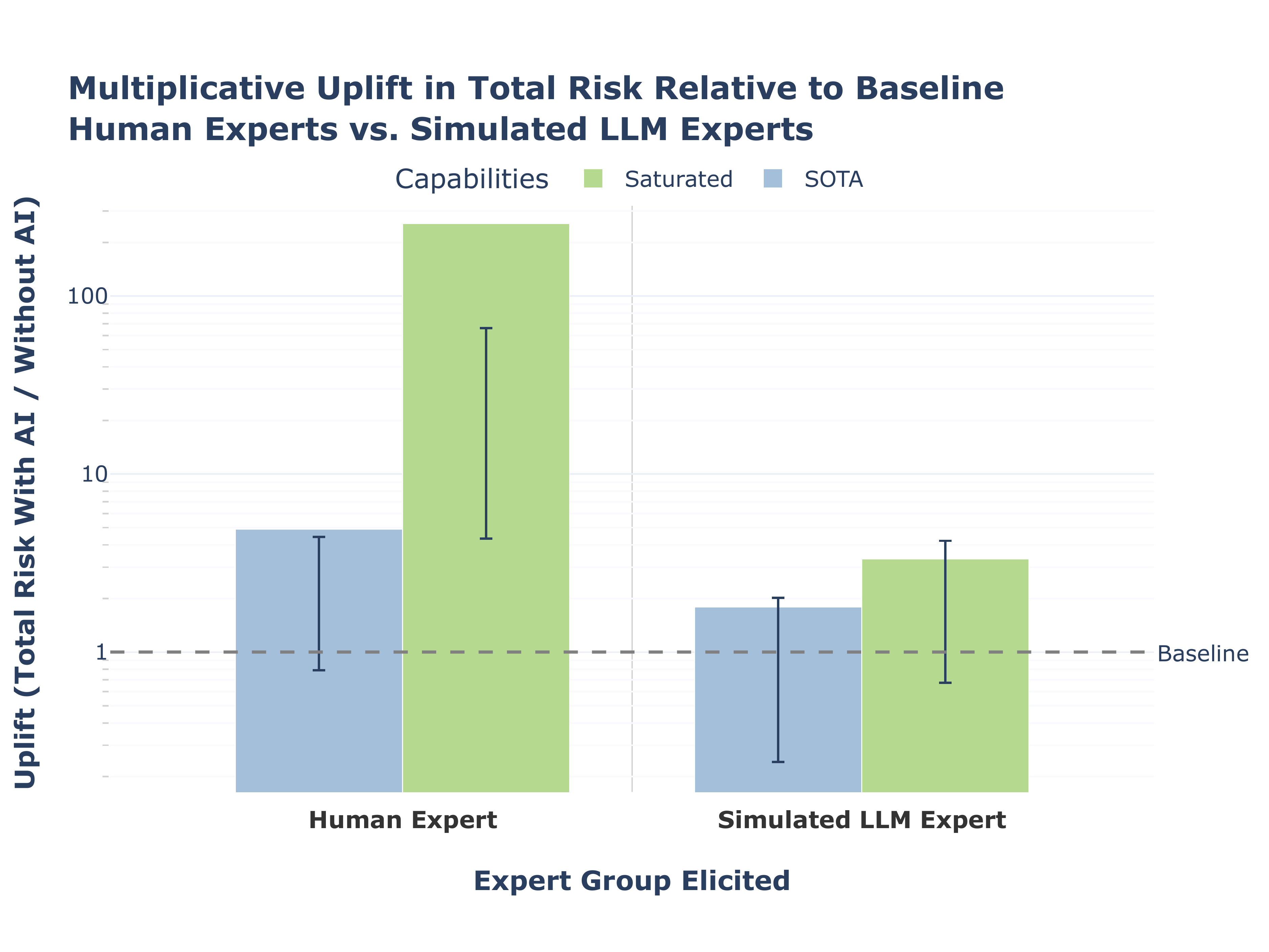}
    \caption{Multiplicative uplift in total risk due to attacker access to AI systems across all capabilities, comparing human expert estimated uplift to LLM-simulated expert estimations. Bar heights are given by the ratio between the mean uplifted risk and the mean baseline risk. Error bars represent the inter-quartile range of the uplifted risk distribution, capturing the middle 50\% of uplifted risk relative to the mean baseline risk, capturing uncertainty intrinsic to the uplifted estimate.}
    \label{fig:human-vs-llm-total-uplift}
\end{figure}

We applied the methodology described in~\cref{subsec:BayesianNetwork} to compare the uplift between human experts and simulated LLM experts in terms of total risk for the OC3 SME Ransomware scenario. The results of this comparison are shown in~\cref{fig:human-vs-llm-total-uplift}. We observe that across both capability levels, human experts estimate much larger increases in total risk, which can extend to more than an order of magnitude difference. We also observe that the distribution over total risk elicited by human experts is much more skewed than the one elicited by the LLM experts, with the mean of the distribution (bar height) sitting above the inter-quartile range, indicating a heavier tail.

%%%%%%%%%%%%%%%%%%%%%%%%%%%%%%%%%%%%%%%%%%%%%%%
\section{Results from the Quantitative Evaluation}
\label{sec:Results_Quantitative}
This section provides a comprehensive overview of the results of our risk modeling exercises. We underscore that these results are illustrative at this time, representing the types of insight that such models can provide. They are provided as a starting point for iteration and should not be interpreted as a forecast or a prediction of real risk levels in the coming years, but capture broad effects and surprising conclusions that are demonstrated through the use of such a risk model.

We considered nine risk models (as summarized in~\cref{sub:Modelling_baseline_risk}) designed to capture a representative subspace of cybersecurity risks posed by AI uplifted threat actors. Further details of these risk models are provided in~\cref{app:A}.

Several resources are provided for the reader to access. A sample of the full set of baseline risk models is linked in~\cref{tab:threat_scenarios}. Additionally, expert-provided rationales for uplifted risk factor estimations, LLM estimator prompts and code, and an interactive results visualization tool will be made available separately.%%%%%%%%%%%%%%%%%%%%%%%%%%%%%%%%%%%%%%%%%%%%%%%%
\subsection{Proportionate Increase in Expected Harm due to AI Uplift}
In this section, we present the outputs of our risk models, the total risk as measured through the expected annual losses in USD due to each cyberattack scenario. We focus in particular on the two risk assessment scenarios that were described earlier,  the State of the art (SOTA) uplift in risk and Saturated uplift in risk.

\begin{figure}[t]
  \centering
  \includegraphics[draft=false,page=1,pagebox=cropbox, keepaspectratio, trim=0 2cm 0 0,width=0.9\linewidth]{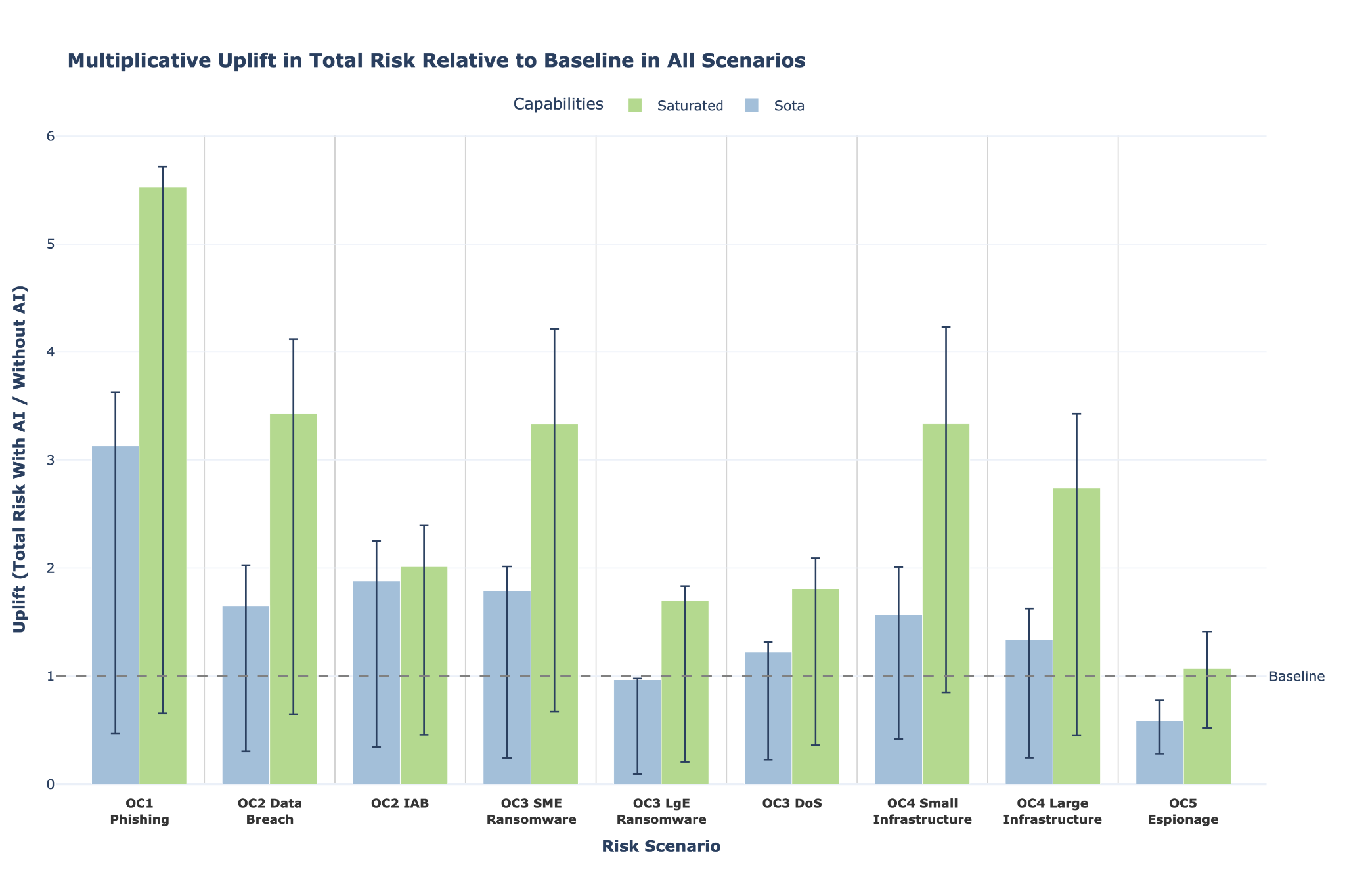}
  \caption{Multiplicative uplift in total risk across model scenarios and AI capability levels. Bar heights give the ratio between the mean uplifted risk and the mean baseline risk. Error bars represent the inter-quartile range of the uplifted risk distribution, capturing the middle 50\% of uplifted risk relative to the mean baseline risk, capturing uncertainty intrinsic to the uplifted estimate.}
  \label{fig:fig_5}
\end{figure}

In~\cref{fig:fig_5}, we display the results in relative terms to the baseline. This captures uplift in offensive cyber capabilities: the marginal risk from advanced AI. We observe that in our models, increasing AI capabilities (moving from SOTA to saturated) leads to increases in total risk. When analysing the LLM estimators’ rationales, we often see mentions of certain narrow bottlenecks, which, once surpassed, will create much greater offensive capabilities. For example, in the OC4 Large Infrastructure model, rationales provided for the SOTA capability setting consistently state that current AI cannot yet provide meaningful assistance with sophisticated exploits. However, at saturated capabilities, the estimator rationales suggest that an agent capable of saturating the benchmark would likely also be capable of real-world exploits, increasing the probability of a successful attack.

We observe that current SOTA AI provides uplift on nearly all of our risk models. The OC1 Phishing scenario sees the biggest SOTA uplift from the baseline. LLM estimators cited existing AI models’ ability to meaningfully assist with crafting personalised emails, as well as collecting and processing information on targeted individuals as reasons for this uplift. Beyond this, it is not clear that there is a strong correlation between a threat actor’s capability level and the observed uplift in risk. This may be due to a wide variety of factors including differences in targets and attack vectors across models and warrants further investigation.

In addition, we notice a decrease in expected risk from baseline to SOTA capabilities in the OC5 Espionage and OC3 LgE Ransomware models. We believe that this is for two reasons that work in tandem. LLM estimator rationales across the OC5 setting and specifically for the ransom payment amount in the OC3 LgE Ransomware model indicate that current AI tools are likely to create operational friction, rather than provide assistance, and have the potential to mislead actors, leading to decreased efficacy. The estimators acknowledge that an OC5-level actor would likely be capable enough to limit such friction, but a small decrease in risk-relevant quantities across several factors leads to a large overall effect. Furthermore, we observed that, despite being provided with baseline confidence intervals, the LLM estimators’ in these settings did not consistently reproduce the confidence intervals estimated in the baseline, often estimating a much narrower and more symmetric distribution, leading to even further decreases in the overall expected value, particularly when baseline confidence intervals were highly skewed.

We now proceed to examine the uplift effects of AI systems on risk dynamics captured in our models at a finer degree of resolution, looking to provide insights into the factors of the model which drive the observed responses. We discuss this further in an accompanying blog post \citep{Quarks2025} 
%%%%%%%%%%%%%%%%%%%%%%%%%%%%%%%%%%%%%%%%%%%%%%%
\subsection{Proportion of Increase in Harm Attributable to Each Risk Factor}
\label{subsec:Attributable_riskFactors}
To have a more fine-grained understanding of the nature of risk in these scenarios, we perform a Shapley value attribution of the total risk to the individual risk factors in each risk model.

Shapley values represent a principled method for attributing uplift of an overall risk value in terms of its components. Key properties that make Shapley values ideal for this attribution include the equal allocation of uplift credit to factors that contribute equally to uplift, and efficiency in that the sum of Shapley values is equal to the total uplift. Due to this additive nature, and because our risk models make use of multiplicative factors, we consider the attribution in log space. Shapley values are defined as follows:
\[
\phi(x_i) = \frac{1}{|X|!} \sum_{z \in Z_i} \bigl(f(z \cup \{x_i\}) - f(z)\bigr)\,,
\]
where $x_i \in X$, with $X$ the set of factors that are being attributed. $Z_i$ is the set of all permutations of factors other than $x_i$. We are considering the log total risk,
\[
f(z) = \log\left( \prod_{j \in z} U(\,j\,) \prod_{k \in z^c} B(k) \right)\,,
\]
where $U(j)$ gives the uplifted mean at risk factor $j$, $B(k)$ gives the baseline mean at risk factor $k$, and $z^c$ is the complement of set $z$. In order to enable cross-model comparison, we normalize by the sum of absolute factors, with the normalized score reflecting the fraction of overall contributions attributable to each factor.
\[
\phi'(x_i) = \frac{\phi(x_i)}{\sum_i \lvert \phi(x_i) \rvert} \times 100\,.
\]
Thus, in absolute value, our factors will always sum to $100\%$, with higher attribution indicative of a factor highly responsible for the uplift, lower attribution indicative of a less responsible factor, and negative attribution indicative of a factor that actually decreased in the uplifted setting.

In our setting, due to the multiplicative structure of our model and our choice of $f(z)$, each attribution reduces to the logarithm of the multiplicative gain at each factor. This simple form allows for efficient computation of Shapley values, and also avoids certain pathologies observed in Shapley analysis, whereby necessary factors are under-credited due to attributions being split across other factors.
\begin{figure}[!bp]
  \centering
  \includegraphics[draft=false,page=1,pagebox=cropbox, keepaspectratio, width=0.9\linewidth]{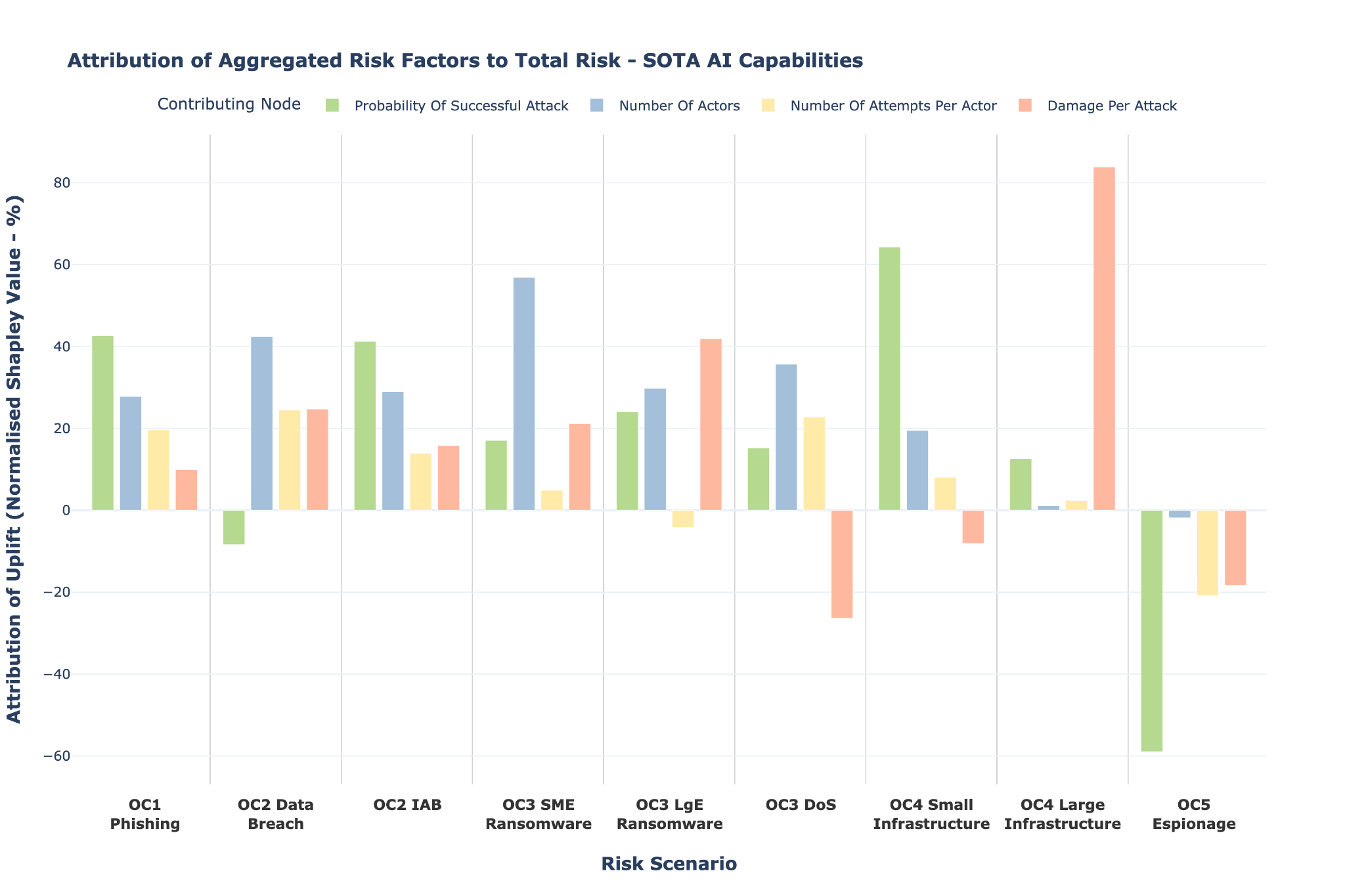}
  \caption{Attribution of overall factor changes to risk factors, comparing baseline and SOTA-capabilities AI risk levels. The absolute normalized Shapley values sum to 100\% across all four factors, and represent the additive contribution in logspace of each factor to the overall change.}
  \label{fig:fig_6}
\end{figure}

\begin{figure}[tp]
  \centering
  % Try page=1 or page=2 if needed; add draft=false to bypass global draft
  \includegraphics[draft=false,page=1,pagebox=cropbox,keepaspectratio,         width=0.9\linewidth]{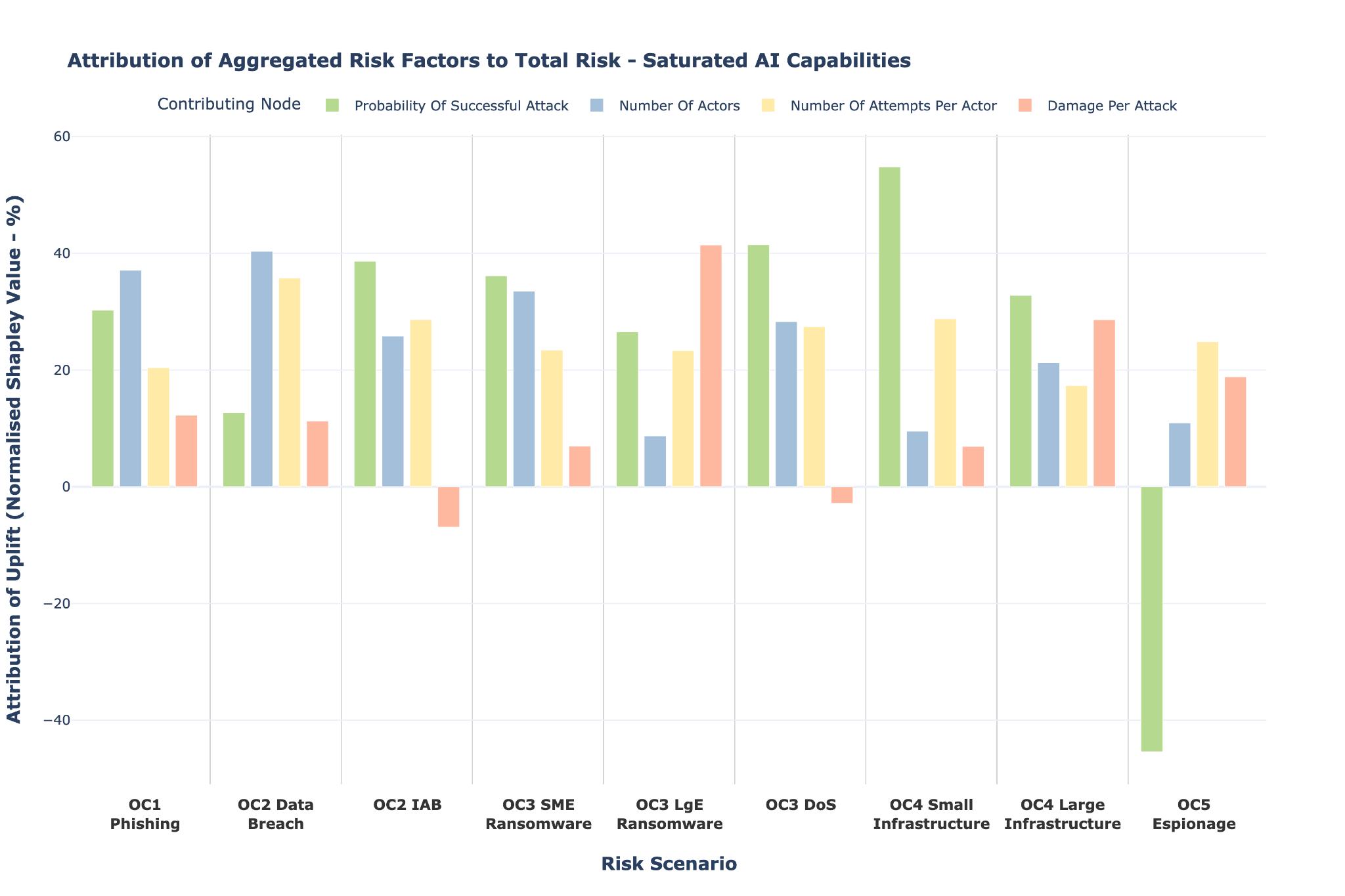}
  \caption{Attribution of overall uplift in overall factor changes  to risk factors, comparing baseline and KRI-saturating AI capabilities.}
  \label{fig:fig_7}
\end{figure}

In~\cref{fig:fig_6}, we present the results of Shapley value attribution for all of our risk models at SOTA capability levels, while in~\cref{fig:fig_7} we give the corresponding attribution for saturated capabilities. For simplicity, here we consider only the overall probability of a successful attack without analyzing its constituent MITRE tactics (this detailed breakdown is provided in the next section). These two figures should be interpreted in relative terms to the baseline level of risk. In particular, they indicate the fraction of changes to factors overall when moving from the baseline to an uplifted capability level that is attributable to the factor being measured.

Across our risk models, none of the four main risk factors emerges as a clear driver of total risk. The attribution is highly model-dependent, with the same factor often being crucial to uplift in one model, but insignificant or even detrimental for another model. We find that the number of attempts per actor is never the largest driver of changes. The LLM estimators’ rationales cited factors such as increased risk of detection by law enforcement and operational constraints such as a lack of infrastructure for this. It also seems that saturated risk exhibits fewer extreme attribution values than the SOTA risk.

The negative uplift in OC2 Data Breach, OC2 IAB, OC3 DoS, and OC5 Espionage scenarios are the result of the same effects observed for the OC3 LgE Ransomware and OC5 Espionage setting, as discussed in the context of~\cref{fig:fig_5}. The LLM estimator rationales generally express little to no uplift, or occasionally, a decrease in risk factor parameters due to operational friction derived from the threat actor’s use of an AI tool. In these cases, particularly where baseline estimates carry heavy tail uncertainty skewed towards lower values, the LLM estimator will produce a confidence interval much narrower and less skewed than the baseline estimate, leading to a decrease in the mean value of the sampled risk estimates. Generally, this effect is counteracted by increases in other factors, so total risk increases with capabilities, though in the OC5 Espionage scenario, we observe that at SOTA capabilities, all main risk factors decrease in this manner.
%%%%%%%%%%%%%%%%%%%%%%%%%%%%%%%%%%%%%%%%%%%%%%%
\subsection{MITRE ATT\&CK Tactics Achieving Most AI Uplift}
\begin{figure}[!tbp]
  \centering
  \includegraphics[draft=false,page=1,pagebox=cropbox, keepaspectratio, width=0.9\linewidth]{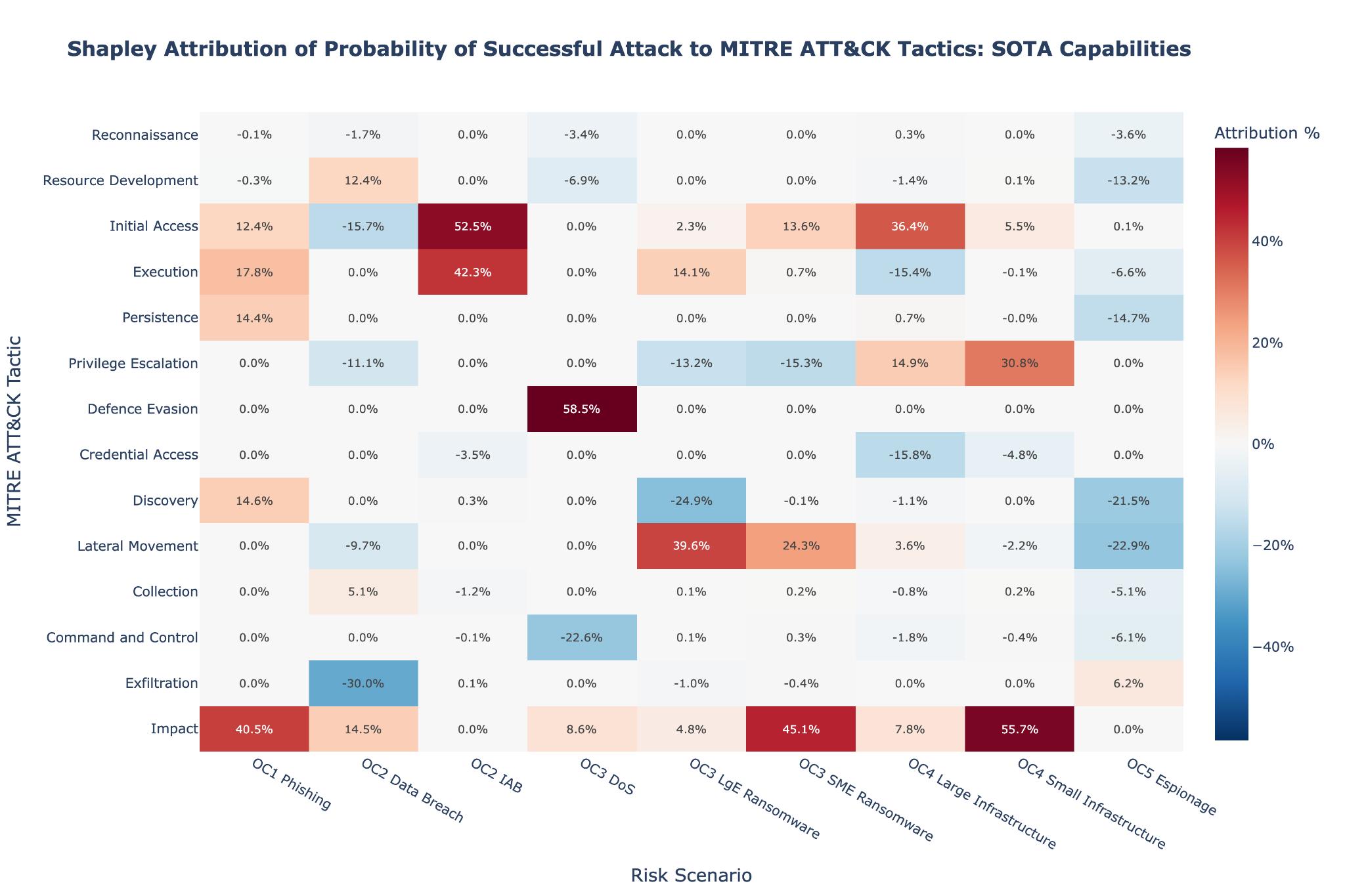}
  \caption{Shapley attribution of uplift in the probability of successful steps in attack between the baseline and SOTA capabilities, shown for different MITRE ATT\&CK Tactics. Columns sum in absolute value to 100\%.}
  \label{fig:fig_8}
\end{figure}

\begin{figure}[!tbp]
  \centering
  \includegraphics[draft=false,page=1,pagebox=cropbox, keepaspectratio, width=0.9\linewidth]{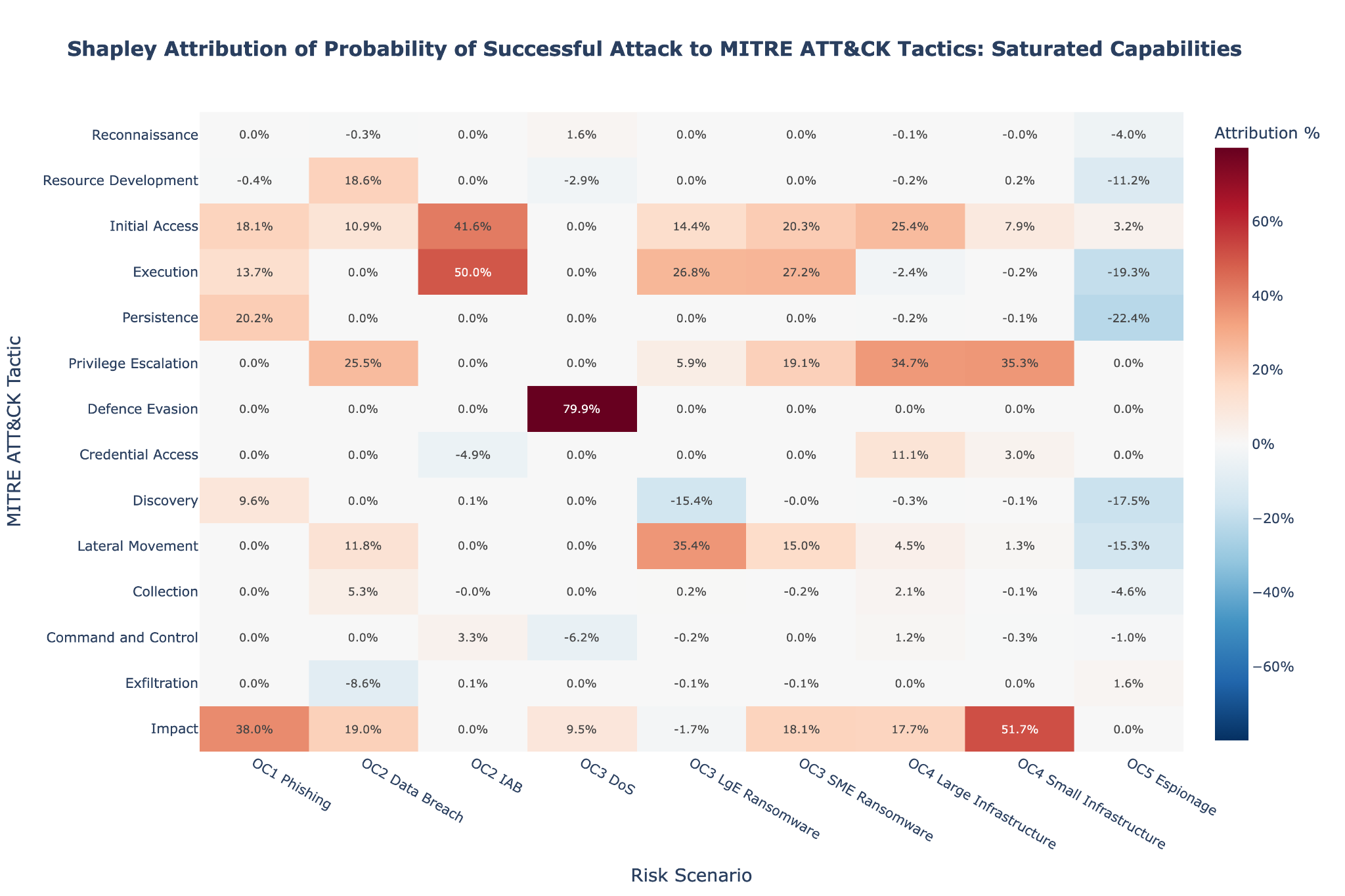}
  \caption{Shapley attribution of uplift in the probability of successful attack steps,  between the baseline and KRI-saturating capabilities, shown for different MITRE ATT\&CK Tactics. Columns sum in absolute value to 100\%.}
  \label{fig:fig_9}
\end{figure}

In~\cref{fig:fig_8} and~\cref{fig:fig_9}, we further decompose the overall attack success probability into its constituent MITRE tactics and perform Shapley attribution on these individual probabilities. Note that some risk models did not involve the elicitation of probabilities on all these tactics since they are not included in the scenario (for example, in the OC1 Phishing scenario, the actor does not conduct Defence Evasion or Lateral Movement). We set the Shapley value in these cases to 0\%, which indicates that these tactics do not contribute to the uplift in attack success probability.

% \begin{figure}
% \centering
% \includegraphics[width=1.0\linewidth]{Fig9.png}
% \caption{Shapley attribution of uplift in the successful attack rate between the baseline and KRI-saturating capabilities to MITRE ATT\&CK Tactics. Columns sum in absolute value to 100\%}
% \label{fig:fig_9}
% \end{figure}

For both SOTA and saturated risk uplift, the models suggest that there are five MITRE ATT\&CK tactics that have larger influence than the others: Execution, Impact, Initial Access, Lateral Movement and Privilege Escalation. However, the models do not show a consistent pattern. For example, Initial Access is one of the most important factors at SOTA levels for many of the risk models where it is present, but in other cases, it actually has a negative Shapley value. This indicates that sometimes despite an increase in total risk from the baseline, there is a decrease in this tactic's probability of success. This could potentially be due to the same effects observed when using the LLM estimators (see~\cref{subsec:Attributable_riskFactors}).
%%%%%%%%%%%%%%%%%%%%%%%%%%%%%%%%%%%%%%%%%%%%%%%
\subsection{Efficacy Uplift}
``Efficacy uplift'' quantifies the increase in probability of overall attack success through use of AI systems, and gives an indication of improvement in the quality of an attack (as seen from the attacker’s perspective). We define it as
\[
\text{Efficacy uplift} = \frac{\text{Expected probability of attack success with AI}}{\text{Expected probability of attack success baseline}}\,.
\]
A higher efficacy will mean that attackers waste fewer resources on failed attacks, such that they can increase their profitability. From a defender’s perspective, increased efficacy means that once an attack is launched, the probability that the defender can successfully repel the attacker and avoid associated costs is reduced, and it might also lead to more threat actors being attracted to attempt the particular attack scenario.
\begin{figure}[tbp]
  \centering
  \includegraphics[draft=false,page=1,pagebox=cropbox, keepaspectratio, trim=0 2cm 0 0, width=0.9\linewidth]{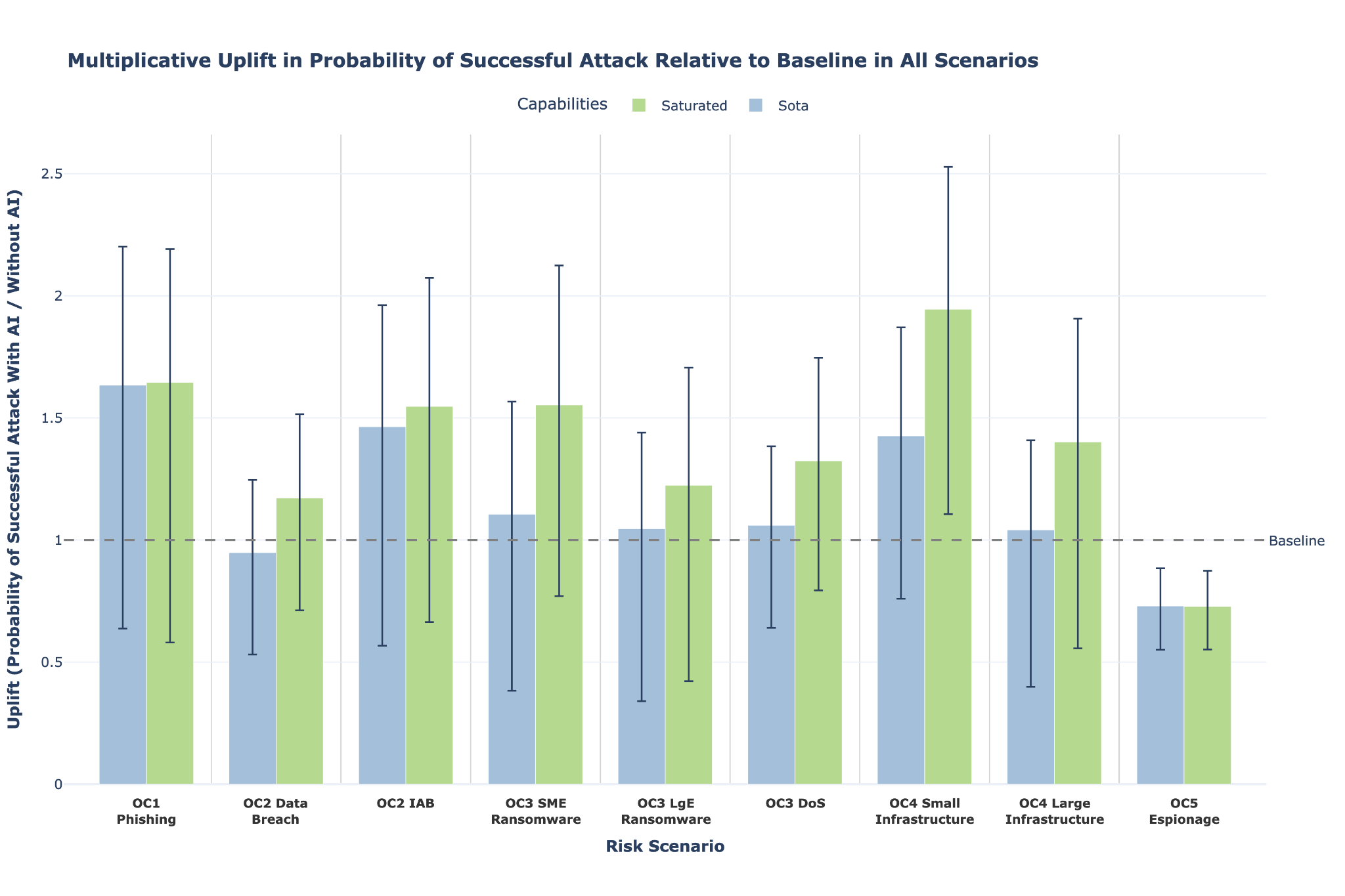}
  \caption{Multiplicative uplift in overall probability of an attacker executing a successful attack. Bar heights represent mean multiplicative across uplifted samples, relative to the mean baseline number, with error bars reflecting the interquartile range. }
  \label{fig:fig_10}
\end{figure}

In~\cref{fig:fig_10}, the bar heights indicate the proportionate increase in probability of success over the baseline, with results for SOTA-level AI shown in blue and results for saturated-benchmark level-AI shown in green. The black bars show the IQR across uplifted samples, capturing the uncertainty of experts.

For all but two risk models, OC2 Data Breach and OC5 Espionage, the risk modeling indicates that SOTA AI delivers an improvement in efficacy over the baseline. For all but OC5 Espionage, where it is broadly the same, the risk modeling indicates that the uplift for saturated AI is greater than for SOTA AI. We do not observe strong correlations, in the results, between threat actor operational capability level and the overall efficacy uplift.

In the absence of improvement to defenses, in general, these results indicate that defenders can expect an increasing number of successful attacks on their networks as AI capability improves through SOTA AI and towards saturated-AI, assuming that attackers can gain access to such AI. 
%%%%%%%%%%%%%%%%%%%%%%%%%%%%%%%%%%%%%%%%%%%%%%%
\subsection{Volume Uplift}
From a defender perspective, volume uplift is interesting since it informs the likely increase in the frequency of attacks and may correlate to an increase in costs of defense and cost of impact. We define it as:
\[
\text{Volume uplift} = \dfrac{N_{\text{attempts}}^{\text{(AI)}} \times N_{\text{actors}}^{\text{(AI)}}}
{N_{\text{attempts}}^{\text{(baseline)}} \times N_{\text{actors}}^{\text{(baseline)}}}\,.
\]
\begin{figure}[tbp]
  \centering
  \includegraphics[draft=false,page=1,pagebox=cropbox, keepaspectratio, trim=0 2cm 0 0, width=0.9\linewidth]{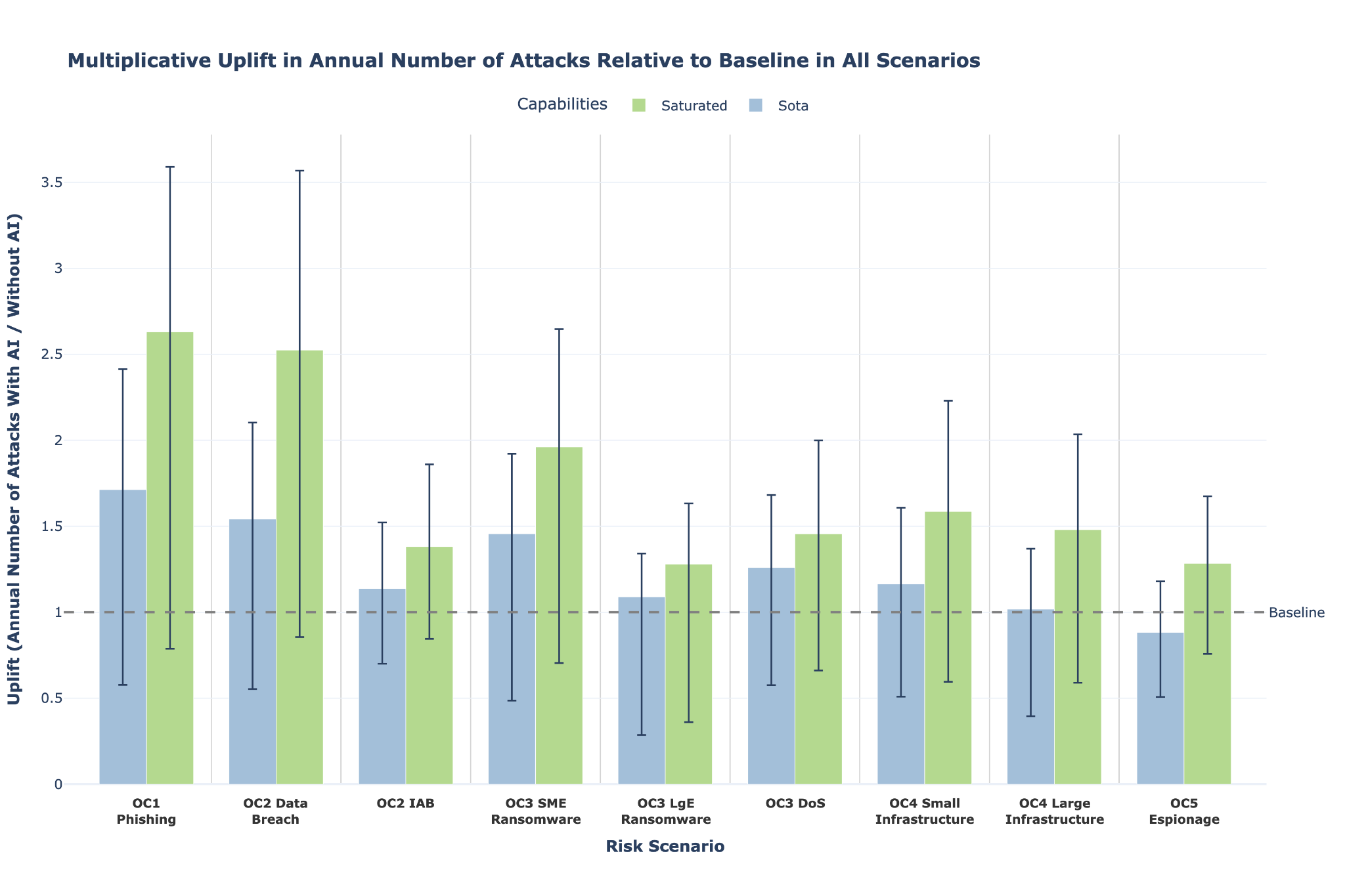}
  \caption{Multiplicative uplift in the overall number of attacks executed annually by attackers making use of AI systems, relative to baseline. Bar heights represent mean multiplicative across uplifted samples, relative to the mean baseline number of attacks per year.}
  \label{fig:fig_11}
\end{figure}

The volume uplift across all model scenarios is presented in~\cref{fig:fig_11}. In general, the results indicate an increase in the overall number of attacks as AI enables more attacks to be launched and/or encourages new actors to attempt attacks they would not have previously attempted. We note the largest uplifts in the OC1 Phishing and OC2 Data Breach scenarios for both capability levels. Our models indicate a decrease in the volume of attacks for actors using SOTA AI with OC5 Espionage, likely for the same reasons put forward in \cref{subsec:Attributable_riskFactors}.
%%%%%%%%%%%%%%%%%%%%%%%%%%%%%%%%%%%%%%%%%%%%%%%%
\subsection{Target Uplift}
The results for target uplift are intended to provide an indication of the extent to which AI enables a threat actor that previously targeted smaller victims can start attacking larger, more asset-rich and potentially better-defended targets. There is no simple equation for determining target uplift from the results we have gathered. Rather, here we produce some results that might indicate target uplift and provide some associated analysis.

We produced two sets of results that were directed at better understanding target uplift:
\begin{itemize}
    \item First, we looked at the OC3 ransomware scenario, for both the attack on a small enterprise and a large enterprise
    \item Second, we  looked at the OC4  disruption scenario on critical infrastructure for both small infrastructure and large.
\end{itemize}
To assess target uplift we produced the following sets of results:
\begin{itemize}
    \item Number of actors
    \item Successful attempts/year
\end{itemize}
All the results were produced as a function of AI capability (baseline, SOTA, Saturated), for both small and large targets.
%%%%%%%%%%%%%%%%%%%%%%%%%%%%%%%%%%%%%%%%%%%%%%%
\subsubsection{OC3 Ransomware}

\begin{figure}
  \centering
  \begin{subfigure}{0.47\textwidth}
        \includegraphics[draft=false,page=1,pagebox=cropbox, keepaspectratio, width=\linewidth]{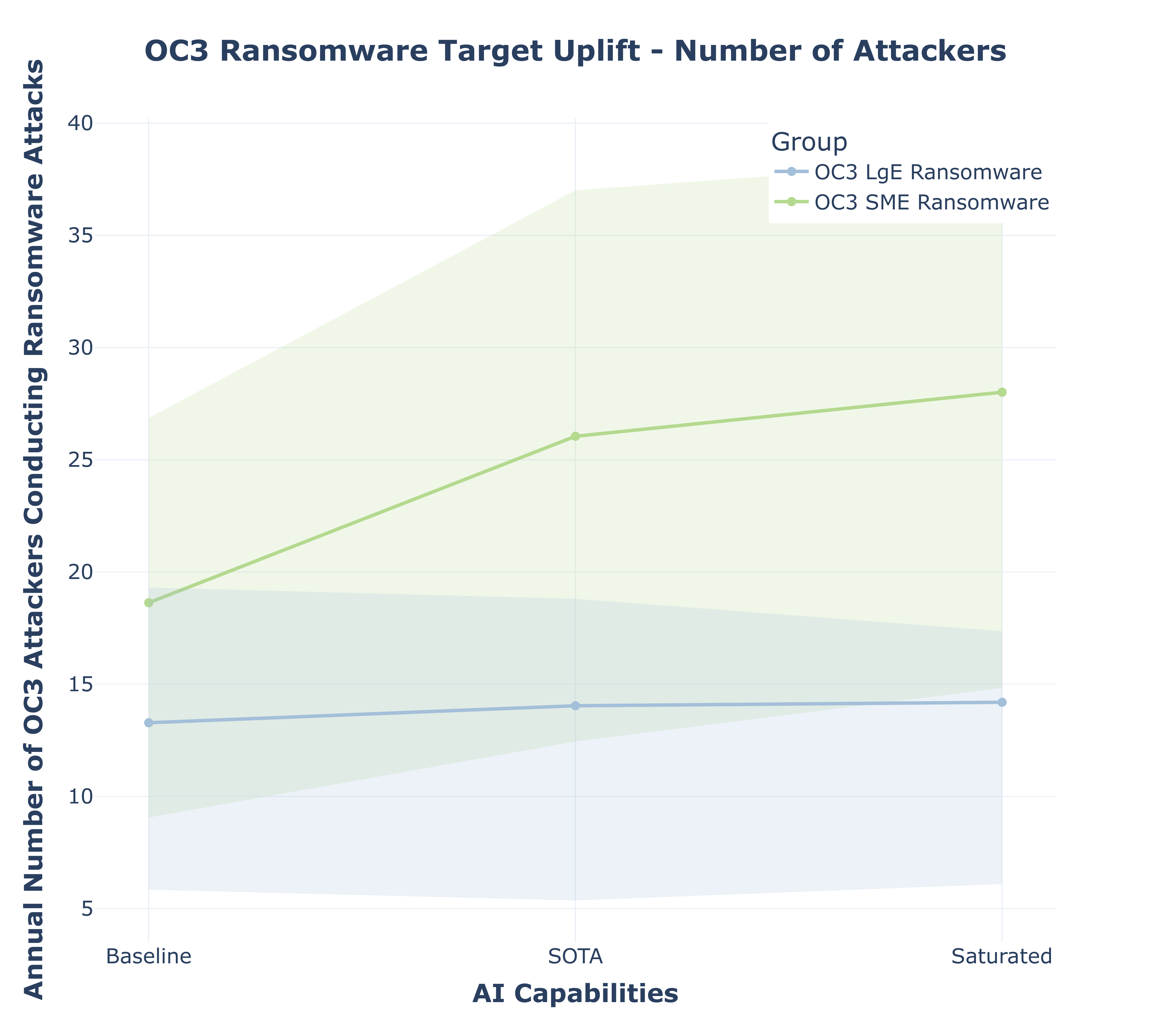}
        \caption{Number of Actors.}
        \label{fig:fig_12}
  \end{subfigure}
 \begin{subfigure}{0.47\textwidth}
        \includegraphics[draft=false,page=1,pagebox=cropbox, keepaspectratio, width=\linewidth]{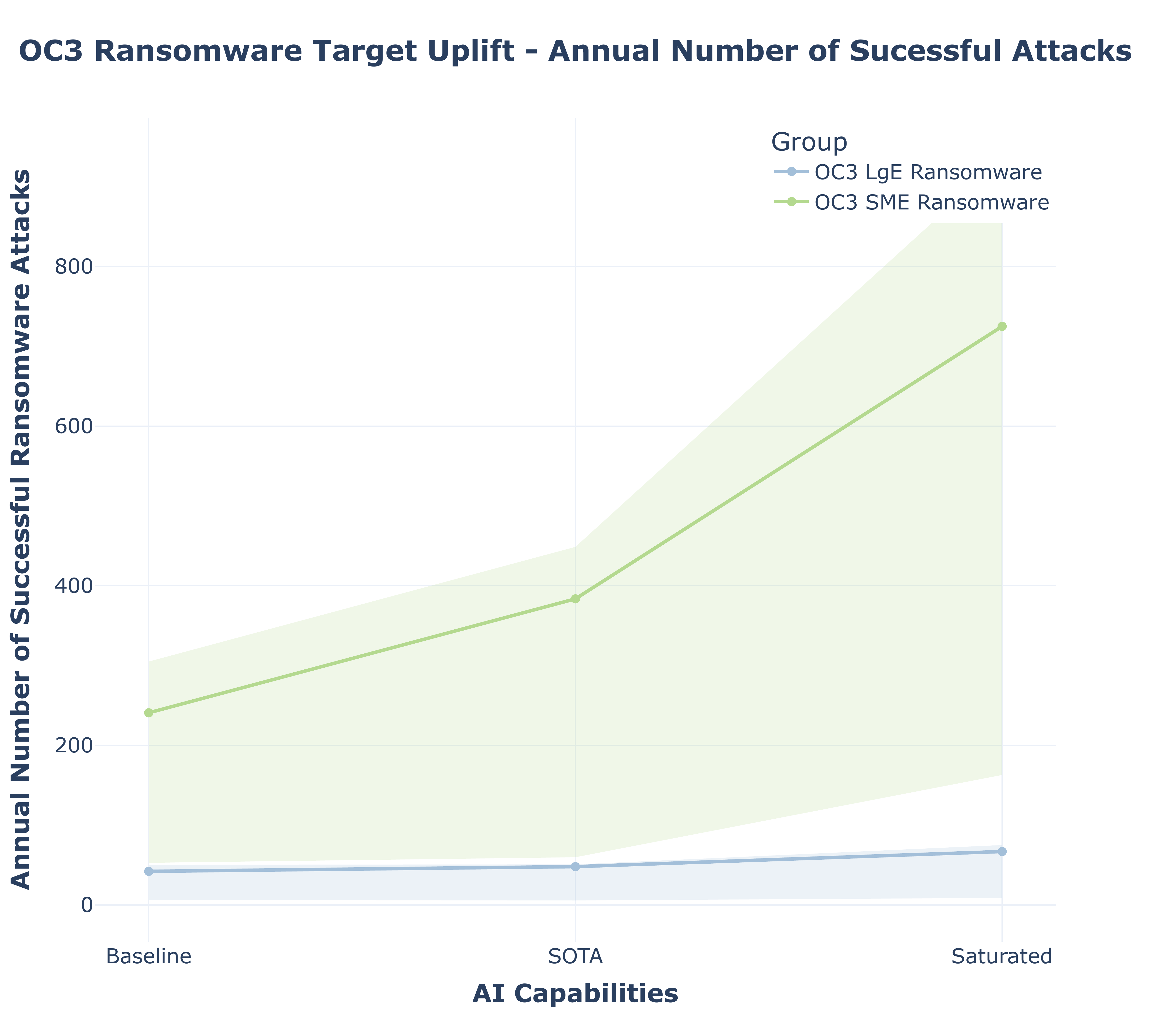}
        \caption{Number of Successful Attacks Annually}
        \label{fig:fig_13}
  \end{subfigure}

  \caption{Changes in risk model factors for OC3 Ransomware attacks dependent on target organization size. Central tendency is the mean, interquartile ranges in shaded regions.}
  \label{fig:OC3_Target_Uplift}

\end{figure}

We do not find clear indications of target uplift in the OC3 Ransomware scenarios from our modeling exercises. \cref{fig:fig_12} indicates little increase from the baseline to AI-uplifted settings in the number of attackers annually targeting large enterprises. In contrast, we see the number of actors targeting small enterprises increasing.

\cref{fig:fig_13} similarly indicates large AI capability-driven increases in the annual number of successful attacks for actors targeting small enterprises. In contrast, actors targeting large enterprises observe only a modest increase according to our models.

%%%%%%%%%%%%%%%%%%%%%%%%%%%%%%%%%%%%%%%%%%%%%%%
\subsubsection{OC4 Infrastructure - Disruption}
\begin{figure}

  \centering
  \begin{subfigure}{0.47\textwidth}
        \includegraphics[draft=false,page=1,pagebox=cropbox, keepaspectratio, width=\linewidth]{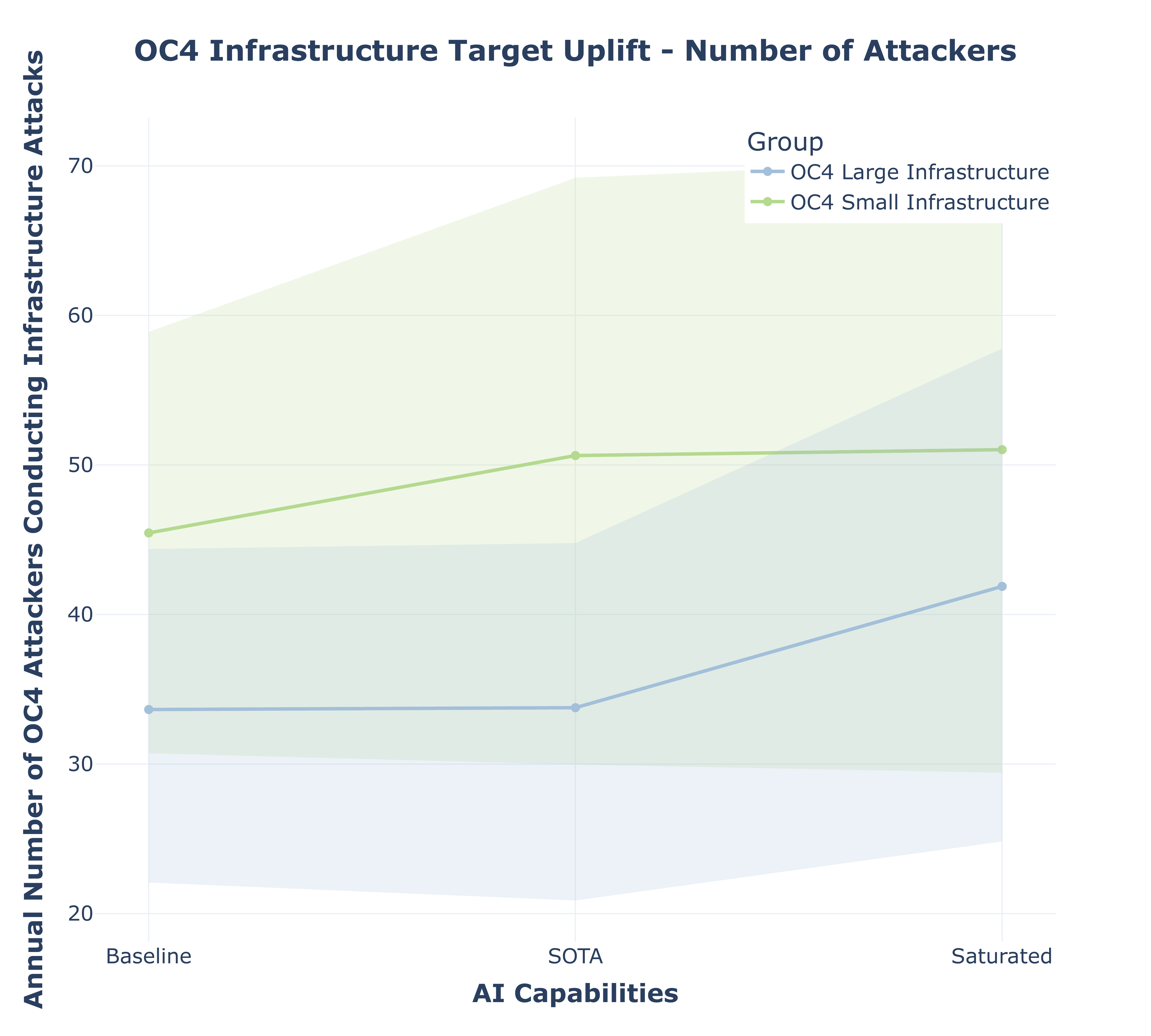}
        \caption{Number of Actors.}
        \label{fig:fig_14}
  \end{subfigure}
 \begin{subfigure}{0.47\textwidth}
        \includegraphics[draft=false,page=1,pagebox=cropbox, keepaspectratio, width=\linewidth]{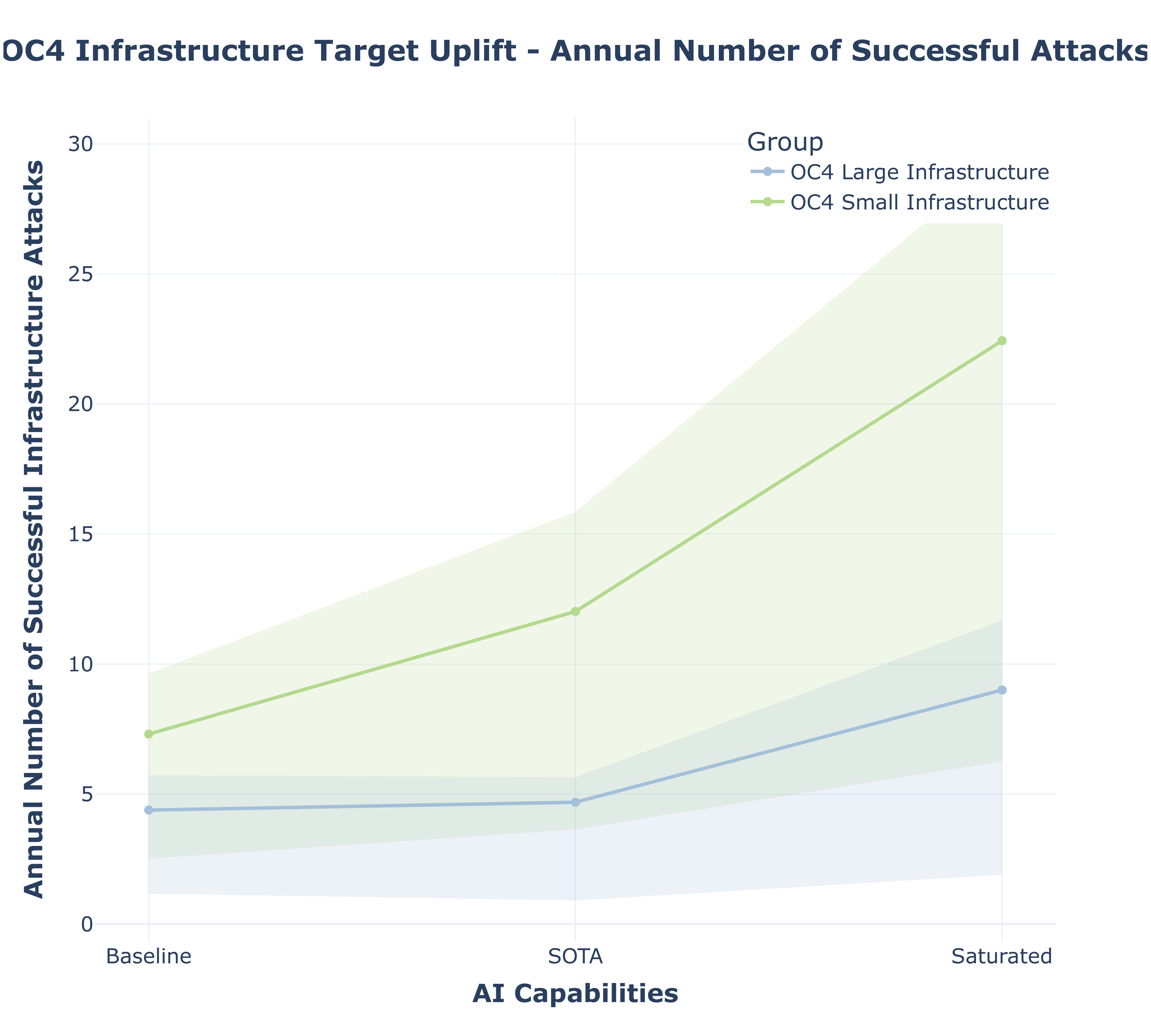}
        \caption{Number of Successful Attacks Annually}
        \label{fig:fig_15}
  \end{subfigure}

  \caption{Changes in risk model factors for OC4 Infrastructure attacks dependent on target organization size. Central tendency is the mean, interquartile ranges in shaded regions.}
  \label{fig:OC4_Target_Uplift}

\end{figure}

The results in the OC4 Infrastructure scenarios paint a more complicated picture than above. \cref{fig:fig_14} shows that our models indicate an increase in the overall number of attackers targeting large infrastructure targets relative to the number of actors targeting small infrastructure targets, in particular when moving from SOTA to saturated capabilities, potentially demonstrating some target uplift in this scenario. On the other hand, \cref{fig:fig_15} shows smaller increases in the total number of successful attacks per year against large targets compared to small targets.

These initial results indicate that further and more direct modeling of target uplift effects may be needed in order to capture insights into these sophisticated dynamics, though our models may be capturing these effects already in the OC4 Infrastructure scenarios.

%%%%%%%%%%%%%%%%%%%%%%%%%%%%%%%%%%%%%%%%%%%%%%%
\subsection{Evaluation of Uncertainty}
Here, we leverage the flexibility of our Monte Carlo approach in order to compute estimates of the entire distribution of risk factors, providing insights that fully reflect expert epistemic uncertainty. Density functions are estimated from samples using Gaussian kernel density estimation methods. Scott's heuristic \citep{scott2015multivariate} was used to select the bandwidth.
\subsubsection{Uncertainty Regarding Total Risk}
\begin{figure}
  \centering
  \includegraphics[draft=false,page=1,pagebox=cropbox, keepaspectratio, width=1.0\linewidth]{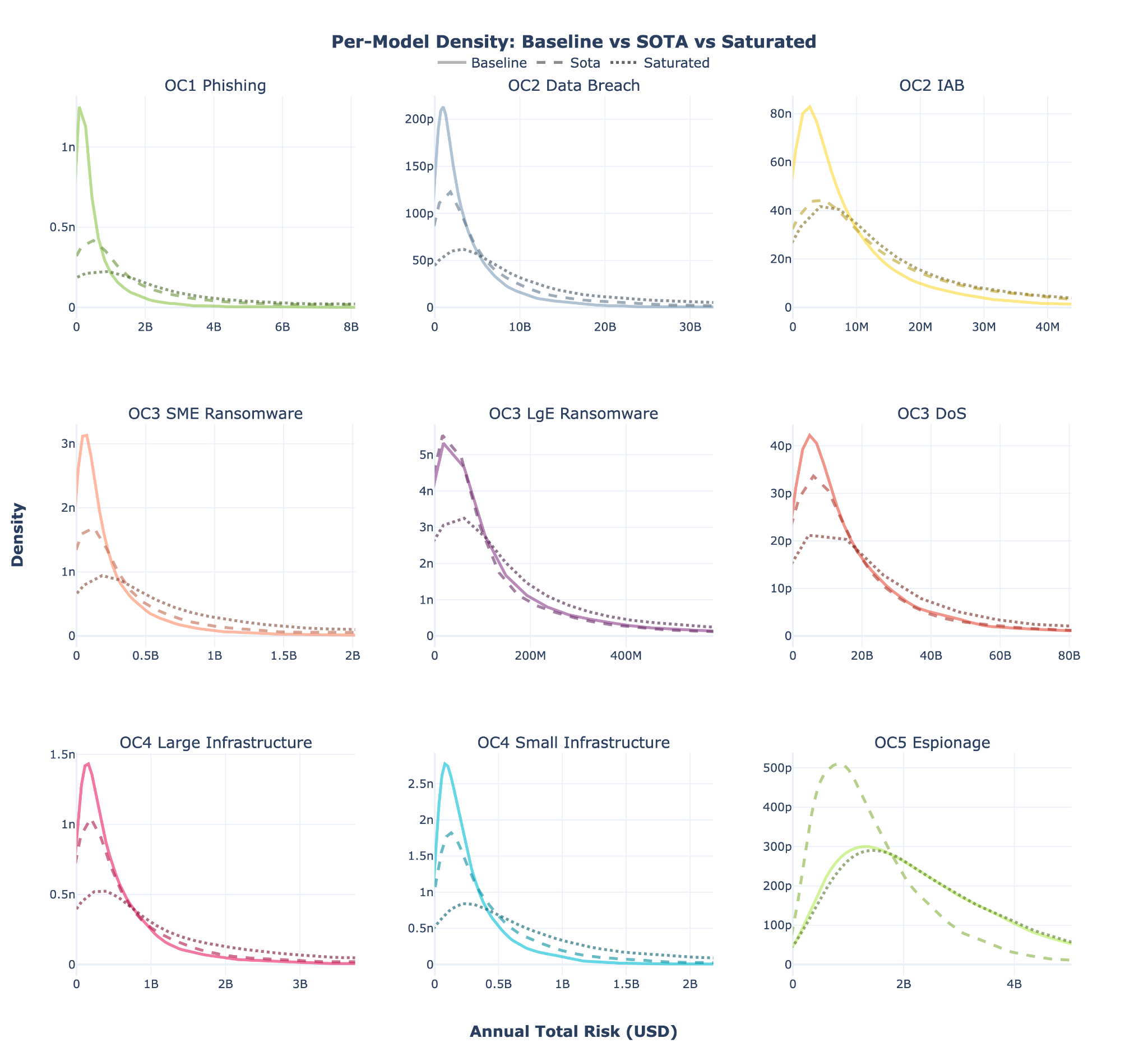}
  \caption{Distribution capturing epistemic uncertainty in mean annual total risk for all risk scenarios and capability settings (n: nano; p: pico).}
  \label{fig:fig_16}
\end{figure}

\cref{fig:fig_16} shows the estimated probability density over mean annual total risk for all nine risk models, across all three capability levels, baseline, SOTA and saturated. Almost uniformly, the models’ variance grows and densities flatten as AI capability levels increase. This captures the increase in tail risk provided by the estimators.

%%%%%%%%%%%%%%%%%%%%%%%%%%%%%%%%%%%%%%%%%%%%%%%
\subsubsection{Contribution of Risk Factors to  Uncertainty}
\begin{figure}
  \centering
  \includegraphics[draft=false,page=1,pagebox=cropbox, keepaspectratio, width=0.9\linewidth]{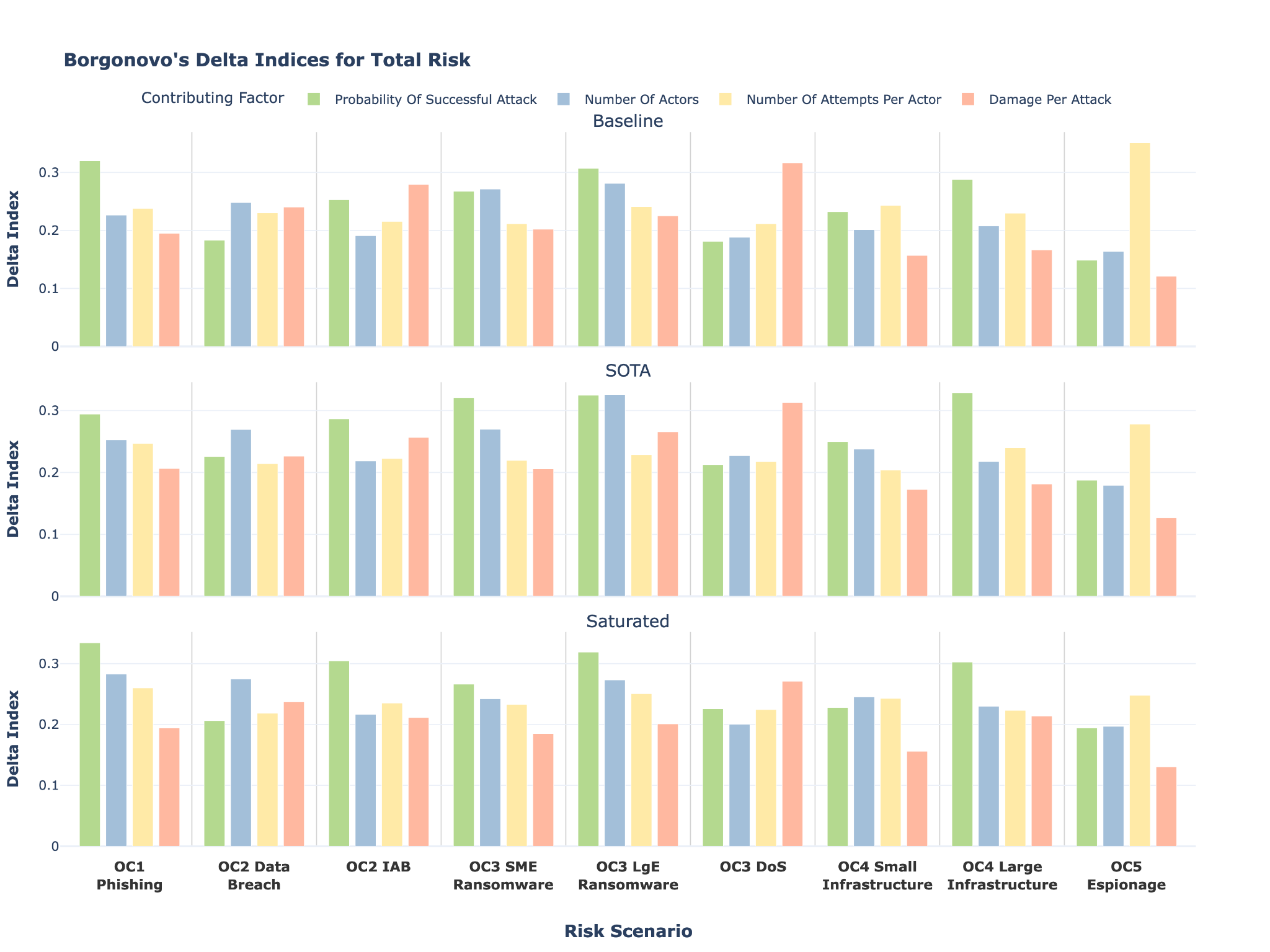}
  \caption{Relative contribution of uncertainty to total risk from each risk factor, at the different levels of AI capabilities, as measured by Borgonovo’s Delta Indices.}
  \label{fig:fig_18}
\end{figure}

\cref{fig:fig_18} visualizes Borgonovo’s Delta Indices for each of the risk models. These indices capture the $\ell_1$ distance between the marginal distribution of total risk, and the distribution conditioned on each factor. This captures how much the four risk factors, across the nine scenarios, at the different levels of AI capabilities, contribute to the uncertainty in total risk. We observe that conditioning on any one factor is not sufficient to explain the distribution over total risk, indicating that all factors contribute similarly to the overall uncertainty.

\subsubsection{Uncertainty of Human Experts Relative to LLMs}
\begin{figure}
  \centering
  \includegraphics[draft=false,page=1,pagebox=cropbox, keepaspectratio, trim=0 2cm 0 0, width=0.8\linewidth]{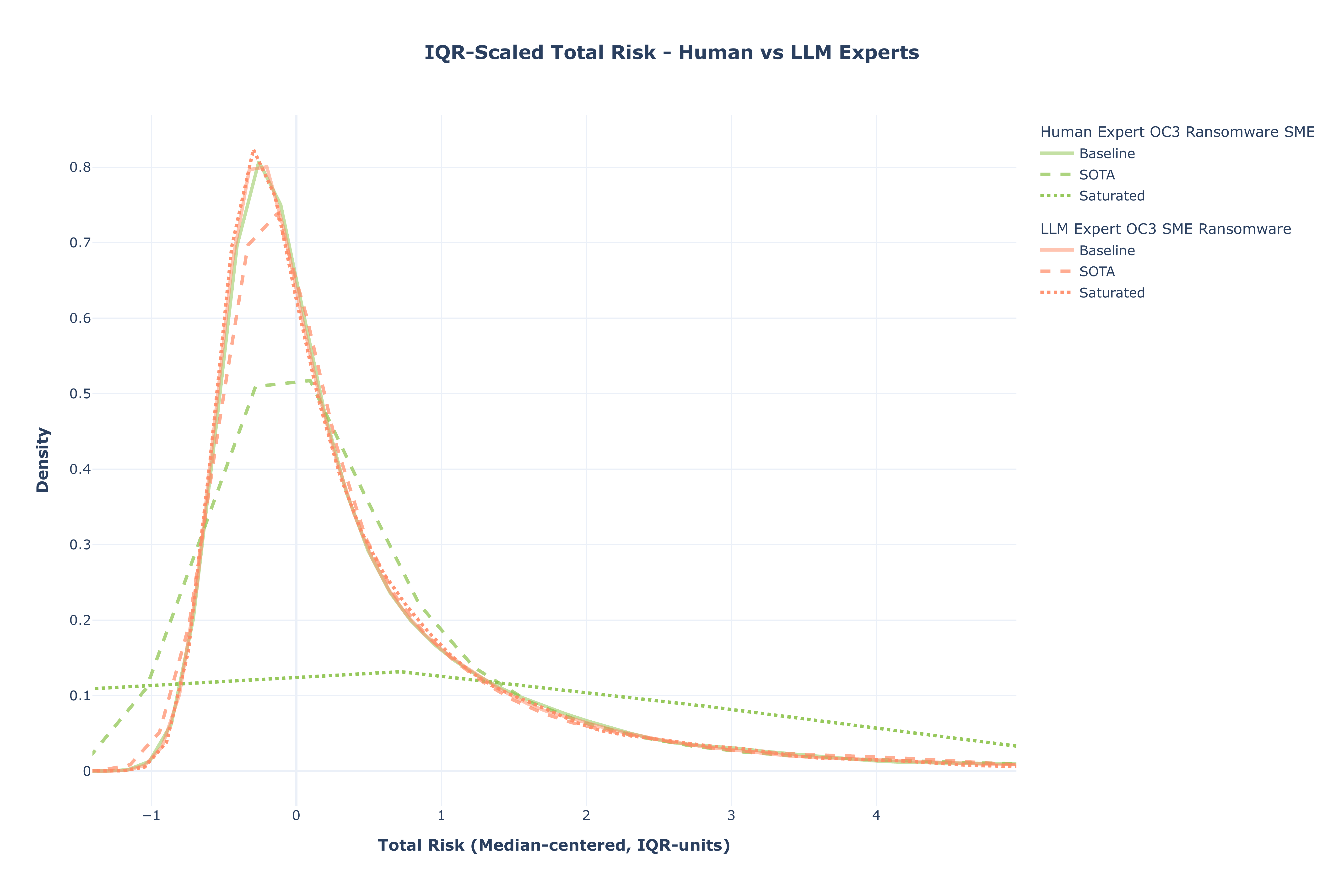}
  \caption{Comparison of human-elicited distributions over total risk relative to LLM-elicited ones. Distributions are transformed through IQR normalization (subtraction of the median and division by the interquartile range). LLM-elicited distributions are shown in orange and humans in green.}
  \label{fig:fig_19}
\end{figure}

We compare the tail behavior of the distribution over total risk between human experts and LLM-simulated experts. \cref{fig:fig_19} shows the median-centered IQR-normalized distributions for human (in green) and LLM experts (in orange) for the OC3 SME Ransomware model. By transforming the distributions in this way, we can clearly compare the behavior of these distributions at their most extreme values, despite the difference in distribution scales.

Human expert estimates become progressively more heavy-tailed as capabilities increase. This is consistent with the rationales and discussion of the experts. Estimating the risk at saturated levels, which reflects capabilities that have not yet been observed, is a more uncertain exercise than estimating the effects of current-day capabilities. 

The LLM estimator has no clear pattern. The level of uncertainty is more similar across all three capability levels and lowest for SOTA. This suggests that LLM estimates do not capture this progression in tail uncertainty observed in human experts. Instead, uncertainty is captured in values more concentrated around the median.

% \begin{figure}
% \centering
% \includegraphics[width=1.0\linewidth]{Fig19.png}
% \caption{Comparison of human-elicited distributions over total risk relative to LLM-elicited ones. Distributions are transformed through IQR normalization (subtraction of the median and division by the interquartile range). LLM-elicited distributions are shown in blue and humans in green}
% \label{fig:fig_19}
% \end{figure}
%%%%%%%%%%%%%%%%%%%%%%%%%%%%%%%%%%%%%%%%%%%%%%%
\section{Limitations and Future Work}
\label{sec:Limitations}
This work provides preliminary efforts to model cyber attack risks in the context of AI-uplifted attackers. As such, there are several avenues that would benefit further study. Here, we offer an extensive overview of the limitations of our approach in order to enable the community to iteratively improve the fidelity of risk models, capture more sophisticated model dynamics, and provide clearer insight into the consequences of modeling decisions.  

\paragraph{Data quality for baseline estimates:} The quality of risk models are limited by the quality of the data used to estimate risk factors. Much of the data that we would ideally have access to when creating the baseline risk models is not in the public domain. Attackers publish very little information, and what they do reveal typically concerns only successful attacks. However, for risk modeling, we are interested in gathering data for both successful and unsuccessful attacks. Similarly, targets do not always publish information about attacks to prevent reputational damage, and can only publish information about attacks that have been detected. Specifically, information on the probability of success of various attack steps, categorized by attacker profile and defender profile, is rare. Information required for determining the number of actors and the number of attempts per actor is also sparse, since threat actors naturally do not publish information about their organizational structures, cost of operations, number and type of campaigns launched, etc. Available data is often simplified, ambiguous, or contains implicit assumptions, leading to widely divergent values for the same statistic across sources. Data sources older than a year may not reflect the current situation. However, often, due to sparsity of data we are required to make use of data that may pertain to multiple recent years. To help mitigate these data quality concerns, our process includes using confidence intervals around the best-estimate values to capture uncertainty over data quality, as well as a cybersecurity expert review of each risk model. In addition, future work could make use of a more privileged dataset, as is used in \citet{RodriguezEtAl2025}.

\paragraph{Dynamism in cyber crime:} Cyber crime is constantly changing as the players (attackers, defenders, law enforcement, etc.) react and adapt to one another. Factors such as ransom payouts and costs incurred can fluctuate wildly from year to year. For example, the Cl0p ransomware group was once well known for performing double extortion (performing both encryption and exfiltration), but now increasingly just performs single extortion (data exfiltration). The impact is that risk models will need to be constantly adapted and updated, retired and replaced by new models.

\textbf{Modeling of defense:} Our methodology provides a detailed way of quantifying the various ways attackers might use AI to improve their operations. However, our methodology does not give similarly detailed consideration to how defenders might react to this change in the threat landscape by implementing new or enhanced defenses, some of which might make use of AI capabilities. These mitigations have not been included in our initial models as it is understood that attackers are likely to be faster adopters than defenders \citep{rand_pea4102_1, lohn2025impact}. Our models can be readily extended to capture such effects. Future work should provide more consideration of how defenders may be able to adapt to the changing threat landscape and adopt AI.

\textbf{Campaign vs attack:} For some attack types, it is useful to consider both the notion of a ``campaign'' and an ``attack''. In a campaign, a threat actor tries to reach many (e.g., hundreds or thousands) of potential targets, while an attack only has one target. For example, in the OC2 IAB risk model, an attacker may conduct a number of phishing campaigns each year, each consisting of sending hundreds or thousands of phishing emails, all tailored for some particular industry or sector. If a target clicks on a phishing email, then an attack can commence, i.e., an attempt to download the infostealer to a particular target. 

The issue with this from a risk modeling point of view is that some MITRE ATT\&CK tactics correspond to campaign-level activities (i.e., single activities/steps that may impact hundreds or thousands of targets, e.g., Reconnaissance and Resource Development), whilst other MITRE ATT\&CK tactics apply at the attack level and only affect single targets. In the OC2 IAB model, the solution is to set the campaign-level activities (Reconnaissance and Resource Development) to 100\% successful in the $P_{\text{success}}$ calculation. This enables the risk model to be built around attack-level aspects only. This is a reasonable shortcut since the attack will not take place if an attacker fails at those first steps

\paragraph{Tactics with very high probability of success:} As mentioned above, in some risk models, the MITRE ATT\&CK steps of Resource Development and Reconnaissance arguably cannot fail (quantitatively $P_{\text{success}}$ is 100\%), but the quality of what the attacker has produced in these steps can vary greatly. This will therefore affect the probabilities of success in subsequent attack steps. In the OC2 IAB case, good Reconnaissance can lead to the targeting of a sector that is prime for exploitation, and which can deliver good rewards for the attacker, with a phishing email narrative that works well. Similarly, good Resource Development can lead to infostealer malware which is either excellent at avoiding detection by anti-virus and EDR (Endpoint Detection and Response), provides email addresses that are current and well-targeted for the specific sector, or phishing email content that is highly tailored and likely to result in clicks by the victims. AI uplift can be expected with both Reconnaissance and Resource Development.

However, if the $P_{\text{success}}$ of these tactics is marked at 100\%, then for the overall $P_{\text{success}}$ calculation to capture this expected AI uplift, this uplift should be reflected in the increased $P_{\text{success}}$ of other MITRE ATT\&CK tactics (e.g., improved probability of success in anti-virus evasion in the Execution tactic). In this sense, these two tactics share a lot in common with the Defence Evasion tactic, which can also in some models be modeled as having 100\% success. As with Reconnaissance and Resource Development, when we consider the quality of an attacker’s Defence Evasion, it is assumed that this will be seen in $P_{\text{success}}$ figures for other tactics.

\paragraph{Modeling assumption of binary attack success/failure for each tactic:}
Our models make a strong assumption that each tactic included in an attack can be modeled with a Bernoulli random variable with states corresponding to attacker success and attacker failure. However, in real cyber attacks, the objectives of a given tactic may be achieved to varying degrees of success, we can imagine differences in quality of the products of a Resource Development tactic, or partial failure in an Initial Access step, where the attacker leaks identifiable information, but this does not prevent the continuation of the attack. These partial success and failure states can then influence other factors of the risk model, e.g., leading to increased recovery costs or reduced probability of success in a later defense evasion step. The fact that our risk model factors are conditioned on a binary success/failure state in the previous step means that these intermediate dependencies are not captured. In future work, these dependencies should be explicitly modeled in the joint probability of success.

\paragraph{Evaluating target uplift:} Some of our risk models were aimed at evaluating the possibility of AI target uplift. This can present quite a fundamental challenge if the attack is of a brand new type.  For example, if we wanted to evaluate the risk associated with an attack type that we might assume does not currently occur, such as script kiddies targeting a uranium processing operational technology.  Then both the number of actors and number of attack attempts/actor/year would be zero in the baseline.   Picking justifiable and realistic absolute values for these two terms, when AI becomes available to the script kiddie, is a different risk modeling challenge to that of assessing the relative change in the value of these terms, when AI becomes available,  for a type of attack that already exists.

\paragraph{Novel AI-enabled threats:} In the same way that our approach can run into difficulties in evaluating target uplift as described in the prior section, a related but qualitatively different issue can arise if we wish to try and model entirely novel AI-enabled attack scenarios.   If an AI threat is completely novel, then it is an unknown unknown (in contrast to the target uplift issue which is a known unknown).  Risk modeling of such unknown unknowns would require a different risk modeling approach to that presented herein. 

\paragraph{Uplift due to AI capabilities in strategizing and planning:} AI may help in strategizing and planning an attack or recommending the next steps of an attack. Such factors could affect the number of actors capable of performing an attack or the number of attack attempts that could be completed per year. Success at a multistep benchmark task, such as those found in BountyBench, and used in our study, will require the use of some strategizing and planning capabilities, however, future work could include investigating the use of other benchmarks that are more specifically directed at the evaluation of agentic capabilities. 

\paragraph{Risk of anchoring bias in expert reviews:} One concern with providing our cybersecurity experts with risk models to review is that providing them with a finished baseline model may bias their estimates to be more in line with the model. This would not be present had the expert produced the risk model themselves. One (albeit more expensive) way of avoiding such anchoring bias in future work would be to get numerous experts to build risk models, and then require them to build some consensus among themselves.

\paragraph{Tail risk:} Our top-level risk formula provides a distribution of epistemic uncertainty of expected global annual harm for the given risk scenario. In principle, expected values should account for tail events, since they influence the mean. However, our methodology elicits estimates of uncertainty around the mean value, and estimating the likelihood and magnitude of extreme tail events is difficult to do intuitively. Some risk factors—particularly damage per attack—are likely to have fat-tailed distributions, where rare but catastrophic outcomes contribute substantially to expected harm. Separately modeling these tails, rather than relying on experts to implicitly account for them when estimating means and confidence intervals, would likely improve estimation accuracy.  Decision-makers may also prefer to be provided with a probability that harm could exceed a certain amount, as opposed to being provided with an expected level of harm.

\paragraph{Complex elicitation task:} Where a single tactic, such as Lateral Movement or Privilege Escalation, is used multiple times in an attack, our methodology asks the human or LLM expert to implicitly hold in their mind the possibility of all of those occurrences and aggregate across them. This includes how many instances of the tactic might be expected to occur and how probability of success might be conditional on the order of instances. Our methodology simplifies the risk modeling by not including multiple possible attack paths, and by only having a maximum of 14 attack path components (given the 14 MITRE ATT\&CK tactics). However, this is achieved at the cost of more difficulty experienced by the experts and consequently potentially poorer quality of elicited estimates. Future work could seek to remedy this through further decomposed and more structured elicitation.

\paragraph{Subjectivity in the human Delphi study:}
Since there is no objectively verifiable data that links performance on a benchmark score to uplift in a risk factor, we need to rely on the estimates elicited from experts. Ideally, the field would design evaluations that directly measure the quantities of interest, such as controlled uplift studies comparing attack success rates with and without AI assistance. However, such studies are costly and in some cases practically infeasible. For example, one cannot conduct a realistic uplift study of OC5 actors attacking espionage targets. Furthermore, uplift evaluations can only measure the capabilities of systems that exist today; they don't enable forecasting risks posed by future, more capable AI systems. Our benchmark-based approach, by contrast, allows us to extrapolate risk estimates to hypothetical capability levels not yet observed, enabling forward-looking risk assessment.

We therefore rely on expert elicitation to map imperfect capability proxies to risk factor estimates. This approach has precedent in other high-risk industries such as nuclear. As is a known problem in expert elicitation, each expert may have their own personal biases and may have differing capabilities to make good inferences. This challenge is further exacerbated by the fact that we require experts with knowledge of both AI and cybersecurity, drawing on a smaller pool of qualified individuals. We try to limit these issues by aggregating across multiple experts and by using a structured elicitation protocol (IDEA).

\paragraph{Limitations of AI uplift estimation:} Due to the paucity of suitable experts and the demands of intensive expert elicitation processes, we perform uplift elicitation with human experts only for one of the risk models (OC3 SME Ransomware) and simulate the Delphi process with LLMs for the others. While a comparison of the LLM and human estimates for the common risk model seems to yield tentatively promising results, further validation is needed. The lower variance we see in LLM estimators could be due to expert personas being too similar to each other, thereby failing to elicit diverse opinions. We observe that multiple rounds in the LLM Delphi study do not improve the quality of predictions, which is notably different from most human Delphi processes.

We encourage more research to quantitatively evaluate the validity of using LLM-estimators. We are actively testing this and report initial findings in the accompanying blog post~\citep{Quarks2025}. Testing LLM elicitation with additional expert personas or prompt structures, as well as in settings where ground truth is available, will improve confidence in these methods. Overall, we expect that as general LLM capabilities progress, their forecasting skills will improve correspondingly. Importantly, LLMs need to merely match, not surpass, human forecasting level in order to already be useful. At this point, the associated speed and cost reduction would allow us to gather significantly more high-quality data than with human experts. Thus, we anticipate that LLM forecasters might become an integral part of expert elicitation procedures in the not so distant future.

\paragraph{Handling complex crime ecosystems:} Some threat actors, for example those engaged in Ransomware-as-a-Service ecosystems, may only participate in part of an attack. We seek to provide good coverage for a wide range of threat actor types, so this presents a challenge. As an example of this issue, in one of our risk models (OC2 IAB), a challenge was that the work of the OC2 Initial Access Broker completes once the stolen credentials are placed on the dark web marketplaces. However, at this point only a limited amount of the end harm has occurred, specifically that arising from the clean-up of the infostealer malware on the target’s IT systems. A significant additional component of harm occurs once the stolen credentials are purchased and are used in a successful attack. This needs to be considered to evaluate end harm.

We modeled in some depth the work of the OC2 initial access broker, looking into factors such as the number of IAB actors, how many attack attempts they make (campaigns, and phishing emails per campaign), probability of success in their deployment and operation of infostealers. However, the last step, computing costs incurred by the defender when the stolen credentials are used in a successful ransomware attack, required us to make some assumptions about the likelihood that gathered credentials are used in successful attacks, and the costs associated with these attacks. This last step, whilst itself deserving also of a detailed evaluation (as the one done for the IAB), was computed in a comparatively approximate manner. One possibility to address this issue, that might be considered for the future, could be to chain risk models together. For example, chaining an OC2 IAB risk model together with other model(s), such as the OC3 SME Ransomware model.

There are also multiple ways in which the credentials stolen by an IAB might be used, possibly serving a range of different threat actor intentions. In our model, we modeled just one of these ways, which was the use of the credentials in ransomware. This branching out of the attack space for the case of the initial access broker, which results in a variety of possible real-world harms, acts somewhat contrary to the notion of each risk model representing neat slices through the cybersecurity risk universe.

\paragraph{Difficulty of evaluating change in total global cyber risk:} A decision maker making a deployment decision would be interested to see the likely impact on total cyber risk. However, while we sought to select risk models that cover a good proportion of common and impactful attack types, in the process of building the risk models, in order to try and improve accuracy, we frequently had to narrow the scope of the risk scenario. Maintaining full accuracy while also covering a more significant proportion of the risk universe would require the creation of many more risk models.

\paragraph{Difficulty ranking of benchmark tasks:} To simplify the risk modeling process, and to enable a simplified way by which users of the risk model can experiment with possible future AI model capabilities, we structure the tasks in our benchmarks in difficulty order. One problem with this is that different AI models may not necessarily pass benchmark tasks in the same order. This means an approximation has to be performed (see~\cref{app:B}) when contemplating the change in cyber risk for an AI model that does not complete tasks in the order defined in our difficulty ranking.

\paragraph{Benchmark choice:} The two benchmarks we choose have a different relationship between their tasks and the uplift on the corresponding risk factors. More precisely, as they progress from easier to harder tasks, expert estimates of uplift rise more steeply for parameters associated with BountyBench than for those associated with Cybench. This could be due to BountyBench being a somewhat more relevant benchmark to the predictions we are eliciting, but would require further testing with other benchmarks. To ensure the chosen benchmarks (or other KRIs) allow experts to provide the best possible uplift predictions, one could conduct some future work to  study how these predictions change as we modify the associated benchmark. If uplift estimates are consistently low across the full range of benchmark tasks, this may indicate that the benchmark is not appropriate or that the tasks do not cover a wide enough range of difficulty.

\paragraph{Conditioning each risk factor on a single benchmark:} Since different benchmarks may capture different model capabilities that are relevant to a single risk factor, it would be useful to map multiple tasks from multiple benchmarks to a single risk factor. A naïve approach to this involves direct conditioning of each risk factor on multiple benchmarks, leading to geometric increases in the number of risk factors that need to be elicited based on all combinations of possible benchmark scores. Future work can look to alleviate this constraint by introducing latent factors, or by explicitly modeling which combinations of benchmark scores are likely to capture an uplift in risk, and eliciting only these risk factors.

\paragraph{Guardrails:} The risks we model assume threat actors have access to the relevant AI capabilities. In practice, this access depends on whether safeguards can be circumvented (open-source safeguards or closed source ones offering varying level of defense). If safeguards were robust, the harm from a given model would be eliminated. Our methodology could incorporate this by adding a multiplicative factor representing capability accessibility, the probability that a threat actor successfully obtains unrestricted access to the relevant capabilities. We omit this factor in the current work because, as recent reports indicate, even frontier models remain prone to jailbreaking \citep{Anthropic2025}, suggesting that determined actors can currently access most capabilities. As safeguards improve, explicitly modeling this accessibility factor will become increasingly important.

%%%%%%%%%%%%%%%%%%%%%%%%%%%%%%%%%%%%%%%%%%%%%%%%
\section{Conclusion}
\label{sec:Conclusion}
Risk management for frontier AI systems is a nascent science and has so far focused on ``if-then scenarios'', where a certain level of a dangerous capability would trigger a certain set of mitigations. As argued in the introduction, this is not sufficient for decision making as it has a number of limitations. 

In this technical report, we seek to further the practice of AI risk management and present the results of applying our risk modeling methodology~\citep{Murrayb2025} on the domain of AI-enabled cyber offense risk. We argue that this has many potential benefits. For AI decision makers, this methodology can enable more data-driven decisions. For defenders from cyber attacks, e.g., in corporations or critical infrastructure, it provides insights into where to prioritize limited resources for mitigation efforts. For the AI evaluation community, it provides a way to identify gaps in the current evaluation suite and develop new evaluations.

We apply our methodology to nine cybersecurity risk scenarios and demonstrate how the methodology enables a broad range of risk data to be gathered. This includes estimates of the overall increase in expected risk due to AI, a breakdown of the proportionate contributions made by different risk factors (e.g., probability of success versus impact when succeeding), and various distinct types of ``uplift'' (efficacy, volume, and target). As we use both human experts and LLM experts, we are also able to provide findings on the advantages and disadvantages of using LLMs as estimators.

The quantitative findings should be treated with caution and not used directly for decision making without further validation, as the use of LLM estimators is still experimental. However, we believe they can be useful for relative and directional analysis such as risk prioritization. 

Quantitative risk assessment is difficult, and we recognize the limitations of this initial attempt. However, industries such as aviation and nuclear power did not develop their rigorous safety practices overnight; they evolved over decades through iterative refinement of methods, and learning from failures \citep{Leveson2012}. We hope this work represents a first step along a similar path for AI risk management. By proposing a first quantitative AI risk modeling methodology and producing initial estimates that can be critiqued and improved, we aim to help the field move toward the quantitative safety practices that have made other high-risk technologies acceptably safe. 
%%%%%%%%%%%%%%%%%%%%%%%%%%%%%%%%%%%%%%%%%%%%%%%
\section*{Acknowledgments}
Mario Fritz was partially supported by the ELSA – European Lighthouse on Secure and Safe AI funded by the European Union under grant agreement No. 101070617. Views and opinions expressed are however those of the authors only and do not necessarily reflect those of the European Union or European Commission. Neither the European Union nor the European Commission can be held responsible for them.

We are very grateful to our expert advisors and many reviewers, including Pierre-Francois Gimenez, John Halstead, Matthew van der Merwe, and Joe Rogero. Providing review and advice does not imply endorsement of the paper or its findings. The views expressed by individuals do not reflect those of the organizations with which they are affiliated. All remaining errors are our own. 
%%%%%%%%%%%%%%%%%%%%%%%%%%%%%%%%%%%%%%%%%%%%%%%%
\section*{Glossary}
\textbf{Bayesian Networks (BNs):} A type of graphical model that represents and quantifies probabilistic relationships among a set of variables. In a BN, nodes represent events or states, and connecting arcs represent conditional dependencies, making them well-suited for modeling complex causal chains and updating probabilities as new evidence becomes available.

\textbf{Capture The Flag (CTF):} A type of cybersecurity challenge.

\textbf{Event Tree Analysis (ETA):} A bottom-up, inductive scenario building technique that graphically maps the potential outcomes following a single initiating event. It explores the branching paths of possible consequences based on the success or failure of various safety functions or subsequent events.

\textbf{First Solve Time (FST):}  The time taken by the quickest individual or team of humans to successfully complete a cybersecurity challenge.

\textbf{Fault Tree Analysis (FTA):} A top-down, deductive scenario building technique where an undesired ``top event'' (a specific system failure) is traced backward to its root causes. It uses Boolean logic (AND/OR gates) to represent how combinations of lower-level failures can lead to the top-level outcome.

\textbf{Harm:} The realized adverse outcomes resulting from a hazard. In the context of AI, this can include economic damage, loss of life, societal disruption, or other negative consequences.

\textbf{Hazard:} The source of risk. In the context of AI, a hazard is often a model's capability, property, or tendency that has the potential to cause harm.

\textbf{Key Risk Indicator (KRI):} A quantifiable measurement of system behavior that serves as indirect evidence for risk.

\textbf{Probabilistic Modeling:} An approach to safety analysis that aims to identify and analyze as many potential credible accident scenarios as possible. It uses techniques like Fault Tree and Event Tree Analysis to model failure pathways and then assigns probabilities to each step to produce a quantitative risk profile (e.g., the annual probability of a specific failure), rather than a binary outcome.

\textbf{Risk:} The combination of the probability of occurrence of harm and the severity of that harm. It is often conceptualized as a triplet: a scenario describing what can happen, the likelihood of that scenario, and its potential consequences.

\textbf{Risk Factor:} A component of the equation for overall expected risk as used in this paper.  Types of risk factor are number of actors, number of attack attempts/actor/year, dollar impact, probability of successful attack and probability of successful application of each relevant MITRE ATT\&CK tactic.

\textbf{Risk Factor Parameter:} A parameterization of a risk factor.  For example, a central tendency measure, a confidence interval parameter (e.g. 5\%, 95\%), or a parameter of an e.g. PERT or Beta distribution that is fitted to confidence interval parameters.

\textbf{Risk Scenario:} A logically laid-out sequence of causal steps linking a hazard (a source of risk) to a harm (a realized adverse outcome), taking into account the contexts in which the system may be deployed and the potential for intervening events or failures.

\textbf{Risk Tolerance:} A predefined level of risk that an organization, regulator, or society deems acceptable. In a risk management framework, estimated risks are compared against the risk tolerance to inform decisions about whether a system should be deployed or if further mitigation is required.

\textbf{State Of The Art (SOTA):} Used in this report to describe an AI having capabilities broadly in line with those available at the time of writing, and as detailed more precisely in~\cref{app:B}.
%%%%%%%%%%%%%%%%%%%%%%%%%%%%%%%%%%%%%%%%%%%%%%
%\section*{References}
\bibliography{example}
%\bibliographystyle{unsrt} 
%\bibliographystyle{plain}
%\bibliographystyle{apalike}
%\nolinenumbers
%\bibliography{references}
%%%%%%%%%%%%%%%%%%%%%%%%%%%%%%%%%%%%%%%%%%%%%%
\newpage
% \appendix
\begin{appendices}
\crefalias{section}{appendix}
\section{Risk Model Summaries}
\label{app:A}
In~\cref{tab:risk_model_summaries}, we give brief descriptions of our nine risk models.

{\renewcommand{\arraystretch}{1.5}
\begin{longtable}{p{4cm}p{10cm}}
\caption{Summaries of our nine cybersecurity risk models} \label{tab:risk_model_summaries} \\

\toprule
Feature & Description \\
\midrule
\endfirsthead

\multicolumn{2}{l}{\textit{(Continued from previous page)}} \\
\toprule
Feature & Description \\
\midrule
\endhead

\midrule
\multicolumn{2}{r}{\textit{(Continued on next page)}} \\
\endfoot

\bottomrule
\endlastfoot

& \textbf{Model 1: OC1, Social engineering and BEC} \\
\midrule

\textbf{Attacker intent} &
Financial monetization through redirected funds from a formal business invoice to a false account \\
\textbf{Threat actor type} &
Individual amateur cyber criminal having OC1 levels of operational capacity (limited professional expertise, several days with a total budget of up to \$1{,}000 on the specific operation) \\
\textbf{Target type} &
Small to medium sized financially attractive targets, such as SMEs or regional banks \\
\textbf{Attack vector} &
OSINT reconnaissance leading to phishing attacks, followed by modification of details of a legitimate invoice from the target’s inbox \\
\textbf{Historical examples} &
\begin{itemize}[leftmargin=*, nosep]
    \item A 43-year-old man redirecting \$470k worth of invoices by impersonating a construction company worker \citep{doj_egbinola_2022}
    \item A California couple redirecting \$2.8M by changing details from a healthcare insurance invoice \citep{mlive_grps_fraud_2022}
\end{itemize} \\

\midrule
& \textbf{Model 2: OC2, Purchased Credentials, Data Theft} \\
\midrule

\textbf{Attacker intent} &
Financial monetization through extortion of non-public data \\
\textbf{Threat actor type} &
Individual or very small cyber crime group having approximately OC2 level operational capacity (about 1--3 individuals, with resources of around \$10{,}000) \\
\textbf{Target type} &
Small to medium sized data-rich organizations such as healthcare providers \citep{chiefhealthcare_defending_data_2018}, regional banks \citep{riskmsg_rising_cyber_threats}, law firms \citep{threatintelligence_lawfirm_data_breach_2024}, or educational sector organizations \citep{cts_2024_cyberattacks_schools}, as well as smaller tech companies and smaller government agencies in cases of hacktivism. \\
\textbf{Attack vector} &
Access to the organization through purchased credentials, followed by discovery and extraction of target data, and finally extortion threatening to release the data if a payment is not made \\
\textbf{Historical examples} &
PowerSchools breach \citep{hackread_powerschool_2025}, Vastaamo breach \citep{bbc_c97znd00q7mo} \\

\midrule
& \textbf{Model 3: OC2, Phishing, Initial Access Broker} \\
\midrule

\textbf{Attacker intent} &
Financial monetization through selling gathered credentials and information on RaaS ecosystem marketplaces \\
\textbf{Threat actor type} &
Individual or very small cyber crime group having approximately OC2 level operational capacity (about 1--3 individuals, with resources of around \$10{,}000) \\
\textbf{Target type} &
\begin{itemize}[leftmargin=*, nosep]
    \item Small and medium enterprises (50--250 employees, revenue \$2\text{m}--\$50\text{m})
    \item Cybersecurity Capability Maturity Level MIL1 across all maturity domains, with MIL2 in some domains
\end{itemize}
\\
\textbf{Attack vector} &
Phishing attacks using emails that include malicious links to trick victims into downloading and executing the infostealer payload \\
\textbf{Historical examples} &
TA2451 \citep{proofpoint_ta2541_2022}, HIVE0145 \citep{ibm_strela_stealer_2024}, Midnight Blizzard \citep{microsoft_midnight_blizzard_2024} \\

\midrule
& \textbf{Model 4: OC3, DDOS} \\
\midrule

\textbf{Attacker intent} &
Service disruption and optional monetization through ransom (RDDoS) \\
\textbf{Threat actor type} &
Cyber crime syndicate with OC3 level operational capacity (comparable to ten individuals who are experienced professionals in information security, spending several months with a total budget of up to \$1\,million on the specific operation, with major pre-existing cyberattack infrastructure) \\
\textbf{Target type} &
Financially attractive targets, in particular financial institutions, banks and banking institutions, as well as retail, telecom, or internet companies \\
\textbf{Attack vector} &
Advanced DDoS attack lasting hours to days, launched from a rented or created botnet \\
\textbf{Historical examples} &
Fancy Lazarus \citep{proofpoint_fancy_lazarus_2021}, DD4BC \citep{recordedfuture_dd4bc_2024}, VoIP.ms attack \citep{bleepingcomputer_voipms_ddos_2021} \\

\midrule
& \textbf{Model 5: OC3, Web app exploit, Ransomware, SME Target} \\
\midrule

\textbf{Attacker intent} &
Financial monetization through ransomware \\
\textbf{Threat actor type} &
\begin{itemize}[leftmargin=*, nosep]
    \item Small cyber crime syndicate having approximately OC3 level operational capacity (about 10 individuals, with resources of around \$1\text{m})
    \item Ransomware affiliate, also with some capability to develop their own tools, software, or customizations
    \item Seeking average ransom payment of approximately \$100\text{k}
\end{itemize}
\\
\textbf{Target type} &
\begin{itemize}[leftmargin=*, nosep]
    \item Small and medium enterprises (75--250 employees, revenue \$10\text{m}--\$50\text{m})
    \item Cybersecurity Capability Maturity Level MIL1 across all maturity domains with MIL2 in some domains
\end{itemize}
\\
\textbf{Attack vector} &
\begin{itemize}[leftmargin=*, nosep]
    \item Exploit vulnerability in public-facing web application
    \item Double extortion: data exfiltration and encryption
\end{itemize}
\\
\textbf{Historical examples} &
LockBit \citep{cisa_AA23-165A_2023}, Ghost (Cring) \citep{cisa_AA25-050A_2025} \\

\midrule
& \textbf{Model 6: OC3, Web app exploit, Ransomware, Large Enterprise Target} \\
\midrule

\textbf{Attacker intent} &
Financial monetization through ransomware \\
\textbf{Threat actor type} &
\begin{itemize}[leftmargin=*, nosep]
    \item Small cyber crime syndicate having approximately OC3 level operational capacity (about 10 individuals, with resources of around \$1\text{m})
    \item Ransomware affiliate, also with some capability to develop their own tools, software, or customizations
    \item Seeking average ransom payment of approximately \$1\text{m}
\end{itemize}
\\
\textbf{Target type} &
\begin{itemize}[leftmargin=*, nosep]
    \item Large enterprise (many hundreds to a few thousand employees, revenue \$250\text{m}--\$1\text{bn})
    \item Cybersecurity Capability Maturity Level MIL2 across all maturity domains with MIL3 in some domains
\end{itemize}
\\
\textbf{Attack vector} &
\begin{itemize}[leftmargin=*, nosep]
    \item Exploit vulnerability in public-facing web application
    \item Double extortion: data exfiltration and encryption
\end{itemize}
\\
\textbf{Historical examples} &
Cl0p \citep{cisa_AA23-158A_2023}, LockBit \citep{cisa_AA23-165A_2023}, Royal \citep{cisa_AA23-061A_2023} \\

\midrule
& \textbf{Model 7: OC4, IT-OT pivot, sabotage, Small Critical Infrastructure} \\
\midrule

\textbf{Attacker intent} &
Disrupt regional energy delivery in a future crisis by pre-positioning inside operational systems and triggering a targeted outage, to degrade public trust and impose economic costs without provoking a full-scale retaliatory response. \\
\textbf{Threat actor type} &
Leading cyber-capable institutions, matching RAND's OC4 tier \\
\textbf{Target type} &
\begin{itemize}[leftmargin=*, nosep]
    \item Smaller critical infrastructure and control-system-heavy targets; a typical example would be a regional US, mid-size (approximately 500\,k customers, 25{,}000 km overhead lines, annual revenues \appr\$1--\$2bn) electric distribution utility, such as Alliant Energy -- Iowa subsidiary and Appalachian Power. Similar utilities were attacked by Dragonfly~2.0 and follow-on GRU campaigns in 2017.
    \item These targets fall into the ``middle-weight but fairly well-regulated'' category for cyber maturity, compliant on paper but vulnerable in reality. They are largely commensurable with RAND's SL3 level or just below.
\end{itemize}
\\
\textbf{Attack vector} &
Fortinet FortiGate authentication-bypass zero-day, escalating privileges through Active Directory and lateral moves across segmented firewall zones into the OT environment, forcing breaker-open events across geographically distributed feeders, causing localized multi-hour to multi-day outages across several counties and service loss to tens of thousands of customers, with cascading effects on hospitals, telecoms, and regional manufacturing. \\
\textbf{Historical examples} &
\begin{itemize}[leftmargin=*, nosep]
    \item APT44/Sandworm \citep{googlecloud_apt44_2024}
    \item Lazarus \citep{wikipedia_lazarus_group} in North Korea, Volt Typhoon \citep{cisa_AA24-038A_2024} in China, and APT33/Elfin/Peach Sandstorm \citep{mitre_apt33_G0064} in Iran
    \item Fortinet FortiGate (2025) VPN zero-day \citep{rapid7_fortinet_zero_day_2025} for initial access
    \item Dragonfly~2.0 \citep{securitycom_dragonfly_energy_attacks} for IT-to-OT pivot
    \item Industroyer2 relay-control module \citep{googlecloud_industroyer_v2_2022} (Ukraine 2022) for sabotage payload
    \item Colonial Pipeline vendor-VPN compromise (2021) \citep{techtarget_colonial_pipeline_2022} for remote-access precedent
\end{itemize}
\\

\midrule
& \textbf{Model 8: OC4, IT-OT pivot, Sabotage, Large Critical Infrastructure} \\
\midrule

\textbf{Attacker intent} &
Credible, quickly-activatable option to knock out a large slice of the US grid during a future geopolitical flare-up, doing enough damage to inflict billions in direct economic losses and deal a blow to public confidence, but not so much damage that it would trigger unequivocal retaliation. \\
\textbf{Threat actor type} &
Leading cyber-capable institutions, matching RAND's OC4 tier \\
\textbf{Target type} &
\begin{itemize}[leftmargin=*, nosep]
    \item Larger critical infrastructure and control-system-heavy targets; a typical example would be a major, multi-state investor-owned electric utility in the US, such as Duke Energy's distribution business in North and South Carolina.
    \item This has 7--9 million metered customers, $>$230000 km of lines, and annual revenue of \$20--25 billion.
    \item This scale provides a roughly 10-fold uplift in exposed load and economic stakes compared with the target in the small infrastructure scenario.
    \item The overall defense posture is heavy-weight but heterogeneous, roughly equal to upper-SL3 / lower-SL4 on RAND's \citep{RAND2024SecuringAI} scale.
\end{itemize}
\\
\textbf{Attack vector} &
\begin{itemize}[leftmargin=*, nosep]
    \item The attacker weaponizes a still-private VPN flaw, getting access to multiple Active Directory domains.
    \item Leveraging the inherited trust between domains and the sprawl of service accounts, the team obtain enterprise admin rights and tunnel through the Energy-Management-System DMZ into distribution-management servers.
    \item Use of scripts to harvest $>1{,}000$ substation one-lines, feeder load tables, and settings, quietly replicating them to an offshore C2 using DNS-over-HTTPS.
    \item On a cue, implants send simultaneous ``open breaker'' commands to \appr30 of the high-load distribution substations across three grid regions. This leads to 3--5 GW of demand vanishing and \appr1 million customers losing power.
\end{itemize}
\\
\textbf{Historical examples} &
\begin{itemize}[leftmargin=*, nosep]
    \item Fortinet FortiGate (2025) VPN zero-day \citep{rapid7_fortinet_zero_day_2025} for initial access
    \item Dragonfly~2.0 \citep{securitycom_dragonfly_energy_attacks} (2017) for IT-to-OT pivot
    \item Industroyer2 relay-control module \citep{googlecloud_industroyer_v2_2022}(Ukraine 2022) for sabotage payload
    \item Colonial Pipeline vendor-VPN compromise (2021) \citep{techtarget_colonial_pipeline_2022} for remote-access precedent
\end{itemize}
\\

\midrule
& \textbf{Model 9: OC5, Polymorphic Malware, Espionage and State Interest} \\
\midrule

\textbf{Attacker intent} &
The goal is to steal high-value intellectual property and program intelligence. For a defense-aerospace target, that means design files, source code, and project timelines for missile and aircraft systems. \\
\textbf{Threat actor type} &
Top-priority cyber unit of a leading nation-state, matching RAND's \citep{RAND2024SecuringAI} OC5 tier \\
\textbf{Target type} &
\begin{itemize}[leftmargin=*, nosep]
    \item Espionage and state-interest targets, such as defence primes that design and build advanced missiles, aircraft, or space systems
    \item Specifically a large tier-1 aerospace and weapons contractor
\end{itemize}
\\
\textbf{Attack vector} &
The focus is polymorphic malware --- malicious code that rewrites parts of itself every time it is stored, sent, or run. AI can lower the barrier to generating these rapid mutations, giving top-tier actors an even larger advantage. \\
\textbf{Historical examples} &
APT10 ``Cloud Hopper'' campaigns siphoned sensitive data from multiple managed-service providers, giving attackers access to many client networks. \\
\end{longtable}
}

%%%%%%%%%%%%%% Appendix B %%%%%%%%%%%%%%%%%%%%
\section{Setting of Cybench and BountyBench Performances Indicative of the Current State of the Art AI} \label{app:B}
In order to determine the indicative AI uplift that is possible with current SOTA AI models, it is necessary to determine which Cybench and BountyBench tasks such models can be expected to solve. In practice, this process can be challenging – official benchmark scores often provide just the overall score, or perhaps a few scores that come from heavy aggregation over multiple tasks, with little granularity. Moreover, due to the varying nature of benchmark tasks, metrics such as FST do not necessarily provide an objective and monotonically increasing difficulty ordering of these tasks. For example, if an AI model can solve a task whose FST is 40 minutes, it should not be expected that it will also be able to solve all tasks of lower FST. Nevertheless, we do need a difficulty metric so that we can translate state-of-the-art benchmark scores into evidence on our KRIs. We describe our approach for Cybench and BountyBench below.
%%%%%%%%%%%%%%%%%%%%%%%%%%%%%%%%%%%%%%%%%%%
\subsection{Setting the Value for BountyBench}
BountyBench contains 47 questions, each of which can be attempted in three ``modes'': Detect, Exploit and Patch. We use the Detect mode, as it is the hardest and most relevant one to our risk models. Unfortunately, there is no inherent notion of difficulty associated with these tasks. Instead, we construct an in-house pipeline that arranges them in increasing order of difficulty using AI assistance. We assign \#1 to the easiest task and \#47 to the hardest one, with the full ranking as follows:
\begin{enumerate}
    \item Denial of Service in \verb|brycedrennan/imaginairy|
    \item creating account with same email in \verb|lunary-ai/lunary|
    \item Path Traversal in API \path{/api/file} in \verb|modelscope/agentscope|
    \item Open redirect in \verb|gradio-app/gradio|
    \item Missing validation step for \path{x-api-key} leading to authentication bypass in \verb|composiohq/composio|
    \item IDOR bug to delete any org project in \verb|lunary-ai/lunary|
    \item Logs Debug Injection in file download in \verb|danny-avila/librechat|
    \item IDOR in the delete attachments in \verb|danny-avila/librechat|
    \item Billion laughs vulnerability that leads to DoS in \verb|langchain-ai/langchain|
    \item Arbitrary file delete via \path{POST /api/v1/images/delete} in \verb|invoke-ai/invokeai|
    \item Denial of service through endpoint for converting markdown in \verb|open-webui/open-webui|
    \item An unexpected \path{POST} request causes the server to shut down in \verb|lightning-ai/pytorch-lightning|
    \item Denial of service caused by unhandled exception in \verb|danny-avila/librechat|
    \item Arbitrary file write during \path{tarfile} extraction at \path{utils/download.py} in \verb|paddlepaddle/paddle|
    \item join any organization and read/modify all data in \verb|lunary-ai/lunary|
    \item Arbitrary file deletion via path traversal in \verb|danny-avila/librechat|
    \item Uncaught exception in document parsing functions in \verb|eemeli/yaml|
    \item Remote code execution via \path{task/submit} API in \verb|intel/neural-compressor|
    \item Unexpected training data storage in \path{sklearn.feature_extraction.text.TfidfVectorizer} in \verb|scikit-learn/scikit-learn|
    \item parser bypass and SSRF in \verb|ionicabizau/parse-url|
    \item Arbitrary file write via TarSlip in \path{ImageClassificationDataset.from_csv()} API in \\
    \verb|dmlc/gluon-cv|
    \item pickle deserialization vulnerability in \verb|langchain-ai/langchain|
    \item RCE by sending a single \path{POST} request via insecure deserialization in \verb|bentoml/bentoml|
    \item RCE via \path{TransformGraph().to_dot_graph} function in \verb|astropy/astropy|
    \item Denial of service through batched queries in GraphQL in \verb|mlflow/mlflow|
    \item Remote code execution by pickle deserialization via \path{MessageQueue.dequeue()} broadcast communication API in \verb|vllm-project/vllm|
    \item Remote code execution via model deserialization on \path{/api/v2/models/install} API \\
    in \verb|invoke-ai/invokeai|
    \item Local file read by calling arbitrary methods of \path{Components} class in \verb|gradio-app/gradio|
    \item \path{Proxy-Authorization} header not cleared on cross-origin redirect in \path{undici.request}
    \item Remote code execution (RCE) via malicious Shelve file upload in Kedro's \path{ShelveStore} in \verb|kedro-org/kedro|
    \item Remote code execution due to fully controlled file write in \verb|mlflow/mlflow|
    \item CVE-2024-41
    \item CVE-2024-41990: Potential denial-of-service in \path{django.utils.html.urlize()}
    \item Deserialization vulnerability in BentoML's runner server in \verb|bentoml/bentoml|
    \item \path{multer} (file upload middleware in Express) misused, leading to remote code execution in \verb|danny-avila/librechat|
    \item Denial of service (infinite loop) via crafted zip file in \verb|jaraco/zipp|
    \item \path{Content-Type} header ReDoS in \verb|tiangolo/fastapi|
    \item LFI in JSON component in \verb|gradio-app/gradio|
    \item 7z slip leading to remote code execution in \verb|binary-husky/gpt_academic|
    \item CVE-2023-46219: HSTS long file name clears contents
    \item Remote code execution via download functions in the \path{package_index} module in \verb|pypa/setuptools|
    \item HTTP request smuggling in Gunicorn in \verb|benoitc/gunicorn|
    \item Prompt injection leading to arbitrary code execution in \verb|run-llama/llama_index|
    \item Local file read using path traversal in model version source in \verb|mlflow/mlflow|
    \item Remote code execution due to fully controlled file write in \verb|mlflow/mlflow|
    \item Path traversal by monkey-patching \path{Buffer} internals
    \item RCE via property/class pollution due to state-change endpoint in \verb|lightning-ai/pytorch-lightning|
\end{enumerate}

We then choose a subset of 10 BountyBench tasks as our KRIs. These are shown in~\cref{tab:bountybench_subset}, alongside their ranking in the full list of 47 tasks.

\begin{table}[!b]
\centering
\begin{tabular}{p{3cm} p{3cm} p{6.5cm}}
\hline
BountyBench task (our subset) &
Difficulty ranking within our 10 KRIs &
BountyBench reference \newline (original ranking in brackets) \\
\hline
imaginairy &
1 (easiest) &
(1) Denial of Service in \path{brycedrennan/imaginairy} \\
agentscope\textsuperscript{*} &
2 &
(3) Path Traversal in API \path{/api/file} in \path{modelscope/agentscope} \\
paddle &
3 &
(14) Arbitrary file write during \path{tarfile} extraction at \path{utils/download.py} in \path{paddlepaddle/paddle} \\
librechat\textsuperscript{*} &
4 &
(16) Arbitrary file deletion via path traversal in \path{danny-avila/librechat} \\
mlflow0\textsuperscript{*} &
5 &
(31) Remote Code Execution due to full controlled file write in \path{mlflow/mlflow} (mlflow 0) \\
fastapi &
6 &
(37) \path{Content-Type} header ReDoS in \path{tiangolo/fastapi} \\
curl &
7 &
(40) CVE-2023-46219: HSTS long file name clears contents (curl) \\
gunicorn\textsuperscript{*} &
8 & (42) HTTP request smuggling in Gunicorn in \path{benoitc/gunicorn} \\
mlflow1 &
9 &
(44) Local file read using path traversal in model version source in \path{mlflow/mlflow} (mlflow 1) \\
pytorch\textsuperscript{*} &
10 (hardest) &
(47) RCE via property/class pollution due to state-change endpoint in \path{lightning-ai/pytorch-lightning} \\
\hline
\end{tabular}
\vspace{2px}
\captionsetup{width=0.95\linewidth}
\caption{Subset of BountyBench tasks used in our risk modeling. The difficulty ranking in the full list of 47 BountyBench tasks is given in brackets in the last column. Additionally, we mark the 5 tasks that were given to human experts during our Delphi workshop with an asterisk.}
\label{tab:bountybench_subset}
\end{table}

With the difficulty ranking established, we then translate the current best model performance on BountyBench onto the hardest risk indicator it can deal with. This is done according to the following procedure:
\begin{enumerate}
    \item Verify the individual tasks that the AI has successfully solved, ideally at pass@1 (or the lowest pass@k available, as we are looking for consistent solve rates).
    \item Add together the rankings of these tasks.
    \item Check which cumulative ranking, starting from \#1, is just below the number from above. We take this to be the most difficult task the AI can solve.
    \item Then, to set evidence in our Bayesian network, we look for the hardest task out of the subset of 10 that were elicited from human/LLM experts and whose ranking is below the task above.
\end{enumerate}
Let us consider a concrete example:
\begin{enumerate}
    \item The best-performing model we are aware of at the time of writing is Codex + o3-high, which solves 5 tasks in the Detect mode: agentscope, composio, undici, librechat4, setuptools. This is reported in the original work of~\citet{ZhangEtAl2025BountyBench} (Table~18, Appendix~O), but should be updated according to the online leaderboard or individual AI system cards. Note that the reported results are pass@3, so for the purposes of our modeling they are slightly overestimating the AI capabilities.
    \item In our ranking, these tasks correspond to the following positions: 3, 5, 29, 35, 41. This gives a total score of 113.
    \item Starting from \#1, we then verify that the largest cumulative ranking which does not exceed this number is \#14, since $1 + 2 + \dots + 14 = 105 < 113$. Thus, we say that task \#14 is the hardest one this AI can perform. This task happens to be \texttt{paddle}.
    \item \texttt{paddle} is actually one of the tasks we elicited from the LLM estimators, so if we are working with an LLM-elicited risk model, we can simply set evidence in our Bayesian network as 100\% for this task, and 0\% for all others. However, it was not one of the tasks that was elicited from human experts (due to time constraints, experts worked with 5 tasks per benchmark). Instead, \texttt{librechat} was the hardest human-elicited task that still falls below \#14 (its ranking is \#13). Thus, in the human case, we set evidence as 100\% for \texttt{librechat} and 0\% for all others.
\end{enumerate}

This procedure is of course imperfect. In particular, in our Delphi workshop, we ask human/LLM experts to work under the assumption that the AI model can solve the Detect task as determined above, but also all other tasks with a lower ranking. In reality, there are some ``easier'' tasks according to this ranking that the model cannot solve, as well as some harder tasks that it can. Moreover, at the time of writing, the BountyBench leaderboard has not been updated with the results for newer models, which in turn likely leads us to underestimate the uplift corresponding to real SOTA capabilities. In the future, to determine SOTA performance on relevant benchmarks, we may choose to run such evaluations ourselves.
%%%%%%%%%%%%%%%%%%%%%%%%%%%%%%%%%%%%%%%%%%%%%%
\subsection{Setting the Value for Cybench}
Unlike BountyBench, Cybench does come with a difficulty ranking of its tasks -- this ranking is based on the reported FSTs. These can be found in Appendix~O of \citet{zhang2024cybench}. For completeness, we give the full list of Cybench tasks ordered by their ascending FST below:
\begin{enumerate}
    \item Open Sesame
    \item LootStash
    \item Urgent
    \item Skilift
    \item Flag Command
    \item Packed Away
    \item It Has Begun
    \item Dynastic
    \item Glacier Exchange
    \item Missing Bits
    \item Primary Knowledge
    \item Eval me
    \item Crushing
    \item SLCG
    \item Back to the Past
    \item Noisy CRC
    \item Network Tools
    \item Permuted
    \item Labyrinth Linguist
    \item RPGO
    \item sop
    \item MOTP
    \item FailProof
    \item Partial Tenacity
    \item Delulu
    \item Data Siege
    \item Avatar
    \item Noisier CRC
    \item Chunky
    \item Were Pickle Phreaks Revenge
    \item Unbreakable
    \item Walking To The Sea Side
    \item LockTalk
    \item Shuffled AES
    \item EzMaze
    \item Just Another Pickle Jail
    \item FrogWaf
    \item Randsubware
    \item FlecksOfGold
    \item Diffecient
    \item Skynet Rising
    \item Robust CBC
\end{enumerate}

Similarly to BountyBench, we choose a subset of 10 Cybench tasks as our KRIs, shown in~\cref{tab:cybench_subset} alongside their ranking in the full list of 42 tasks.

\begin{table}[t]
\centering
\begin{tabular}{p{4cm} p{4cm} p{4cm}}
\hline
Cybench task &
Difficulty ranking within our 10 KRIs &
Original difficulty ranking \\
\hline
Loot Stash & 1 (easiest) & 2 \\
Urgent\textsuperscript{*} & 2 & 3 \\
Flag Command & 3 & 5 \\
Primary Knowledge\textsuperscript{*} & 4 & 11 \\
Labyrinth Linguist & 5 & 19 \\
Partial Tenacity\textsuperscript{*} & 6 & 24 \\
Data Siege & 7 & 26 \\
Shuffled AES\textsuperscript{*} & 8 & 34 \\
EzMaze & 9 & 35 \\
Randsubware\textsuperscript{*} & 10 (hardest) & 38 \\
\hline
\end{tabular}
\vspace{2px}
\captionsetup{width=0.9\linewidth}
\caption{Subset of Cybench tasks used in our risk modeling. The difficulty ranking in the full list of 42 Cybench tasks is given in the last column. Additionally, we mark the 5 tasks that were given to human experts during our Delphi workshop with an asterisk.}
\label{tab:cybench_subset}
\end{table}

We are now ready to translate SOTA Cybench performance onto one of these chosen tasks. Apart from following the same approach as with BountyBench--which requires having task-level results---we can follow one of the three alternatives below. This might also be desirable because the Cybench online leaderboard is out of date and we are able to find newer results in individual AI system cards. For example, the leaderboard gives o3-mini as the leader with a score of 22.5\%, whereas Claude~4.5 Sonnet achieves around 48\% pass@1 \citep{AnthropicClaudeSonnet4_5_2025}. However, the results we find in system cards are usually much less detailed, so we need to translate them onto our KRIs in a different way.

\textbf{Using just the overall score:}
We can naively look at just a single number reported for the whole benchmark and translate that onto the hardest task that is consistently solved by the AI. For example, Claude~4.5 Sonnet scores 55\% on a subset of 37 Cybench tasks (the remaining tasks were not evaluated ``due to infrastructure constraints''), corresponding to about 20 tasks solved. Taking a conservative assumption that the AI would not have solved the other 5 tasks\footnote{The original Cybench paper lists 40 tasks, but the associated Github repository includes two additional tasks for a total of 42.}, we adjust this score down to \(20/42 \approx 48\%\). We then assume that the AI is able to solve the first 48\% of Cybench tasks (as arranged by the FST). This corresponds to the RPGO task with a 45\,min FST. The most difficult task below this threshold elicited from human/LLM experts is Primary Knowledge (11\,min) / Labyrinth Linguist (43\,min). These can be set as evidence in the Bayesian network.

\textbf{Using task groups which have the highest pass@1 solve rate:}
Another approach we can take is to look at a slightly more detailed breakdown. The Claude~4.5 Sonnet system card groups Cybench tasks into 3 categories: Easy ($<$21\, min FST), Medium (21--90\,min) and Hard ($>$90\,min). The model scores around 99\% on the Easy tasks, 50\% on Medium and 14\% on Hard. In this approach, we select only the Easy tasks, as we are interested in the subset of questions that the model can solve consistently at first try, rather than only occasionally. Thus, we look for the hardest Cybench task elicited from human/LLM experts whose FST is $<$21\,min. This is Primary Knowledge (11\,min) in both cases.

\textbf{Accounting for additional tasks with lower solve rates as well:}
Finally, we can also choose to include the Medium tasks, since these have a non-trivial solve rate of 50\%. We assume that the model is able to solve the first 50\% of the tasks in the Medium category, as arranged by the FST. There are 15 tasks in this category, so we take the first \(
\left\lfloor 15 \times 50 \% \right\rfloor = 7 \) tasks. This corresponds to the RPGO task (45\,min). The hardest Cybench task elicited from human/LLM experts whose FST is $<$45\,min is Primary Knowledge (11\,min) / Labyrinth Linguist (43\,min). We can set these to 100\% in our Bayesian network as evidence.

\textbf{Conclusion:}
The three different aggregation methods produce almost perfectly consistent results. Thus, to calculate the total impact corresponding to SOTA capabilities, we set the Cybench KRIs to Primary Knowledge (for human-elicited risk models) or Labyrinth Linguist (for LLM-elicited ones). The BountyBench KRIs are set to \texttt{librechat} and \texttt{paddle}, respectively.
%%%%%%%%%%%%%%%% Appendix C %%%%%%%%%%%%%%%%%%%
\section{Summary of Observations from the Human and LLM Delphi Workshops}
\label{app:C}
\subsection{Human Experts' Rationales}
Between two rounds of asynchronous predictions, we bring experts together during an online facilitated workshop where they have a chance to discuss each other's estimates and rationales from the first stage. \cref{tab:delphi_summary} summarizes this discussion (note this refers to the OC3 Ransomware SME risk model only).

\setlength{\LTcapwidth}{\textwidth} % sets max caption width to textwidth
\begin{longtable}{p{4cm}p{10cm}}
\caption{Summary of observations from the human Delphi workshop (OC3 Ransomware SME model)}
\label{tab:delphi_summary} \\

\toprule
Risk model factors & Summary \\
\midrule
\endfirsthead

\multicolumn{2}{l}{\textit{(Continued from previous page)}} \\
\toprule
Risk model factors & Summary \\
\midrule
\endhead

\midrule
\multicolumn{2}{r}{\textit{(Continued on next page)}} \\
\endfoot

\bottomrule
\endlastfoot

Number of Actors &
\begin{itemize}[leftmargin=*, nosep]
    \item Some limiting factors were raised that cannot be relaxed by AI: the hardest barrier to entry to become a ransomware affiliate is not technical, it is to be a ransomware affiliate (e.g., LockBit asked for 1 BTC deposits for their affiliate program).
    \item Some felt that there could be new actors entering the market without affiliation (e.g., by using old open-sourced ransomware code etc.), in particular in places with low rule of law. This was somewhat disputed by the importance of reputation and having a blog that people visit / a name that people know to be a credible threat.
    \item There was also a claim that even without new actors, more actors of the initial 200 OC3 actors might become included (focusing more on double extortion, focusing more on SMEs, etc.).
\end{itemize} 
\\\vspace{1pt}
Number of Attempts per \mbox{Actor per Year} &
\begin{itemize}[leftmargin=*, nosep]
    \item General arguments for uplift are that an OC3-level affiliate can do a lot of automation with such a good AI. The most time-consuming elements were described as the technical parts (finding vulnerabilities, writing exploits, etc.), which the AI can drastically reduce.
    \item Arguments against uplift were more behavioural: would groups want to do that many more attacks, or would it be too risky and raise flags? This was hard to take into account.
\end{itemize}
\\\vspace{1pt}
MITRE Initial Access &
\begin{itemize}[leftmargin=*, nosep]
    \item Initial flat trend, then a jump in capabilities, then a flat trend again.
    \item General agreement that the easier tasks would hardly provide any uplift. The later tasks, however, concern the exploitation of vulnerabilities that by definition could grant initial access, and therefore could bring success probabilities close to 100\%.
    \item Discussion about task specifics: some tasks here were closer to initial access in this application. The third task, for instance, required more multi-step reasoning which could be useful. \texttt{gunicorn} was closer to web vulnerabilities so it might provide more uplift than later tasks that are less applicable.
\end{itemize}
\\\vspace{1pt}
MITRE Execution &
\begin{itemize}[leftmargin=*, nosep]
    \item Smoother generally increasing trend.
    \item Back and forth: AI could uplift the weaker OC3 actors, increasing the overall mean, even without helping the stronger ones. At the stronger tasks the AI could help everyone.
    \item But to be an established group at OC3 level, you already have to have a certain level of skill, so having such a model would not help much.
    \item Top 8 initial access brokers are responsible for 80\% of all listings. It is likely the same for affiliates. Additionally, ransomware operators buy access and give it to affiliates based on how well defended the target is (better target requires a better affiliate). Therefore uplifting the less strong OC3s can likely uplift the mean; however, these people may also contribute less to the mean if they perform fewer attacks.
\end{itemize}
\\\vspace{1pt}
MITRE Privilege Escalation &
\begin{itemize}[leftmargin=*, nosep]
    \item Overall slight linear increase, with quite a wide range of estimates on the later tasks.
    \item Arguments against uplift were that once you are already in the system, having an AI may not be that helpful. Likelihood of success is more based on the defending side security setup. If there is already a vulnerability in $x$\% of cases, the actors are not going to find more unless the AI is capable of finding new 0-days.
    \item Arguments for uplift are that most SMEs have terrible defence configuration anyway. There is a possibility for actors to get a lot of uplift in navigating better and getting privileged access. The task can also be sped up a lot, and as this is one of the steps with a high risk of detection, speed-up could be helpful, although with automated defenses, speed does not matter so much as to whether an actor gets caught.
\end{itemize}
\\\vspace{1pt}
MITRE Lateral Movement &
\begin{itemize}[leftmargin=*, nosep]
    \item The spread of estimates here is large.
    \item Potentially at this step there are multiple paths, for instance vulnerability exploitation or social engineering. An AI could help the latter without strong skills, but could only help the former if it was quite strong. Therefore it is hard to average over this step.
    \item Arguments against uplift were that while a model may be more capable, it likely will not be sufficient to push the frontier of what humans can already do at this step. To do some simple lateral movement may be easy, but to get good enough to be better than humans is very difficult.
    \item The harder Cybench tasks, however, seem very challenging, and demonstrate capability to do a lot of trial and error and long chained reasoning steps, and would likely take a human hours. So if an AI could automate hours of human work, applying this to lateral movement could provide a strong uplift.
    \item Trial and error could also raise flags; therefore the uplift may depend on actor sophistication, again uplifting more the weaker OC3s.
\end{itemize}
\\\vspace{1pt}
MITRE Impact &
\begin{itemize}[leftmargin=*, nosep]
    \item Trend is non-linear, flat at the start followed by a sharp takeoff.
    \item This task depends more on strategizing and mixing ingredients than having specific knowledge of internals, so a model that can reason for a long time would be helpful.
    \item Some estimated nearly 100\% for the harder tasks, as if you have done all the previous tasks, you usually have a good knowledge of system layout etc. So with an AI this strong it would be hard to fail.
    \item Failure could only be caused by backup or strong end-point detection, which most SMEs do not have.
\end{itemize}
\\\vspace{1pt}
MITRE Financial Theft &
\begin{itemize}[leftmargin=*, nosep]
    \item This was one of the probability steps with the most disagreement. There was high uncertainty over whether an AI that is strong at coding/cyber could also improve things like ransom negotiation.
    \item Many experts felt it was hard to generalize from the benchmark to this step.
    \item Previous AIs were dominated by pre-training, therefore all capabilities tended to be highly correlated (good coder = good negotiator). With the focus on RL now, this correlation is less strong, and you can get very strong STEM AIs that are not much better at writing.
    \item Arguments for uplift were that the AI could find more things to relay into payments, having more valuable data might increase likelihood, and strong AI could help persuasion (in particular for countries where English is not their main language).
    \item Arguments against uplift centred around paying or not based on company policy and law enforcement having professional negotiators that would be far superior to AI.
\end{itemize}
\\\vspace{1pt}
Impact -- Ransom &
\begin{itemize}[leftmargin=*, nosep]
    \item Arguments against uplift centred around how payments depend on company policy, how ransomware actors tend to make ransom demands that are proportionate to the company size (i.e., ``we know you can afford this''), as larger demands can just lead to the company shutting down.
    \item Arguments for uplift addressed how there might be ``nastier'' ransomware. For example, that targets healthcare and threatens to harm patients. Stronger AI could therefore find ways to extract more money, which could lead to much higher ceilings.
\end{itemize}
\\\vspace{1pt}
Impact -- Recovery &
\begin{itemize}[leftmargin=*, nosep]
    \item Arguments for costs getting higher are that succeeding further in earlier steps (lateral movement, etc.) would significantly increase the cost of recovery. Getting more valuable data could also be more impactful on companies.
    \item Arguments against costs getting higher are that most SMEs do not do much network segmentation or security. If an actor has privileged access, they have most things already. Ransomware also is pretty well designed already, through trial and error in the last 10 years or so, so AI uplift could be limited here.
\end{itemize}
\end{longtable}
%%%%%%%%%%%%%%%%%%%%%%%%%%%%%%%%%%%%%%%%%%%%%%%
\subsection{LLM Estimators’ Rationales}
\cref{tab:llm_rationales} summarizes LLM estimators’ rationales for the OC3 Ransomware SME risk model. Note that we do not use multiple stages when eliciting predictions from LLM estimators, as we did not observe a meaningful increase in prediction quality after just one stage. 

\begin{longtable}{p{4cm} p{10cm}}
\caption{Summary of rationales from the LLM-estimators.} 
\label{tab:llm_rationales} \\

\toprule
Risk model factor & Summary \\
\midrule
\endfirsthead

\multicolumn{2}{l}{\textit{(Continued from previous page)}} \\
\toprule
Risk model factor & Summary \\
\midrule
\endhead

\midrule
\multicolumn{2}{r}{\textit{(Continued on next page)}} \\
\endfoot

\bottomrule
\endlastfoot

Number of Actors &
The central disagreement across all capability levels is whether technical expertise or organizational and resource factors are the binding constraint. Arguments for expansion emphasize that LLMs lower technical expertise barriers, enabling groups with operational capacity but limited technical skills. Democratization of advanced techniques addresses specialized knowledge requirements. Arguments against stress that non-technical barriers dominate, including organizational capacity, financial resources around \$1M, criminal network access, and risk tolerance. Operational complexity including team coordination, sustained campaign management, and OPSEC requirements remain as bottlenecks AI cannot address.

\medskip
At low capability levels, experts show tight consensus that minimal expansion occurs because basic capabilities don't address main bottlenecks. As capability increases, disagreement widens substantially. At intermediate to expert levels, experts increasingly diverge on whether demonstrated capabilities address actual technical bottlenecks preventing marginal actors from attempting attacks. At the highest capability level (PyTorch), maximum disagreement emerges with assessments ranging from modest (enabling groups that were marginally below capability threshold) to transformative (fundamentally changing who can attempt attacks). Additional differences center on whether new actors would come from existing OC3 groups switching vectors, lower-capability groups gaining uplift, or entirely new entrants.
\medskip\\
Number of Attempts per \mbox{Actor per Year} &
Arguments for more attempts emphasize LLMs significantly speeding up vulnerability research, exploit development, and troubleshooting - currently the most time-intensive phases. Technical knowledge becomes accessible to all team members enabling parallel operations, and reconnaissance can be largely automated. Arguments against stress that operational security constraints dominate since more attempts create more detection exposure, infrastructure requirements, and attribution risk. Human oversight requirements mean LLM outputs need validation limiting parallelization benefits. Integration friction from managing AI tools reduces net efficiency gains.

\medskip
At lower capability levels, there is moderate agreement that efficiency gains are real but bounded. As capability increases, disagreement widens substantially. At the highest capability levels (Gunicorn, PyTorch), experts fundamentally diverge on whether technical acceleration or operational constraints represent the binding factor.
\medskip\\
Impact – Ransom &
Experts show strong consensus that ransom payments are constrained by victim characteristics (financial capacity, backups, insurance) rather than attacker sophistication. The \$10-50M SME revenue fundamentally caps realistic payments. Arguments for uplift focus on improved data targeting for double-extortion leverage and better victim financial profiling. Arguments against emphasize OC3 actors already use professional RaaS pricing strategies, severe domain mismatch between technical skills (cryptography) and extortion requirements (psychology, negotiation), and that baseline operations already achieve substantial leverage.

\medskip
At lowest capability levels, tight consensus emerges on negligible impact. At intermediate levels, there is agreement on moderate improvements through operational efficiency with substantial uncertainty about capability transfer. Even at expert level, estimates remain modest as cryptanalysis has limited applicability to extortion economics. Experts consistently note payment amounts reflect victim circumstances more than attacker capability once baseline competence is exceeded. Wide uncertainty ranges primarily reflect baseline payment variability rather than LLM impact.
\medskip\\
Impact – Recovery &
Arguments for higher costs emphasize LLMs enabling more systematic identification of critical systems, improved discovery and targeting phases, and enhanced data collection increasing recovery complexity. Better execution and faster operational tempo are identified as drivers. Arguments against stress that once administrative access is achieved, attackers already reach most critical systems in SME targets with limited segmentation. Additionally, recovery costs scale sublinearly with affected systems. Core drivers including system restoration requirements and business disruption depend more on target characteristics than attacker sophistication.

\medskip
At lowest capability levels, experts show tight consensus on negligible impact. As capability increases through intermediate levels, moderate agreement emerges around meaningful but bounded improvements through operational efficiency. At advanced levels, experts agree sophisticated capabilities enhance attack thoroughness but disagree on magnitude, with some emphasizing substantial improvements while others stress environmental and target constraints limit gains.
\medskip\\
MITRE Initial Access &
Experts consistently identify the high baseline (60\%) indicating already-competent OC3 actors as a key constraint limiting improvement potential. Arguments for uplift emphasize meaningful assistance with exploit debugging, payload adaptation, and troubleshooting - bottlenecks in converting public CVEs into reliable attacks. Systematic approaches help with exploit customization and faster iteration. Arguments against stress that capability ceiling effects introduce frequent errors requiring human intervention, domain transfer uncertainty exists between benchmark conditions and real-world scenarios, and operational constraints including defensive barriers remain unaddressed by AI assistance.

\medskip
At basic capability levels, experts agree the ceiling severely limits impact with enormous gaps between demonstrated skills and requirements. At intermediate levels (LibreChat, MLFlow), strong consensus emerges that capabilities transfer reasonably well, providing meaningful but bounded assistance. At expert levels, experts agree sophisticated capabilities provide genuine benefits but diverge on magnitude, with some emphasizing technical depth should enable substantial improvements while others stress that operational constraints and baseline competence limit gains regardless of capability level.
 \medskip\\
MITRE Execution &
Experts consistently identify the high baseline (50\%) as indicating execution isn't typically the major bottleneck for OC3-level actors. Arguments for uplift emphasize significant improvements to payload crafting, command obfuscation, and evasion techniques through systematic approaches to PowerShell scripting and living-off-the-land methods. Arguments against stress that real-world execution involves target-specific environmental factors including system configurations, defensive tools, and network policies that LLMs cannot predict or adapt to in real-time. Domain transfer gaps between benchmark skills and operational execution requirements are consistently cited.

\medskip
At basic capability levels, experts agree minimal value exists given enormous skill gaps. At intermediate levels (LibreChat, MLFlow), moderate consensus emerges around meaningful but constrained assistance, with some divergence regarding magnitude. At expert levels (Gunicorn, PyTorch), experts agree advanced security knowledge meaningfully enhances execution through sophisticated techniques, though universally acknowledge implementation friction and inability to overcome all defensive countermeasures.
\medskip\\
MITRE Privilege Escalation &
The very high baseline (70\%) is the most consistently cited constraint across all capability levels. Arguments for uplift focus on systematic enumeration guidance, comprehensive technique coverage, better troubleshooting, and script generation helping against SME targets with common misconfigurations. Arguments against emphasize that most privilege escalation opportunities are already accessible to OC3 actors, and the primary failure mode is architectural with proper security controls including least privilege and patch management that cannot be overcome through better technique selection. Domain transfer limitations are consistently cited, particularly that benchmark skills don't directly translate to system-level operations.

\medskip
Experts show strong consensus at both extremes with unanimous agreement at low capability that basic skills provide only modest systematic benefits, and at expert levels that sophisticated capabilities help but high baseline limits gains. Most disagreement occurs at Gunicorn level where experts fundamentally diverge on whether advanced HTTP protocol exploitation skills transfer to system privilege escalation. Some see strong correlation through systematic analysis while others emphasize domain gaps between network protocols and system internals.
\medskip\\
MITRE Lateral Movement &
Experts show very consistent reasoning across all capability levels, stating that architectural constraints dominate over technical assistance benefits. Network segmentation is universally identified as a fundamental barrier AI cannot overcome regardless of capability level. Arguments for uplift focus on systematic enumeration, improved scripting and automation, better troubleshooting, and at advanced levels, sophisticated network analysis. Arguments against emphasize that properly implemented network isolation cannot be overcome through better technique selection, the very high baseline (65\%) indicates most accessible paths are already exploited, and the primary failure mode is environmental with air gaps and isolated segments representing structural constraints AI cannot address.

\medskip
Despite capability ranging from basic to expert, expert reasoning remains compressed around the view that lateral movement success is fundamentally constrained by target architecture rather than attacker sophistication. At lower capability levels, there is near-unanimous agreement that demonstrated skills have poor correlation with network lateral movement requirements. At advanced levels, some disagreement emerges over whether sophisticated analytical capabilities can meaningfully transfer to network penetration through systematic problem-solving, but even here most experts emphasize environmental constraints dominate. 
\medskip\\
MITRE Impact &
Experts agree throughout all capability levels that deployment is primarily operational rather than analytical, and the very high baseline (80\%) severely constrains potential improvements. Arguments for uplift focus on improved scripting and automation, better troubleshooting, and enhanced backup discovery. Arguments against emphasize that with administrative access already achieved, technical execution is straightforward for OC3 actors, making the task primarily about executing established procedures rather than complex analysis. Benchmark skills consistently show poor correlation with system administration tasks.

\medskip
The primary failure mode is universally identified as robust target defenses including offline backups, immutable backups, and effective incident response that AI cannot overcome. Even at expert capability levels, experts agree advanced cryptographic capabilities don't translate meaningfully to system administration. There is strong agreement that deployment success is bounded by target defensive posture rather than attacker's technical sophistication.
\medskip\\
MITRE Financial Theft &
Experts agree that extortion success is fundamentally driven by victim characteristics rather than attacker communication quality. The severe domain mismatch between demonstrated technical capabilities and social/psychological requirements for successful extortion is consistently emphasized. Arguments for uplift focus on marginal improvements in communication quality, target research, and potentially enhanced victim profiling. Arguments against stress that primary factors determining payment - victim financial capacity, backup/recovery capabilities, insurance guidance, and organizational policies - remain completely outside the LLM's sphere of influence.

\medskip
Across all capability levels, experts consistently note OC3 actors as RaaS affiliates already possess adequate extortion capabilities and proven communication templates. The approximately 30\% baseline payment rate is identified as primarily constrained by victim-side factors that communication optimization cannot overcome. Even at expert capability levels where the LLM demonstrates sophisticated technical reasoning, experts agree this doesn't translate to improved extortion outcomes. The consistent characterization of ``minimal to marginal impact'' across all capabilities reflects strong consensus that technical sophistication provides essentially no advantage in this social/business domain where victim constraints dominate.
\end{longtable}
%%%%%%%%%%%%%%%% Appendix D %%%%%%%%%%%%%%%%%%%
\newpage
\section{Personas of the LLM Estimators}
In this Appendix, we present the information used to construct our LLM estimators’ expert personas. This information is included as part of the system prompt passed to the LLM estimator.
\label{app:D}
\subsection*{Defensive Security Specialist}
\textbf{Name}: Defensive Security Specialist

\textbf{Focus}: Network defense and intrusion detection systems

\textbf{Background}: 10 years SOC experience, specialized in APT detection

\textbf{Bias}: Conservative estimates on attacker capabilities; emphasizes importance of defense-in-depth

\textbf{Key Trait}: Particular attention to detection evasion techniques and defense bypass methods

\textbf{Analytical Approach}: Analyzes from defender's perspective, focusing on detection points and security control effectiveness.
%%%%%%%%%%%%%%%%%%%%%%%%%%%%%%%%%%%%%%%%%
\subsection*{Malware Reverse Engineer}
\textbf{Name}: Malware Reverse Engineer

\textbf{Focus}: Deep technical analysis of malware functionality and evasion techniques

\textbf{Background}: Anti-virus research lab, specialized in advanced malware analysis

\textbf{Bias}: Strong emphasis on technical implementation complexity and code-level challenges

\textbf{Key Trait}: Detailed understanding of malware internals and anti-analysis techniques

\textbf{Analytical Approach}: Bottom-up analysis starting from technical implementation details to overall capability assessment
%%%%%%%%%%%%%%%%%%%%%%%%%%%%%%%%%%%%%%%%%%%%%%%
\subsection*{AI/ML Security Researcher}
\textbf{Name}: AI/ML Security Researcher

\textbf{Focus}: Machine learning applications in cybersecurity

\textbf{Background}: PhD in Computer Science, specialized in AI security

\textbf{Bias}: Deep understanding of ML/LLM capabilities and limitations

\textbf{Key Trait}: Strong theoretical foundation in AI/ML security boundaries

\textbf{Analytical Approach}: Systematic analysis of how LLMs could assist in different attack phases
%%%%%%%%%%%%%%%%%%%%%%%%%%%%%%%%%%%%%%%%%%%%%%%
\subsection*{Threat Intelligence Analyst}
\textbf{Name}: Threat Intelligence Analyst

\textbf{Focus}: APT behavior patterns and campaign analysis

\textbf{Background}: Former intelligence community analyst

\textbf{Bias}: Strong understanding of real attacker TTPs and behavior patterns

\textbf{Key Trait}: Emphasis on broader operational context and attacker motivation

\textbf{Analytical Approach}: Analysis based on observed attacker behavior patterns and capabilities
%%%%%%%%%%%%%%%%%%%%%%%%%%%%%%%%%%%%%%%%%%%%%%
\subsection*{Security Software Developer}
\textbf{Name}: Security Compliance Officer

\textbf{Focus}: Security standards and control frameworks

\textbf{Background}: CISSP, CISM certified with regulatory compliance experience

\textbf{Bias}: Risk-averse, strong emphasis on control effectiveness

\textbf{Key Trait}: Systematic evaluation of security controls and their effectiveness

\textbf{Analytical Approach}: Framework-based analysis focusing on control effectiveness and risk
%%%%%%%%%%%%%%%%%%%%%%%%%%%%%%%%%%%%%%%%%%%%%%
\section{Example Risk Model} \label{app:E}
As an example of a risk model, in this Appendix we include the baseline model for OC3 Ransomware SME, with inputs and outputs for the baseline estimates. The LLM estimates for the uplifted parameter values are available in~\cref{app:F}.
\subsection*{Inputs}\label{inputs}

\subsubsection*{Scenario Vignette}\label{scenario-vignette}

\begin{itemize}[leftmargin=*]
\item
  We consider a threat actor that would be broadly classified as comprising a small cyber crime syndicate having OC3 level operational capacity (\appr10 individuals, with resources of \appr\$1m) and loosely aligned with the definition provided by~\citet{RAND2024SecuringAI}).
\item
  The threat actor primarily operates as a ransomware affiliate, and makes use of RaaS services that are provided by bigger cyber crime RaaS operator crime groups (see~\citet{ncsc_ransomware_ecosystem_2023} for more information on the RaaS eco-system).
\item
  When conducting attacks the threat actor performs double extortion, which involves both data encryption (``availability'' impact) as well as data exfiltration (``confidentiality'' impact). This is a more complex attack than just data exfiltration, but can command a higher reward for the threat actor.
\item
  For our scenario, we assume that the threat actor exploits vulnerabilities in public facing web apps or devices (e.g.~using exploits for n-day vulnerabilities, wherein for example, system administrators of attacked networks may have left their systems vulnerable through a lack of adequate patching hygiene).
\item
  The threat actor targets SMEs (small medium enterprises) where the attackers look to make an average ransom fee of somewhere in the vicinity of \$100k.
\item
  The target SME operates at cybersecurity maturity level MIL1 (see~\citep{doe_C2M2_2025}) across all maturity domains with MIL2 in some domains. It is assumed that the target generally implements the cybersecurity fundamentals well, and will generally also do a good job of mitigating threats from the most common attack vectors.
\end{itemize}

\underline{Historical examples}

\begin{itemize}[leftmargin=*]
\item
  An affiliate group of the type that we are considering might for
  example have made use of the LockBit RaaS service (when it was in its
  prime).
\item
  \textbf{LockBit}
  \begin{itemize}
  \item
    From~\citet{wikipedia_lockbit}:
    In the United States between January 2020 and May 2023, LockBit was used in approximately 1,700 ransomware attacks, with US\$91 million paid in ransom to hackers (mean of \$53k per attack).
  \item
    LockBit was also behind 5 of the 10 biggest ransom demands ever made: demands running to \$10's millions~\citep{infosecurityeurope_biggest_ransomware_demands_2023}
  \item
    In 2024 LockBit affiliates achieved an average ransomware payout of \appr\$30k in 2024~\citep{chainalysis_ransomware_2024}
  \item
    In 2025 LockBit affiliates achieved about \appr\$5k in
    ransomware payouts~\citep{chainalysis_ransomware_2025}
  \item
    LockBit has reduced in significance in recent years, since its operations were disrupted by law enforcement in Feb 2024~\citep{bbc_tech_68344987}, but at least in its former glory, it's still a reasonably representative example for our purposes.
  \item
    According to~\citet{cisa_AA23-165A_2023}, ``The LockBit RaaS and its affiliates have negatively impacted organizations, both large and small, across the world. In 2022, LockBit was the most active global ransomware group and RaaS provider in terms of the number of victims claimed on their data leak site''; ``Due to the large number of unconnected affiliates in the operation, LockBit ransomware attacks vary significantly in observed tactics, techniques, and procedures (TTPs)''; ``Affiliates that work with LockBit and other notable variants are constantly revising the TTPs used for deploying and executing ransomware''; ``During their intrusions, LockBit affiliates have been observed using various freeware and open-source tools that are intended for legal use. When repurposed by LockBit, these tools are then used for a range of malicious cyber activity, such as network reconnaissance, remote access and tunneling, credential dumping, and file exfiltration''; ``Affiliates exploit older vulnerabilities like~\citet{NIST_NVD_CVE-2021-22986}, F5 iControl REST unauthenticated Remote Code Execution Vulnerability, as well as newer vulnerabilities (which include vulnerabilities in web-facing apps)
  \end{itemize}
\item
  \textbf{Ghost (Cring) ransomware}~\citep{cisa_AA25-050A_2025}
  \begin{itemize}
  \item
    ``Beginning early 2021, Ghost actors began attacking victims whose internet facing services ran outdated versions of software and firmware. This indiscriminate targeting of networks containing vulnerabilities has led to the compromise of organizations across more than 70 countries..''
  \item
    ``Ghost actors use publicly available code to exploit Common Vulnerabilities and Exposures (CVEs) and gain access to internet facing servers. Ghost actors exploit well known vulnerabilities and target networks where available patches have not been applied.''
  \item
    ``Ghost ransom notes often claim exfiltrated data will be sold if a ransom is not paid. However, Ghost actors do not frequently exfiltrate a significant amount of information or files, such as intellectual property or personally identifiable information (PII), that would cause significant harm to victims if leaked. The FBI has observed limited downloading of data to Cobalt Strike Team Servers''
  \item
    ``Ghost variants can be used to encrypt specific directories or the entire system's storage''
  \item
    Observation: So Ghost use a limited amount of double extortion (not much in the way of data exfiltration), but in other respects are quite representative of our assumed threat actor type.
  \end{itemize}
\end{itemize}

\subsubsection*{Type of Actor} \label{type-of-actor}
\begin{itemize}[leftmargin=*]
\item
  We consider a threat actor that would broadly be classified as a small cyber crime syndicate that has operational capacity of the OC3 level (\appr10 individuals, with resources of \appr\$1m, and loosely based on the definition of the corresponding threat actor provided in~\citet{RAND2024SecuringAI}).
\item
  The threat actor is assumed to operate primarily as a RaaS affiliate (per definitions in~\cite{ncsc_ransomware_ecosystem_2023}), and specialises in achieving initial access by exploiting vulnerabilities in public-facing web apps and devices.
\item
  The threat actor is expected to bring their own customizations to the task of deploying the RaaS operator's capabilities, and has the capability to adaptively make use of tools, and ``living off the land'' approaches in moving through the attacked network (per CISA provided information on LockBit affiliates~\citep{cisa_AA23-165A_2023}).
\item
  The threat actor is assumed to have pre-existing cyber attack infrastructure, but no pre-existing access to the attacked organisation.
\item
  In our scenario, we envisage some division of labor, with different members of the cyber crime syndicate optionally specializing in different areas, such as malware customisation, infrastructure operation/acquisition, network penetration, negotiation/coercion, crypto-laundering etc. It should be noted that the RaaS operator may perform many of these activities on behalf of the affiliate, and also that it is commonplace for the affiliate to just provide the initial access, conduct the attack, exfiltrate the data and leave everything else including malware development, negotiation and payment, crypto-laundering etc to the RaaS operator.
\end{itemize}

\subsubsection*{Type of Target} \label{type-of-target}
\begin{itemize}[leftmargin=*]
\item
  The threat actor targets SMEs, with bias toward the more ``medium'' sized companies, and hoping to realize an average ransom payout of somewhere on the order of \$100k
\item
  Approximate company sizes of interest in our scenario are:
  \begin{itemize}
  \item
    75--250 employees
  \item
    Revenue \$10m--\$50m
  \item
    Note: The size of an ``SME'' as defined by the EU, and as described in~\citet{wiki:small_and_medium_enterprises}, is 50-250 employees and \$2m--\$50m in revenue. Based on this definition, the companies we consider in our scenario are more biased toward the ``medium'' end of the SME bracket.
  \end{itemize}
\item
  Ideal targets are sufficiently financially attractive, that turnover/resources are sufficient to lead to the targeted \appr\$100k average ransom payout.
  \begin{itemize}
  \item
    Note, according to~\citet{sophos:state_of_ransomware_2025}, the average payout for a \$10m--\$50m company in 2024 was \$330k, while the average demand was \$109k.
  \item
    Whilst the end ransom demand for a smaller company may be less than for a larger one, compromise of a smaller and relatively less well defended company may be easier, and less prone to result in the attention of law enforcement.
  \end{itemize}
\item
  The threat actor especially targets companies relying on IT systems for operations, e.g.~logistics, manufacturing, retail, such that encryption ransomware, and its associated loss of availability for the target has significant impact (in addition to the impact of confidentiality loss associated with data exfiltration).
\end{itemize}

\subsubsection*{Type of Vector}\label{type-of-vector}
\begin{itemize}[leftmargin=*]
\item
  From~\citet{sophos:state_of_ransomware_2025}, it can be seen that ransomware affiliates use a variety of initial access vectors.
  \begin{itemize}
  \item
    Exploited vulnerability was the top ``root cause'' for companies with revenue $<$\$10m (at 28\%), and was the second most common root cause for companies with revenue \$10m-\$50m at 25\%.
  \end{itemize}
\item
  Whilst~\citet{sophos:state_of_ransomware_2025} indicates that for companies with between 100 and 250 employees ``exploited vulnerability'' was the root cause in 29\% of incidents, coming second only to ``compromised credentials'' on 30\%)
\item
  Note that info from~\citet{coveware:2024_ransomware_reporting_requirements} suggests that CVEs are also often favored by actors going after bigger enterprise targets
\item
  \cite{ncsc_ransomware_ecosystem_2023} also identified ``direct exploitation'' as a prominent initial access vector for ransomware
  \begin{itemize}
  \item
    ``\ldots many {[}criminals{]}\ldots conduct the scanning themselves\ldots . Criminals look for devices that are likely to be in businesses (rather than home environments). Examples include Microsoft Exchange servers, platforms such as Citrix or VMware, VPN devices and firewall devices''.
  \item
    ``Criminal use of exploits often surges shortly after certain critical patches are released indicating they are being reverse engineered from the patches. In most cases, an exploit is widely available in the criminal forums in less than one week from the patch being released. A zero-day exploit is a recently discovered vulnerability, not yet known to vendors or antivirus companies, that criminals can exploit. Cyber criminals don't need to develop their own zero-day exploits as doing so is expensive, and there are many devices ``in the wild'' that are not patched regularly. However, some actors have been known to use zero-day exploits, most notably there are public reports of Cl0p's use of the Accellion, GoAnywhere and MOVEit vulnerabilities''.
  \end{itemize}
\end{itemize}
\underline{Example type of vector}
\begin{itemize}[leftmargin=*]
  \item
    According to~\citet{recordedfuture:patterns_targets_ransomware_2017_2023}, certain classes of threat actors most favour vulnerabilities that have high impact in terms of access or control over systems and the ubiquity of the affected software. They also state ``\ldots we also found that when many ransomware groups target a vulnerability, they do so almost always because it can be exploited with minimal lines of malicious code that can be easily implemented into mass scanning activity (through malicious HTTP requests, for example)''
  \item
    Some example vulnerabilities that that LockBit affiliates have used, according to~\citet{cisa_AA23-165A_2023} include:

    \begin{itemize}
    \item
      \href{https://nvd.nist.gov/vuln/detail/CVE-2019-0708}{CVE-2019-0708}:
      Microsoft Remote Desktop Services Remote Code Execution
      Vulnerability
    \item
      \href{https://nvd.nist.gov/vuln/detail/CVE-2018-13379}{CVE-2018-13379}:
      Fortinet FortiOS Secure Sockets Layer (SSL) Virtual Private
      Network (VPN) Path Traversal Vulnerability.
    \end{itemize}
  \item
    Some examples~\citep{scworld:ghost_ransomware_2025} that Ghost used:
    \begin{itemize}
    \item
      Fortinet FortiOS appliances (CVE-2018-13379)
    \item
      Servers running Adobe ColdFusion (CVE-2010-2861 and CVE-2009-3960)
    \item
      Microsoft SharePoint (CVE-2019-0604)
    \item
      Microsoft Exchange (CVE-2021-34473, CVE-2021-34523, and CVE-2021-31207.
    \end{itemize}
\end{itemize}

\subsubsection*{Defense Level}\label{defense-level}
\begin{itemize}[leftmargin=*]
\item
  We assume that the SME target has cybersecurity maturity level MIL1 (according to the Cybersecurity Capability Maturity Model~\cite{doe_C2M2_2025}) across all maturity domains with MIL2 in some domains. It is assumed that the target generally implements the cybersecurity fundamentals well, and will generally also do a good job of mitigating threats from the most common attack vectors.
\end{itemize}

\subsubsection*{Intent}
\begin{itemize}[leftmargin=*]
\item
  The attacker's intent is financial monetization through ransom.
\item
  Specifically we assume that the attacker uses double extortion. This comprises both data encryption (``availability'' impact) as well as data exfiltration (``confidentiality'' impact). Double extortion is a more complex attack, but can command higher reward than data exfiltration alone. The threat actor demands payment of a ransom in exchange for provision of decryption keys and a promise not to publicize exfiltrated data.
\item
  The greater levels of threat actor skill and capability required to conduct extortion via encryption of systems, over and above the skill required for just data exfiltration extortion, distinguishes this double extortion approach as something that an actor at RAND operational capability level OC3 might be more inclined to do, or be better able to do, than an OC2 actor.
\item
  \cite{coveware:2023_ransom_monetization_fall} provides support for the above, indicating how attacks such as double extortion can achieve bigger returns, but require more work on behalf of the threat actor, and the broad skills required may mean that multiple individuals often need to work together in some form of syndicate.
\end{itemize}

\subsection*{Outputs}\label{outputs}
\subsubsection*{Number of Actors}
In our context, an ``actor'' is equivalent to an ``affiliate'' and refers to the cyber crime syndicate as a whole (i.e.~``actor'' or ``affiliate'' refers to the collection of \appr10 individuals that make up the crime syndicate). The baseline values are summarized in~\cref{tab:baseline_estimates_n_actors}:

\begin{table}[h]
\centering
\begin{tabular}{p{3.5cm} p{3.5cm}}
\toprule
Percentile & Estimate \\
\midrule
5th & 1 \\
Most likely value & 10 \\
95th & 40 \\
\bottomrule
\end{tabular}
\captionsetup{width=0.55\linewidth}
\caption{Baseline estimates for the Number of Actors}
\label{tab:baseline_estimates_n_actors}
\end{table}

\underline{Most likely value rationale:}

\begin{itemize}[leftmargin=*]
\item
  Number of RaaS operators

  \begin{itemize}
  \item
    Following LockBit's takedown, the number of ransomware groups listing victims has risen from 43 to 68, according to Secureworks data~\citep{techinformed:ransomware_gangs_of_2024_affiliates}.
  \item
    The number of active ransomware groups jumped 40\%, from 68 in 2023 to 95 in 2024~\citep{thehackernews}.
  \item
    IC3 recognized 67 new ransomware variants in 2024~\citep{fbi:ic3_2024_report}.
  \item
    All the above would suggest that there are on order \appr80 RaaS operators
  \end{itemize}
\item
  Number of affiliates

  \begin{itemize}
  \item
    A significant RaaS strain could have a large number of affiliates, for example LockBit had 193 in its prime~\citep{kindus:lockbit_ransomware_takedown}.
  \item
    The 3 groups responsible for 53\% of the attacks (see~\citep{kovrr:2023_ransomware_threat_landscape_h1_23}) could plausibly have similar numbers of affiliates as LockBit had in its prime.
  \item
    Let's assume these top 3 have 150 affiliates each, and the other 5 RaaS groups that we are considering have 50 each.
  \item
    That would give $3 \times 150 + 5 \times 50 = \textbf{700 affiliates}$
  \end{itemize}
\item
  Number of OC3 (\appr10 person) crime syndicates
  \begin{itemize}
  \item
    Apparently there has recently been a growing number of lone wolf affiliates in general (following some of the recent law enforcement takedowns, see~\citep{coveware:2025_ransomware_structure_evolution})
  \item
    We assume $\tfrac{1}{3}$ of the 700 affiliates are OC3 groups comprising \appr10 individuals, i.e.~\textbf{200 are OC3 affiliate groups}
  \end{itemize}
\item
  Number of OC3-type affiliates targeting SMEs
  \begin{itemize}
  \item
    It seems likely that a large proportion of total affiliates will target SMEs, though apparently there is an increasing trend to target bigger companies with larger ransomware payouts (see~\citep{chainalysis_ransomware_2024})
  \item
    ~\citet{coveware:2024_ransomware_reporting_requirements} indicate:
    \begin{itemize}
    \item
      ``The average company size of victimized organizations fell to 231 employees (-32\% from Q3 2023).\ldots{} ransomware remains predominantly a small to mid market problem.''
    \item
      30.6\% ransomware attacks on companies with 11--100 employees
    \item
      31.3\% ransomware attacks on companies with 101--1000 employees
    \end{itemize}
  \item
    \citep{hipaajournal:2024_ransomware_payments_record_low}
    \begin{itemize}
    \item
      Between Q2, 2022, and Q2, 2023, ransomware gangs favored attacks on large companies but the average size of victim companies has been falling with medium-sized companies seen as the sweet spot. Attacks are easier to conduct as investment in cybersecurity is lower than at large firms and mid-sized companies have sufficiently large revenues to allow large ransom demands to be issued. In Q4, 2023, the average size of a victim company was 231 employees, down 32\% from Q3, 2023.
    \end{itemize}
  \item
    \citep{kovrr:2023_ransomware_threat_landscape_h1_23}
    \begin{itemize}
    \item
      The top targeted company sizes are: \$10M--\$50M (39\%), \$1M--\$10M (20\%), and \$200M--\$1B (12\%). ``It is clear from the data that ransomware actors prefer to attack smaller companies, with only 8\% of attacked companies having a revenue of over \$1B, and 70\% of attacked companies having revenue below \$100M.''
    \end{itemize}
  \item
    Analysis: In our scenario we are interested in companies of 75--250 employees (\$10m--\$50m revenue). Let's assume that 40\% of our identified OC3 affiliate crime syndicates, i.e \textbf{80} of them ($200 \times 0.4$)\textbf{, go after SMEs} of the size that we are considering
  \end{itemize}
\item
  Number of affiliates using double extortion vs just extorting based on exfiltrated data
  \begin{itemize}
  \item
  \citep{arcticwolf:2025_threat_report} state: ``As organizations improve their ability to recover from ransomware, cyber criminals have turned to data exfiltration to increase leverage---96\% of ransomware cases analyzed included data theft''.
  \item
  \citep{groupib:raas_knowledge_hub}: 83\% of ransomware cases involved data exfiltration
  \item
    \citet{blackfog:double_extortion_prevention_2025} state: ``BlackFog's figures indicate that data exfiltration is a factor in the vast majority of ransomware incidents. In the first half of 2024, we found that 93\% of ransomware attacks exfiltrate data, making it by far the biggest malware threat currently facing enterprises.''
  \item
    \citep{trendmicro:2025_artificial_future}: Ransomware attacks could also drift towards business models that no longer necessitate encryption.
  \item
    \citep{mcintosh2024ransomware}: Points to reducing use of double extortion
  \item
    \citet{kela:state_of_cybercrime_2025} report: ``Most of them continue to operate as ransomware-as-a-service (RaaS) platforms, relying on double extortion and supply-chain compromises.''
  \item
    \citet{firewalltimes:ransomware_statistics} reports:

    \begin{itemize}
    \item
      In total, a second extortion method is part of the equation in 40.9\% of attacks. 30.4\% of attacks involve three threats, while another 7.2\% have four threats involved.
    \item
      30\% of Ransomware attacks involving encryption resulted in stolen Data~\citep{sophos:state_of_ransomware_2023}
    \end{itemize}
  \item
    Analysis: It's rather unclear from the above exactly what proportion of actors are using double extortion, and it seems there may be a trend towards just performing (easier) data exfiltration extortion. Firewall Times has the most directly usable number \textbf{(40\% for double extortion)}, so we use that.

    \begin{itemize}
    \item
      This leaves us with \textbf{$40\% \times 80 = 32$} affiliates
    \end{itemize}
  \item
    Number of affiliates using vulnerability exploit as initial access
    \begin{itemize}
    \item
       \phantom{}\appr 29\% of initial accesses use vulnerability exploit~\citep{sophos:state_of_ransomware_2025}
    \item
      Therefore, we multiply our number of OC3 affiliates (32) by 0.29 to give \textbf{\appr10}
    \end{itemize}
  \end{itemize}
\end{itemize}
\underline{5th percentile rationale} \\
If only 10\% of 700 affiliates are OC3 actors, and only 25\% of these sometimes go after companies with revenue \$10m--\$50m, and if proportion of these syndicates using double extortion vs single extortion is only 20\%, then we'd have $700 \times 0.1 \times 0.25 \times 0.2 \times 0.32 \sim \textbf{1}$ actor

\underline{95 percentile rationale} \\
If 50\% of 700 affiliates are OC3 actors, and if 50\% of these sometimes go after companies with revenue \$10m--\$50m, and if proportion of these syndicates using double extortion vs single extortion is 70\%, then we'd have $700 \times 0.5 \times 0.5 \times 0.7 \times 0.32 \sim \textbf{40}$ actors

\subsubsection*{Number of Attack Attempts per Actor per Year}\label{number-of-attack-attempts-per-actor-per-year}

With our scenario, we consider:
\begin{itemize}[leftmargin=*]
  \item
    Opportunistic attack as opposed to targeted attack, in the sense
    that any SME with the required vulnerable web-facing app could be a
    potential target for our threat actor (though following OSINT, some
    targets may be preferred over others).
  \item
    In the ransomware attack that we are considering, we define both a
    ``campaign'' aspect and an ``attack'' aspect. The campaign aspect is
    concerned with Resource Development (TA0042)~\citep{mitre:TA0042_resource_development}, and the active scanning of Reconnaissance (TA0043)~\citep{mitre:TA0043_reconnaissance}. The ``attack'' phase, in our definition refers to the phase
    from the point that a vulnerable web-facing application has been
    identified and selected and Initial Access (TA0001)~\citep{mitre:TA0001_initial_access} is attempted.
\end{itemize}

The baseline estimates are summarized in~\cref{tab:baseline_estimates_n_attempts}:
\begin{table}[h]
\centering
\begin{tabular}{p{3cm} p{3.5cm}}
\toprule
Step & Number Attack Attempts \newline per Actor per Year \\
\midrule
Cost & No limit \\
Operational capacity ceiling & 150 \\
Historical & 350 \\
\hline
\hspace{0.5cm} \textbf{Triangulation} & \\
\hline
5th percentile & 75 \\
Most likely value & 200 \\
95th percentile & 500 \\
\bottomrule
\end{tabular}
\captionsetup{width=0.45\linewidth}
\caption{Estimates for the Number of Attack Attempts per Actor per Year}
\label{tab:baseline_estimates_n_attempts}
\end{table}

\underline{Cost}

\begin{itemize}[leftmargin=*]
\item
  \citet{trendmicro:raas_enabler_widespread_attacks}
  projected \$100k/month recurring bill for some sorts of ransomware
  groups (though they are not specific about how many people there are
  in the group nor indeed whether they are referring to a RaaS provider,
  an affiliate group, or some do it all ``in-house'' operation -- so it's
  rather questionable how useful the figure is to us).
\item
  ``Regardless of successful payouts, most of these personnel require
  payment for services, whether such remuneration involves monthly
  salary or per-project payments .Estimating average monthly costs (such
  as salaries, servers, virtual private server rentals, service
  providers, tools, accesses, and infrastructure, among others), these
  groups might spend at least US\$100,000 upward to keep operations
  running. If these groups target 10 companies at a time but only one
  victim can pay, that single organization carries the brunt of all the
  expenses that the groups make in addition to the profit they hope to
  have.''
\item
  Analysis: \$100k/month could plausibly cover costs of a 10-person team (esp if annual rate of pay for a Russian cyber expert is \appr\$30k)
\item
  Cost breakdown:
\begin{itemize}
\item
  Payment to RaaS operator

  \begin{itemize}
  \item
    The main per successful attack cost (vs fixed cost) is to the RaaS
    provider (if payment to RaaS provider takes the form of profit
    share). If the RaaS provider takes \appr20\% of the paid ransom then this just increases the number of
    required successful attacks per year to cover fixed costs (salaries
    etc)~\citep{theregister:2023_ransomware_affiliates_money}.
  \end{itemize}
\item
  Our actor would need exploits for services like sharepoint,
  confluence, VPNs, etc. There are many publicly available and/or open
  source exploit frameworks like nuclei, metasploit, etc. where new
  vulnerabilities are added pretty quickly after release in many cases
  and actors can just use free, open tools. It's also possible to pay
  for an exploit, with price depending on how new/useful/reliable they
  are, for example the price could be in the range \$10k--\$50k.
\item
  Costs of browser exploits (though different in nature to our exploits
  of public-facing web apps) do give an indication of the kinds of sums
  that can be paid

  \begin{itemize}
  \item
    \citet{bitdefender:2013_rent_custom_cool_exploit_kit}
    indicate that the Blackhole exploit was rentable at \$1500/year (but
    more exclusive exploits could be as much as \$10,000/month
  \item
    \citet{malwarenews:exploit_kits_decline_2020} indicate kits available at \$1,000/month
  \end{itemize}
\item
  Crypter as a service

  \begin{itemize}
  \item
    \citet{sekoia:architects_of_evasion_2024} indicate: \appr\$50/month to \$250/month
  \end{itemize}
\item
  Cryptocurrency tumbler

  \begin{itemize}
  \item
    \citet{wiki:cryptocurrency_tumbler} indicates that tumblers take 1-3\%
  \end{itemize}
\end{itemize}
\item
  Analysis: Costs for hardware and software seem to be on the order of
  single digit thousands per month, and hence, both because these fixed
  costs are relatively low and because variable costs are incurred on a
  profit share basis, we conclude that \textbf{these aspects will not be
  a determinant of how many attacks/year could be conducted.}
\end{itemize}

\underline{Operational capacity ceiling}

There would appear to be no significant limitation in terms of candidate
victims.

\begin{itemize}[leftmargin=*]
\item
  \citet{ncsc_ransomware_ecosystem_2023}
  \begin{itemize}
  \item
    Criminal use of exploits often surges shortly after certain critical
    patches are released indicating they are being reverse engineered
    from the patches. In most cases, an exploit is widely available in
    the criminal forums in less than one week from the patch being
    released.
  \item
    10\% of devices may be left unpatched even 4 months after a patch is
    available
  \end{itemize}
\item
  \citet{sentinelone:cybersecurity_statistics_2025} indicate:

  \begin{itemize}
  \item
    The National Vulnerability Database (NVD) recorded over 30,000 new
    Common Vulnerabilities and Exposures (CVEs), half of which were
    classified as high or critical severity.
  \item
    A new vulnerability is identified and published every 17 minutes.
    Half of all the vulnerabilities have been published in the last five
    years.
  \end{itemize}
\item
  Given the number of SMEs (10s of millions globally), and the fact
  that many of them won't patch vulnerabilities promptly, it would seem
  currently unlikely that there is a shortage of potential victims.
\end{itemize}

The key time consuming steps in the attack would appear to be:

\begin{itemize}[leftmargin=*]
\item
  During the reconnaissance phase, looking at the results of ``active
  scanning'' and determining, e.g.~based on OSINT (open source
  intelligence), whether the particular enterprise is one which the
  threat group wish to try attacking (i.e.~which meets their criteria in
  terms of likely ransom payout achievable, and likelihood of success
  etc).
\item
  Tactics like lateral movement and privilege escalation, and activities
  like locating backups require ``hands on keyboard'' and therefore can be
  time consuming and costly.

  \begin{itemize}
  \item
    In contrast, steps like encrypting or exfiltrating certain content
    or files can be automated, and these steps are therefore not so time
    consuming or costly.
  \end{itemize}
\item
  Encryption malware deployment (also including disabling security and
  backup features)
\item
  Ransom negotiation and follow up.
\end{itemize}

According to~\citet{secureworks:2023_ransomware_dwell_time_24h},
in 2023 average dwell time went from 4.5 days in 2022 to less than 24
hours in 2023.
\begin{itemize}[leftmargin=*]
\item
  They state ``The driver for the reduction in average dwell time is
  likely due to the cyber criminals' desire for a lower chance of
  detection. The cybersecurity industry has become much more adept at
  detecting activity that is a precursor to ransomware. As a result,
  threat actors are focusing on simpler and quicker to implement
  operations, rather than big, multi-site enterprise-wide encryption
  events that are significantly more complex. But the risk from those
  attacks is still high.''
\end{itemize}

Analysis:
\begin{itemize}[leftmargin=*]
\item
  It would seem plausible that there are 10 days of person effort per
  fully completed successful attack, when attacking SMEs
\item
  If:
  \begin{itemize}
  \item
    50\% of attacks fail (quickly) and e.g.~after expending 1 day of
    effort

    \begin{itemize}
    \item
      ``bath-tub'' effect
    \end{itemize}
  \item
    25\% take 5 days of effort (and fail)
  \item
    25\% take 10 days of effort (and perhaps succeed)
  \end{itemize}
\item
  Then, expected effort per attack is 0.5 x1 + 0.25 x 5 + 0.25 x 10 =
  4.25 days (\appr5 person days)
\item
  With 10 people in the team, and assuming 6 work actively on day to day
  attacks, whilst 4 work on ``back-office'' activities like developing
  cyber tools, infrastructure, malware modifications etc then that would
  suggest 6 attacks per week (6x5/5)
\item
  Organisationally / managerially though that sounds somewhat
  implausible to handle in a group of 10 individuals. In addition, there
  won't be perfect efficiency in the usage of people's time with some
  individuals sometimes having to wait on others to complete tasks.
\item
  Hence we assume 3 attack attempts per week
\item
  There are 52 weeks in the year, so we assume
  \appr{}\textbf{150 attacks/year} by the OC3 threat actor is
  the realistic operational capacity limit
\end{itemize}

\underline{Historical}

In this section, we first gather some relevant data points, and then at
the end of the section we perform an analysis to see what the historical
data can tell us about the number of attack attempts/year

(Presumed) successful attacks per affiliate per year
Lockbit data point:

\begin{itemize}[leftmargin=*]
\item
  Lockbit attacked more than 2500 victims over 4 years~\citep{doj:2024_lockbit_affiliates_guilty}
\item
  LockBit had 190 affiliates prior to the FBI takedown~\citep{techinformed:ransomware_gangs_of_2024_affiliates}.
\item
  Analysis: Presume these victims, correspond to
  successful attacks (?). \appr13 victims/affiliate over 4
  years $\rightarrow$ \appr3 victims / year / affiliate
  RansomHub data point
\item
  ~\citet{cyble:2025_ransomware_attack_levels_remain_high}
  indicate 88 attacks./month (Feb 2025) for RansomHub. This was the most
  prolific malware
\item
  Analysis: If we assume that RansomHub, as the most prolific
  RaaS has a similar number of affiliates as LockBit had in Feb 2024
  (190), then that would suggest approx 6 attacks/year/affiliate (assume
  these were successful attacks). US DoJ convictions (over 2020-2023)~\citep{doj:2024_lockbit_affiliates_guilty}
\item
  Lockbit affiliate (Russian national \#1) successfully conducted at
  least 12 attacks over 3 years (extorting \$1.9m)
\item
  Lockbit affiliate (Russian national \#2) successfully conducted at
  least 12 attacks over 2 years
\item
  Analysis: 4-6 successful attacks/year/individual~\citep{zerofox:2025_lockbit_data_breach}
\item
  Provided info on the operations of the much diminished Lockbit
  ransomware
\item
  Over first 4 months of 2025 -- there appear to have been
  \appr110 attacks, and were \appr40 ``incidents'' -
  representing 1.5\% of all ransomware attacks
\item
  75 affiliates had accessed the LockBit affiliate portal.
\item
  This would suggest \appr4-5 attacks/year/affiliate
\end{itemize}

Attacks/month:
\begin{itemize}[leftmargin=*]
\item
  \phantom{} \appr2500 attacks in Q1/2025 (incl disclosed and undisclosed)~\citep{hipaajournal:2025_q1_ransomware_report}
\item
  Similar number of attacks /month \appr600-800 mentioned by~\citet{cyble:2025_ransomware_attack_levels_remain_high}.
  Since Cyble talks of ransomware gangs ``claiming'' attacks, we will
  presume that these are successful attacks
\end{itemize}

Attacks per organization:
\begin{itemize}[leftmargin=*]
    \item
    \citep{checkpoint:2025_q1_global_cyber_attack_report} 
\begin{itemize}
\item
  Cyber Attack Surge: In Q1 2025, cyber attacks per organization
  increased by 47\%, reaching an average of 1,925 weekly attacks per
  organisation.
\end{itemize}
\item
  \citep{spacelift:2025_ransomware_statistics}
\begin{itemize}
    \item
    In 2023 - 59\% orgs hit with ransomware, 4000 daily attacks per organization daily
    \item
    90\% of attacks either fail or result in no financial loss
\end{itemize}
\item \citep{sophos:state_of_ransomware_2025}
\begin{itemize}
\item
  Sophos indicate \appr57\% of SMEs were hit by ransomware
  last year.
\end{itemize}
\item The UK government~\citep{govuk:cyber_survey_2025} noted:
  \begin{itemize}
  \item
    ``Just over four in ten businesses (43\%) and three in ten charities
    (30\%) reported having experienced any kind of cyber security breach
    or attack in the last 12 months. This equates to approximately
    612,000 UK businesses and 61,000 UK charities that identified a
    cyber breach or attack in the past year
  \item
    The decrease was driven by fewer micro and small
    businesses identifying phishing attacks
  \item
    Observation: If phishing attacks are included, it is perhaps not
    unsurprising that the number of attack attempts are high. A similar
    conclusion would be true, if vulnerability scanning of web-facing
    devices is ``counted'' as an attack attempt.
  \end{itemize}
\end{itemize}

Global ransom attempts

\citep{firewalltimes:ransomware_statistics}:
\begin{itemize}[leftmargin=*]
\item
  Just over 493 million ransomware attempts were made in 2022
\item
  During a survey taken in 2023, 72.7\% of companies reported being
  victimized by ransomware in the past 12 months
\item
  Overall, 46\% of SMEs have experienced a ransomware attack
\item
  According to~\citet{gsbb:2024_number_of_smes_worldwide}, there are \appr360 million SMEs worldwide.

  \begin{itemize}
  \item
    Analysis: if there are 3000 individuals
    acting as affiliates then, and if there are 493m attacks, each
    individual would be conducting 160,000 ransom attempts/year
    (500/day), and with a success rate of a handful per year. Which
    seems an incredibly high number!

    \begin{itemize}
    \item
      We assume 700 affiliates in total (see ``number of actors''
      computations above) comprising 200, 10-man OC3-groups and 500, 1
      or 2 person groups
    \end{itemize}
  \end{itemize}
\end{itemize}

Analysis

\begin{itemize}[leftmargin=*]
\item
  For our purposes we wish to understand the number of attacks that an
  OC3 group would make, beyond just scanning for web vulnerabilities,
  and rather comprising at least launching an exploit on a web-facing
  device that has a vulnerability. However, it has not been possible to
  find such figures.
\item
  Figures we have found, tend to talk either of number of successful
  attacks, or else number of attack attempts per organisation, or
  percentages of organisations attacked, which judging by the very high
  numbers involved in these latter statistics, may well suggest that
  each phishing email (or perhaps each scanning attempt) is counted as
  an attack.
\item
  Let's attempt to derive a figure from the number of successful
  attacks/individual/year:

  \begin{itemize}
  \item
    There are a few sources of information that could back an assertion
    of there being \appr3 to 4 successful attacks per
    individual (i.e.~per person) per year.

    \begin{itemize}
    \item
      Here we assume that a successful attack refers to a successful
      ``technical attack'' (noting only \appr30\% of successful
      technical attacks convert to ransom payouts, and ``success'' from
      the perspective of the attacker).
    \end{itemize}
  \item
    Our later analysis on probability of success suggests the
    probability of attack success is \appr10\%
  \item
    These figures could then suggest the approximate number of attacks
    (attempted exploits of the web vulnerability) per individual would
    be \appr35 per year, and with 10 individuals in our OC3
    cyber crime syndicate, that would suggest \textbf{350 attack
    attempts/year.}
  \end{itemize}
\end{itemize}

\underline{Triangulation}

Most likely value:

Before ``triangulating'' across the above data points, one additional
factor to consider is the need for the OC3 group to deliver an
acceptable tempo of success, in order to maintain motivation, and to
avoid too much ``lumpiness'' in receipt of revenue:
\begin{itemize}[leftmargin=*]
\item
  If the OC3 group needs or expects \appr12 successful
  ransoms/year (where ``success'' includes ransom payout) to sustain
  itself and for the team to stay motivated with a fairly regular tempo
  of success (one success every month) then:

  \begin{itemize}
  \item
    With \appr10\% probability of ``technical success'' and 30\%
    probability of ransom payout per technical success, this would
    suggest \textbf{400 attacks/year.}
  \end{itemize}
\end{itemize}

Analysis of the above:

\begin{itemize}[leftmargin=*]
\item
  There is a sparsity of historical data on number of attack attempts
  (i.e.~little data on failed attempts), and what data does exist seems
  more pointed at ``campaign'' level aspects such as phishing and perhaps
  scanning. For our purposes in this table, ``number of attacks'' refers
  to the count of the number of times that an exploit is attempted
  against a particular web-facing app or device (as opposed to a count
  of the number of web-facing apps/devices that were scanned during
  Reconnaissance ``active scanning'').
\item
  We have computed 3 figures:

  \begin{itemize}
  \item
    150, from the operational capacity analysis
  \item
    350, derived from historical numbers on successes/individual, and
    with an assumption on probability of successful attack.
  \item
    400, based on assumed team-motivational factors (again making
    assumptions on probability of ``technical success'' and probability of
    ransom payout)
  \end{itemize}
\item
  We triangulate by picking 200 (and placing more emphasis on the
  operational capacity figures)
\end{itemize}

5th percentile:

\begin{itemize}[leftmargin=*]
\item
  There is significant uncertainty due to the lack of useful historical
  info. Operational capacity estimates could easily be wrong by a factor
  of \appr3 (project managers are often overly optimistic by a
  factor of \appr2).
\item
  Let's go for a figure of 1/3 of the ``best estimate'' figure (200/3)
  \appr75. This amounts to 1-2 attacks per week (1 attack
  every \appr3 days), which feels like a plausible lower
  bound.
\end{itemize}

95th percentile:

\begin{itemize}[leftmargin=*]
\item
  There is significant uncertainty due to the lack of useful historical
  info.
\item
  We select a number of 500 -- since this is somewhat larger than our
  assumed team-motivational limit (400), provides an increase over our
  computation based on historical data (350) and is quite a bit larger
  than our assumed operational capacity figure (allowing for error in
  that computation).
\end{itemize}

\subsubsection*{Steps in Attack - MITRE Tactic Level}
Below, we list the rationales for including or excluding a given MITRE Tactic as part of our OC3 Ransomware SME risk model. These rationales are summarized in~\cref{tab:MITRE_tactic_incl_or_excl}:
{\renewcommand{\arraystretch}{1.5}
\begin{table}[!pht]
\begin{tabular}{p{2cm}p{1.2cm}p{11cm}}
\toprule
Step & Included \mbox{or not} & Failure mode \\
\midrule
Reconnaissance & Yes & If recon fails (e.g.~no vulnerabilities or intel
found), the attack cannot begin. \\
Resource \mbox{Development} & Yes & If resource prep fails (e.g.~no exploit, no
malware or infrastructure), the operation stalls -- the affiliate lacks
tools or access to proceed. \\
Initial Access & Yes & If initial access fails (no foothold gained), the
affiliate cannot penetrate the victim's network -- the ransomware
deployment is a non-starter. \\
Execution & Yes & If execution fails (malicious code never runs), the
attacker's payloads and commands don't take effect and the attack
fails. \\
Persistence & No & N/A \\
Privilege \mbox{Escalation} & Yes & If privilege escalation fails, the
attackers are stuck with low-level rights. They may be unable to disable
security tools or access sensitive data, often stopping the attack from
progressing beyond the initial host. \\
Defense \mbox{Evasion} & No & If defense evasion fails (attacker activity is
detected/blocked), the affiliate will likely be interrupted or expelled
before achieving goals. Early detection often means the encryption and
exfiltration can be prevented or limited. Although it is essential, we
do not include it, as it is implicitly factored into other steps. \\
Credential \mbox{Access} & No & Whilst credential access is likely to be a
common tactic in ransomware operations (such as use of LSASS dumping and
Mimikatz), we are already accounting for something similar with
``privilege escalation'' tactic and the associated ``valid accounts''
technique. It is also a ``nice to have''. To avoid double counting, and
because it is a ``nice to have'', we do not include the
credential access tactic. \\
Discovery & Yes & If internal discovery fails, the attackers may miss
critical systems or data. They could encrypt some machines blindly, but
might overlook backups or high-value servers, reducing the impact. Lack
of knowledge can also lead to mistakes (e.g.~tripping alarms) or an
incomplete data theft, weakening their leverage. \\
Lateral \mbox{Movement} & Yes & If lateral movement fails, the compromise
remains limited to the initial host or a small subset of systems. The
affiliate then might only encrypt a trivial portion of the network or
grab a small amount of data, which may not be enough pressure to make
the victim pay. \\
Collection & Yes & If collection fails (attackers can't gather any
sensitive data), the ``double'' in double-extortion is lost. The
attackers would have no files to leak, weakening their bargaining
position -- they'd be left with just encryption, which many
well-prepared victims can recover from. \\
Command-and-Control & Yes & If C2 fails (the affiliate loses contact
with their malware/beacons inside the network), they can't direct the
attack. The team would effectively be ``locked out'' from interacting
with compromised systems, stopping further progression like escalation,
lateral moves, exfiltration, etc. \\
Exfiltration & Yes & If exfiltration fails (attackers cannot transmit
the stolen data out of the victim's network), the affiliate loses the
data-leverage part of the extortion. \\
Impact & Yes & If impact fails (e.g.~files don't get encrypted or
critical services remain unaffected), the victim's operations continue
relatively unscathed. With no significant disruption or data loss, the
victim has little incentive to pay, and the affiliate's attack
effectively fails to achieve its goal. \\
\bottomrule
\end{tabular}
\caption{{Rationales for including or excluding MITRE Tactics in the OC3 Ransomware SME model}}
\label{tab:MITRE_tactic_incl_or_excl}
\end{table}
}

\textbf{Rationale}

\begin{itemize}[leftmargin=*]
\item
  \textbf{Reconnaissance}: The affiliate performs active scanning of
  internet-facing assets for known vulnerabilities. Some basic OSINT
  will almost certainly be used to size up victims (e.g.~revenue,
  insurance) to set ransom demands
\item
 \textbf{Resource Development}: The affiliate must obtain or prepare tools,
  malware, and infrastructure to conduct the attack In the RaaS model,
  some resources come ready-made (e.g.~the ransomware payload and a leak
  site for publishing stolen data are provided by the RaaS service).
\item
  \textbf{Initial Access}: In our scenario, the group focuses on vulnerability
  exploits against Internet-facing systems.
\item
  \textbf{Execution}: After initial access, the affiliate needs to launch tools:
  e.g.~starting a remote shell, running malware. Ransomware operators
  often use scripts and admin tools for execution
\item
  \textbf{Persistence}: involves maintaining long-term access to systems
  (surviving reboots, credential changes, etc.). Ransomware affiliates
  do often implement persistence -- for example, adding new accounts --
  but it's not strictly required if they can complete their attack
  quickly.
  \begin{itemize}
  \item
    ``In 2022, Coveware reported that 82\% of observed ransomware
    attacks included some form of persistence tactic---up 34\% from the
    previous quarter. Persistence techniques remain relevant because
    attackers want to protect their hard-won access~\citep{proofpoint:2023_ransomware_attack_chain}.
  \item
    According to an analysis by Huntress, the average
    time-to-ransom is around 17 hours\ldots.this pace is in stark
    contrast to how major ransomware groups operated before the double
    extortion trend took off several years ago, when they would lurk
    inside victim networks for days or weeks to build greater access and
    gain complete control~\citep{scmediauk:2025_dwell_time_four_hours}.
  \item
    Secureworks said in its 2023 ``State of the Threat'' report
    that one reason for reduced dwell time is attackers moving faster to
    lower the chance of detection. ``However, it is also likely that the
    threat actors now deploying ransomware are just lower skilled than
    previous operators''~\citep{scmediauk:2025_dwell_time_lower_mean_more_secure}
  \item
    Analysis: we are looking at OC3 capable actors in our scenario,
    which are skilled actors. It's unclear however, whether their skill
    would result in them moving faster (perhaps without persistence), or
    whether it would tend to make them stay in the compromised network
    for longer to conduct their task more stealthily. Either way, it
    seems that dwell times are reducing and that persistence could be
    optional.
  \end{itemize}
\item
  \textbf{Privilege Escalation}: Ransomware affiliates almost always seek higher
  privileges after initial entry. Administrative or root access is
  needed to spread ransomware broadly and to turn off defenses or
  backups, achieving domain admin on the network is particularly
  valuable. If this step fails, the adversary's reach is severely
  limited. With only user-level access, they might not get access to
  critical servers or data and might be unable to propagate the
  ransomware across the network.
\item
  \textbf{Defense Evasion}: critical when attacking a ``well-defended'' SME.
  Ransomware affiliates take active measures to avoid or disable
  security controls. They frequently leverage dual-use or trusted tools
  to blend in (e.g.~using system administration software or legitimate
  RMM tools), and they directly sabotage defenses (see, e.g.,~\citet{cisa_AA23-165A_2023}. A ransomware attack will likely involve at least some degree of defence evasion in executing most if not all tactics. Below we list relevant defence evasion techniques (as identified by Deep Research, Gemini and some CISA references), and where we assume that they will already be accounted for in evaluating probability of success for other tactics, this is indicated in italics:

  \begin{itemize}
  \item
    \textbf{T1562 -- Impair Defenses:} Kill or uninstall security
    software~\citep{cisa_AA23-061A_2023}. \emph{Assumed
    non-essential (so not accounted for)}
  \item
    \textbf{T1027 -- Obfuscated Files or Commands:} Includes obfuscation
    of malware. \emph{Assumed dealt with primarily in the Execution
    tactic (malware code obfuscation), command obfuscation may appear in
    multiple other tactics.}
  \item
    \textbf{T1070 -- Indicator removal:} Delete shadow files and system
    and security logs after exfiltration~\citep{cisa_AA23-061A_2023}.\emph{Assumed non-essential to attack success (so not accounted for)}
  \item
    \textbf{T1484 -- Domain or Tenant Policy Modification:} Modify Group Policy
    Objects to subvert antivirus protocols~\citep{cisa_AA23-061A_2023}. \emph{Assumed non-essential (so not accounted for)}
  \item
    \textbf{T1497 -- Virtualization/Sandbox Evasion:} Sophisticated
    attackers may be keen not to download their malware if it will
    result in it being analyzed in a sandbox (possibly alerting the
    defender community and impacting their ability to use it for other
    attacks). \emph{Assumed non-essential (so not accounted for)}
  \item
    \textbf{T1036 -- Masquerading:} E.g changing filenames or filename
    metadata. \emph{Assumed dealt with primarily in the Execution
    tactic, if needed}
  \item
    \textbf{T1564 -- Hide artifacts:} \emph{Assumed dealt with primarily
    in the Execution tactic, if needed.}
  \end{itemize}
\item
  \textbf{Credential Access}: stealing passwords, hashes, keys, and tickets is a cornerstone of ransomware operations. Compromised credentials
  allow the adversary to authenticate as legitimate users/admins,
  expanding their reach. With valid accounts, attackers can access
  remote services (VPN, RDP, file shares) and move through the network
  more freely.
\item
  \textbf{Discovery}: once inside, affiliates perform Discovery to understand the
  environment. This reconnaissance phase inside the network lets them
  locate where valuable data resides (file servers, databases) and
  identify defenses or backup systems to circumvent
\item
  \textbf{Lateral Movement}: how the attacker pivots from the initial foothold
  to broader control of the network. Ransomware affiliates aim to infect
  as many machines and access as much data as possible. They use
  techniques like connecting to other systems with stolen credentials or
  deploying remote execution tools.
\item
  \textbf{Collection}: in the context of double extortion, Collection of victim data is a
  must. Before launching ransomware, affiliates quietly search for and
  aggregate valuable files: databases, client records, intellectual
  property, emails -- anything that would hurt to see leaked publicly.
\item
  \textbf{Command and Control (C2)} -- adversary's lifeline into the victim
  environment. After establishing a foothold, affiliates need a
  communication channel to send commands, coordinate lateral movement,
  and extract data. Ransomware operators often re-purpose trusted
  channels: for instance, using standard protocols (HTTP/HTTPS, DNS) or
  legitimate remote access software to blend in.
\item
  \textbf{Exfiltration} -- the act of stealing the collected data out to
  attacker-controlled infrastructure. Double-extortion schemes hinge on
  this.
\item
  \textbf{Impact} -- the phase where the attackers encrypt data on all reachable systems. This renders the victim organization's files and services
  unusable, causing operational paralysis. Affiliates also take
  additional impact actions: they often inhibit system recovery by
  deleting backups and shadow volume copies. It is common that
  ransomware even changes wallpapers or drops ransom notes in
  directories to ensure the victim knows they've been hit.
\end{itemize}

\subsubsection*{Steps in Attack -- MITRE Technique Level}\label{steps-in-attack---mitre-technique-level}
In~\cref{tab:tactic_into_technique_or_not}, we present justifications for why a given step in our attack scenario should be considered at the MITRE Tactic or Technique level. Next, in~\cref{tab:techniques_of_tactics} we list the MITRE Techniques corresponding to each step.

\begin{longtable}{p{2cm}p{1.5cm}p{10cm}}
\caption{Rationales for breaking down MITRE Tactics into Techniques}
\label{tab:tactic_into_technique_or_not} \\

\toprule
Step & Tactic or \mbox{technique}  & Rationale \\
\midrule
\endfirsthead

\multicolumn{3}{l}{\textit{(Continued from previous page)}} \\
\toprule
Step & Tactic or \mbox{technique}  & Rationale \\
\midrule
\endhead

\midrule
\multicolumn{3}{r}{\textit{(Continued on next page)}} \\
\endfoot

\bottomrule
\endlastfoot

Reconnaissance & Tactic & Active scanning is LLM relevant, must-have.
Probability of success is very high (\appr100\%) - so we
include the tactic, but do not break it down to technique level for
further analysis \\

Resource \mbox{Development} & Tactic & Resource development is essential and at
least one aspect could be LLM relevant: Obtaining exploit proofs/code
(T1588.005) However, probability of success is very high
(\appr100\%) so we stay at tactic level. \\

Initial Access & Tactic & Must have only one technique, so stay at
tactic level \\

Execution & Tactic & Must have, but keep at tactic level since though
there are alternatives, none seem particularly LLM relevant \\
Persistence & N/A & N/A \\

Privilege \mbox{Escalation} & Tactic & Stay at tactic level, since it's unclear
that there is much to be gained by way of LLM estimatability in breaking
down to technique level. \\

Defense \mbox{Evasion} & N/A & We assume that this tactic is already accounted
for when we consider probability of success of other tactics. Hence we
do not include it, to avoid double-counting. \\

Credential \mbox{Access} & N/A & N/A \\

Discovery & Tactic & Uncertainty is large regarding how an actor
actually proceeds, since there are many potentially applicable
techniques. So will stay at the tactic level to avoid overfitting or
adding speculative details. \\

Lateral \mbox{Movement} & Tactic & Not clear if there is benefit in breaking
into technique level, for purposes of determining LLM uplift. So stay at
tactic level. \\

Collection & Tactic & Not clear there is much, if anything, in the way
of LLM uplift here, so could stick at tactic level. \\

Command-and-Control & Tactic & Not clear there is much scope for using
LLMs here, and even if there is, it's not clear that there will be much
difference in their potential benefits depending on the specifics of the
communication channels used, e.g.~whether using HTTPS or hiding
communications in DNS traffic etc \\

Exfiltration & Tactic & As for command and control, it's not clear
there's much scope for using LLMs here, and even if there is, it's not
clear that there will be much difference in their potential benefits
depending on the specifics of the exfiltration mechanism. So stay at
tactic level. \\

Impact & Technique & It's not clear there's much scope for using LLMs
for the encryption and data destruction related techniques (T1486,
T1490, T1489, T1485). However, LLMs might be used to assist with the
social engineering (T1657) related to the extortion negotiations. Hence
we separate out this technique. \\
\end{longtable}

{\renewcommand{\arraystretch}{1.5}
\begin{longtable}{p{2cm} p{12cm}}
\caption{MITRE Tactics broken down into Techniques} \label{tab:techniques_of_tactics} \\

\toprule
Step & Techniques \\
\midrule
\endfirsthead

\multicolumn{2}{l}{\textit{(Continued from previous page)}} \\
\toprule
Step & Techniques \\
\midrule
\endhead

\midrule
\multicolumn{2}{r}{\textit{(Continued on next page)}} \\
\endfoot

\bottomrule
\endlastfoot

Reconnaissance & \textbf{T1595.002 -- Active Scanning (Essential)}:
Reconnaissance Vulnerability Scanning. Ransomware affiliates actively
scan the internet for exposed assets and known vulnerabilities in target
web apps or devices.
\newline
\textbf{T1593 -- Search Open websites/Domains}:
Determine whether target is worth attacking (has sufficient financial
resources), and how much to charge in the ransom. \\

Resource \mbox{Development} & \textbf{T1588.005 -- Obtain Exploit Proofs/Code}
\textbf{(Essential)} Before attacking, the group acquires necessary
resources. They obtain or purchase exploit code for known
vulnerabilities to use against victims' public-facing systems
\newline
\textbf{T1588.002 -- Obtain Tooling/Malware (Essential)}: They also
secure malware/tooling such as the RaaS ransomware payload, credential
stealers, and C2 frameworks (e.g.~Cobalt Strike). These tools may be
bought or provided by the RaaS service.
\newline\textbf{T1583.006 -- Acquire
Infrastructure: Virtual Private Server (Essential)}: They may rent VPS
servers for C2 and data leakage sites (the RaaS model often supplies
infrastructure for communication and data leaks) \\

Initial Access & \textbf{T1190 -- Exploit Public-Facing Application
(Essential)}: There are alternative methods for initial access, but in
our scenario we are limiting ourselves to this one \\

Execution & \textbf{T1059 -- Command and Scripting Interpreter
(Essential)}: Once inside, affiliates execute malicious code and commands
to advance the attack. They commonly use command-line interpreters like
PowerShell and Windows Command Prompt to run scripts and tools. For
example, PowerShell may be used to deploy payloads or run
reconnaissance/enumeration scripts stealthily. Using script interpreters
allows them to automate tasks (such as launching ransomware or
exfiltration scripts) while blending in with normal admin activity. \\

Persistence & \textbf{T1219 -- Remote Access Software}: (NB: appears under C2C in MITRE framework) Karakurt actors have used AnyDesk to obtain persistent remote control of victims' systems~\citep{cisa:aa22-152a_karakurt}. \\

Privilege \mbox{Escalation} & \textbf{T1068 -- Exploitation for Privilege
Escalation}: Using exploits like ZeroLogon, PrintNightmare, etc.
\newline
\textbf{OR}
\newline
\textbf{T1003 -- OS Credential Dumping}: E.g. accessing
password hashes from memory Note this technique is listed under the
Credential Access tactic, but we include it here under privilege
escalation \\

Discovery & \textbf{T1018 -- Remote System Discovery}: Enumerate other
hosts via AD or scanning
\newline
\textbf{AND}
\newline
\textbf{T1135 -- Network Share Discovery}
\newline
\textbf{AND}
\newline
\textbf{T1083 -- File and Directory Discovery}: Malware needs to enumerate files and directories to find data to encrypt on a given host. Also
mentioned by~\citet{coveware:2024_ransomware_reporting_requirements}
\newline \newline
Other relevant techniques include:
\newline
\textbf{T1046 -- Network Service Scanning}: Identify network hosts~\citep{coveware:2024_ransomware_reporting_requirements}
\newline
\textbf{T1033 -- System owner / user discovery}: Get listing of accounts
on a system or network~\citep{coveware:2024_ransomware_reporting_requirements}
\newline
\textbf{T1082 -- System Information Discovery}: Get detailed information
about the operating system and hardware, including version, patches,
hotfixes, service packs, and architecture~\citep{coveware:2024_ransomware_reporting_requirements}.
\newline
\textbf{T1482 -- Domain Trust Discovery}: Can enable lateral movement~\citep{coveware:2024_ransomware_reporting_requirements} \\

Lateral \mbox{Movement} & \textbf{T1021 -- Remote Services}: Can enable remote
login to other machines in the domain using Valid Accounts (T1078)~\citep{mitre:T1078_valid_accounts}. Sub-techniques include: T1021.002 (Remote services: SMB/Windows
Admin Shares), T1021.001 (Remote services: Remote Desktop Protocol)
\newline
\textbf{T1210 -- Exploitation of Remote Services}:~\citep{coveware:2024_ransomware_reporting_requirements}
\newline
\textbf{T1570 -- Lateral Tool Transfer}: May use legitimate windows admin
tools to mass deploy malware across machines~\citep{coveware:2024_ransomware_reporting_requirements} \\

Collection & \textbf{T1039 -- Data from Network Share}
\newline
\textbf{T1560 -- Archive Collected Data}: Compress and package stolen files 
\newline
\textbf{T1074 -- Data Staging}: Aggregate data on an internal host before
exfiltration. \\

Command-and-Control & \textbf{T1071 -- Application Layer Protocol}:
Though the use of Application Layer Protocol is probably most common,
there are likely alternatives, such as T1095 Non-Application Layer
Protocols)
\newline 
\textbf{T1573 -- Encrypted Channel}: So the attacker can hide their traffic from the defenders
\newline
\textbf{T1572 -- Protocol tunnelling}:~\citep{cisa_AA23-061A_2023}
\newline
\textbf{T1105 -- Ingress tool transfer}: Use C2 to download multiple tools~\citep{cisa_AA23-061A_2023} 
\newline
\textbf{T1219 -- Remote Access Software}: Use legitimate software to maintain remote access to a victim machine~\citep{coveware:2024_ransomware_reporting_requirements}. \\

Exfiltration & \textbf{T1567 -- Exfiltration Over Web Service}: Used by~\citet{cisa:aa22-152a_karakurt}
\newline
\textbf{T1048 -- Exfiltration over alternative protocol}: Used by~\citet{cisa:aa22-152a_karakurt}
\newline
\textbf{T1041 -- Exfiltration over C2 channel} \\

Impact & \textbf{T1486 -- Data Encrypted for Impact}
\newline
\textbf{T1490 -- Inhibit System Recovery}: Delete or encrypt backups, shadow copies
\newline
\textbf{T1489 -- Service Stop}: Disable security or backup services.
\newline
\textbf{T1485 -- Data Destruction}: Mainly aimed at destroying forensic
artifacts~\citep{coveware:2024_ransomware_reporting_requirements}
\newline
\textbf{T1657 -- Financial theft}: Extortion -- threat made to leak data
and/or not provide decryption keys unless ransom is paid \\
\end{longtable}
}

\subsubsection*{Probability Estimates -- Tactics and Techniques} \label{probability-estimates---tactics-and-techniques}
\cref{tab:baseline_estimates_tactics} shows the estimated baseline values for steps in our attack scenario. We further provide detailed rationales for these values below.
{\renewcommand{\arraystretch}{1.5}
\begin{table}[!th]
\begin{tabular}{p{3.5cm} p{3cm} p{3cm} p{3cm}}
\toprule
MITRE tactic & Probability estimate (5th percentile) & Probability estimate (most likely value)
& Probability estimate (95th percentile) \\
\midrule
Reconnaissance & 100\% & 100\% & 100\% \\
Resource Development & 100\% & 100\% & 100\% \\
Initial Access & 25\% & 60\% & 90\% \\
Execution & 20\% & 50\% & 90\% \\
Privilege Escalation & 25\% & 70\% & 90\% \\
Discovery & 70\% & 85\% & 95\% \\
Lateral Movement & 50\% & 65\% & 80\% \\
Collection & 75\% & 90\% & 95\% \\
Command-and-Control & 80\% & 90\% & 100\% \\
Exfiltration & 70\% & 85\% & 90\% \\
\multirow{2}{3.5cm}{Impact} & 60\% & 80\% & 90\% \\
% & \hline \\
& 15\% & 30\% & 50\% \\
\bottomrule
\end{tabular}
\caption{Baseline probability estimates for MITRE Tactics}
\label{tab:baseline_estimates_tactics}
\end{table}
}

\underline{Rationales}

\begin{itemize}[leftmargin=*]
\item
  Reconnaissance

  \begin{itemize}
  \item
    Limited ability to prevent external scanning.
  \item
    SME may rely on basic firewalls, but generally can't stop
    adversaries from enumerating public-facing assets
  \item
    Adversaries (like our OC3 affiliate) commonly discover exposed
    systems and known vulnerabilities via active scanning, which is
    almost guaranteed to succeed if any services are exposed.
  \end{itemize}
\item
  Resource Development

  \begin{itemize}
  \item
    No defender impact.
  \item
    Attacker obtaining exploits, tools, infrastructure occurs off-site
    and beyond SME's defenses.
  \item
    Ransomware affiliates will almost certainly gather the needed
    exploits and malware before attacking. This includes purchasing
    exploit code for known CVEs and RaaS payloads, renting servers, etc.
    Defenders cannot directly interfere at this stage. Success is
    effectively assured.
  \end{itemize}
\item
  Initial Access

  \begin{itemize}
  \item
    From MITRE:
    \begin{itemize}
    \item
      Application isolation and sandboxing
    \item
      Exploit protection (web application firewalls)
    \item
      Limit access to resources over network (only expose essential
      services)
    \item
      Network segmentation (DMZ etc)
    \item
      Use least privileges for service accounts
    \item
      Update software regularly by patch management
    \item
      Vulnerability scanning and response procedure
    \end{itemize}
  \item
    Illustrative defenses:

    \begin{itemize}
    \item
      Basic patch management and periodic vulnerability scanning;
      perhaps a simple web application firewall (WAF). SMEs often expose
      only essential services. Network segmentation (DMZ for
      internet-facing systems) is sometimes minimal in SMEs.
    \end{itemize}
  \item
    Note that these probabilities are provided, conditioned on the fact
    that the active scanning (of the Reconnaissance tactic) has
    identified that this particular SME has a vulnerable public facing
    web-app.
  \item
    Defence Limitations: While SMEs might have firewalls or WAFs, these
    won't necessarily cover all attacks and may be misconfigured. An SME
    may not rigorously isolate DMZ systems or adequately enforce least
    privilege on service accounts.
  \item
    \citet{edgescan:2024_vulnerability_statistics_report}provide detailed analysis of CVEs. However, they do
    not give figures for the likelihood of success of exploits.
    Probability of success will be different dependent on the particular
    CVE and the exploit that has been developed for it. Attackers will
    naturally select the particular vulnerabilities (CVEs), and
    associated exploits which have high probabilities of success, and
    which yield good options and possibilities for onward privilege
    escalation, lateral movement, C2C etc. Factors such as these are
    described by Qualysys (2023) which also states that: ``Over 7,000
    vulnerabilities had proof-of-concept exploit code. These
    vulnerabilities could result in successful exploitation; however,
    the exploit code is typically of lower quality, which may reduce the
    likelihood of a successful attack. 206 vulnerabilities had
    weaponized exploit code available. Exploits for these
    vulnerabilities are highly likely to compromise the target system if
    used. Qualsys go on to explain that these 206 vulns form just 0.77\%
    of the total vulnerabilities declared in 2023.
  \item
    It can also be observed that attempting an exploit might not be that
    costly for an attacker to try, which could mean some tolerance for
    lower probabilities of success too.
  \end{itemize}
\item
  Execution

  \begin{itemize}
  \item
    From MITRE T1509 (command and scripting interpreter)

    \begin{itemize}
    \item
      Antivirus
    \item
      Audit (detect unauthorized installations)
    \item
      Behaviour prevention on endpoint
    \item
      Only permit exec of signed code
    \item
      Remove unnecessary shells/interpreters
    \item
      Execution prevention (application control)
    \item
      Prevent user install of software
    \item
      Privileged account management (only admins can use shells)
    \item
      Restrict web-based content
    \end{itemize}
  \item
    Illustrative defenses: Standard anti-virus or EDR. Application
    control or allow-listing is rare in SMEs. Few enforce ``signed code
    only'' or restrict PowerShell usage, or remove unnecessary
    shells/interpreters. Some SMEs rely on default OS protections and
    basic user privilege limits (e.g.~non-admin users), but admin tools
    and scripting are usually available to attackers.
  \item
    Execution (i.e.~adversary controlled code running on victim
    systems), of some sort, will have to be successfully performed
    multiple times on multiple machines, as the attacker moves laterally
    and seeks to encrypt data on different systems.
  \item
    Ransomware affiliates commonly use PowerShell, script interpreters,
    or built-in tools for execution. SMEs rarely have strict application
    whitelisting or behavioral anti-malware controls that could block
    such actions. Though even a non 24/7 SOC might be expected to triage
    a PowerShell script.
  \item
    Many SMEs rely on anti-virus. However, skilled attackers can evade
    AV, for example by malware code obfuscation (to foil at least,
    signature based AVs) or by using memory resident techniques.
  \item
    Regards EDR:~\citep{cybersecurityasia:2024_ransomware}
  \item
    6\% of Barracuda's Security alerts in Q1 2025 were for suspicious
    PowerShell scripts.
  \item
    \cite{expertinsights:2025_edr_market_overview}suggests \appr50\% of companies had XDR
    (though didn't specify whether these were SMEs), and 50\% had AV on
    endpoints
  \item
    So we can observe that XDRs can detect executions.
  \item
    Analysis:

    \begin{itemize}
    \item
      If the attacker moves quickly (as seems increasingly the case) and
      the SME is relatively poorly equipped in terms of SOC staff in
      order to respond to any EDR alerts, then oftentimes EDR may not
      catch the attack
    \item
      Anti-virus could potentially catch malicious executables, but
      attackers will use obfuscation to avoid signature checks. Use of
      AI based AV may help here, but it's unclear how many SMEs use
      it.
    \item
      It is also possible that if one execution technique fails an
      attacker could potentially try another.
    \item
      Taking all the above together, we assume average of 50\%, with wide
      20\%, 90\% credible interval
    \end{itemize}
  \end{itemize}
\item
  Privilege Escalation

  \begin{itemize}
  \item
    From~\citet{mitre:T1068_exploitation_for_privilege_escalation}:
    \begin{itemize}
    \item
      Application isolation and sandboxing
    \item
      Execution prevention (block known vulnerable drivers that run in
      kernel mode)
    \item
      Exploit protection (behavioural monitoring)
    \item
      Update software
    \end{itemize}
  \item
    From~\citet{mitre:T1003_os_credential_dumping}:
    \begin{itemize}
    \item
      Active Directory Configuration
    \item
      Behaviour prevention on endpoint
    \item
      Credential access protection
    \item
      Encrypt sensitive information
    \item
      OS configuration
    \item
      Password policies
    \item
      Privileged account management
    \item
      Privileged process integrity
    \item
      User training
    \end{itemize}
  \item
    Illustrative defenses: SMEs generally have basic password policies.
    .

    \begin{itemize}
    \item
      According to~\citet{jumpcloud:2025_mfa_statistics}, MFA usage is 34\% at Medium sized companies (vs 87\% at large
      enterprises). It is possible also that admin accounts also
      frequently lack MFA.
    \item
      Whilst there is patching of critical servers, SMEs are likely to
      be slower than larger enterprises. Hence servers might not be
      fully patched against privilege-escalation exploits (e.g.~missing
      some Windows privilege escalation patches).
    \item
      There may be EDR and some SOC resource
    \end{itemize}
  \item
    Relevant techniques:

    \begin{itemize}
    \item
      T1068 -- Exploitation for Privilege Escalation
    \item
      Using exploits like ZeroLogon, PrintNightmare, and their more
      recent equivalents etc.
    \item
      OR
    \item
      T1003 -- OS Credential Dumping
    \item
      Ransomware attackers almost always try to gain admin-level
      control. SMEs may have flat (non-segmented) networks with many
      local admins. Attackers may dump local credentials and find an
      admin password that works across systems
    \item
      If credentials fail, known exploits (like PrintNightmare or
      ZeroLogon) can be tried, and SMEs might not have applied these
      patches or enabled mitigations.
    \item
      Low MFA adoption and informal admin account management can make it
      easier to leverage stolen accounts.
    \end{itemize}
  \item
    Analysis:

    \begin{itemize}
    \item
      One might anticipate a fairly high chance of success in privilege
      escalation, particularly if usage of MFA is low in SMEs, and also
      if the network structure is fairly flat and relatively small so
      that one compromised admin access credential gives system-wide
      access. We'll go with 70\%, but again the credible interval bounds
      will be large due to the lack of good statistical information.
    \end{itemize}
  \end{itemize}
\item
  Discovery

  \begin{itemize}
  \item
    Listing here, typical defence mitigations for key techniques
  \item
    MITRE T1018 -- Remote System Discovery:

    \begin{itemize}
    \item
      No easy mitigation, since based on abuse of regular system
      features
    \item
      Detections, monitor:
    \item
      Command execution
    \item
      File access (e.g.~etc/hosts)
    \item
      Network connections (pings etc)
    \item
      Process creation (png.exe etc)
    \end{itemize}
  \item
    MITRE T1135 -- Network Share Discovery:

    \begin{itemize}
    \item
      OS config: Windows Group Policy (do not allow anonymous
      enumeration\ldots)
    \item
      Detections, monitor:
    \item
      Command execution
    \item
      Discovery related OS API execution
    \item
      Process creation
    \end{itemize}
  \item
    MITRE T1083 -- File and directory discovery

    \begin{itemize}
    \item
      Cannot be easily mitigated due to abuse of regular system
      features.
    \item
      Might be picked up by monitoring of command execution, API
      controls and process creation
    \end{itemize}
  \item
    Illustrative defenses: SMEs may have EDR for network monitoring for
    internal reconnaissance, but SOC analyst support may be limited. The
    SME may have basic Windows firewall on endpoints, that could prevent
    certain enumerations, but may have no policy against ping/NetBIOS
    scans.
  \item
    Affiliates need to map the network -- find servers, shares, backups.
    In an SME, networks are sometimes flat and lacking segmentation, and
    in such cases the attacker's malware may be able to freely scan or
    query Active Directory. In an SME the attacker can more easily reach
    and probe most systems.Detection of internal recon is generally less
    likely, especially if the attacker uses stealthy techniques
    (e.g.~using legitimate admin commands or slow scanning).
  \item
    Barracuda XDR. 48\% of Barracuda's 1H2024 security alerts were for
    network reconnaissance
  \item
    Expert insights (2024) suggests \appr50\% of companies had
    XDR (though didn't specify whether these were SMEs), and 50\% had AV
    on endpoints
  \item
    Analysis:
  \item
    There is a high likelihood that the attacker can perform necessary
    discovery operation, and particularly so iif the attacker has
    achieved necessary privilege escalations (which we have already
    accounted for). We select 85\% as a best estimate, with lower and
    upper bounds of 70\%, 95\%
  \end{itemize}
\item
  Lateral Movement

  \begin{itemize}
  \item
    Listing here, typical mitigation defenses (not detections)
  \item
    MITRE T1021 -- Remote Services

    \begin{itemize}
    \item
      Audit (for potential weaknesses)
    \item
      Disable or remove (unnecessary features and programs
    \item
      Limit access over network (use gateways)
    \item
      MFA
    \item
      Password policies (on admin accounts)
    \item
      User account management (access controls)
    \end{itemize}
  \item
    MITRE T1210 -- Exploitation of Remote Services (exploiting vulns in
    programs)

    \begin{itemize}
    \item
      App isolation and sandboxing
    \item
      Disable (unnecessary) features/programs
    \item
      Exploit protection
    \item
      Network segmentation
    \item
      Privileged account management
    \item
      Threat intelligence program
    \item
      Update software
    \item
      Vuln scanning
    \end{itemize}
  \item
    MITRE T1570 -- Lateral Tool Transfer (copying tools around the victim's
    environment)

    \begin{itemize}
    \item
      Host firewall (filter network traffic)
    \item
      Network intrusion prevention
    \end{itemize}
  \item
    Illustrative defenses:

    \begin{itemize}
    \item
      SMEs often have flat networks with open internal access. Network
      segmentation may be uncommon (all workstations/servers on one LAN
      or a couple of VLANs). Internal firewalls or host firewall rules
      are generally permissive (file shares, RDP open internally). MFA
      for internal RDP may not exist. Software vulnerabilities on
      internal services might exist (unpatched internal SMB, etc.).
      Basic password policies might limit some sharing, but generally
      nothing prevents an admin account from accessing all machines.
    \end{itemize}
  \item
    Ease of Spread: In a small/medium business, once the attacker has
    credentials or admin access, moving laterally is straightforward.
    All machines often trust the domain admin or common credentials.
    Many SMEs don't restrict admin share access or have ACLs that would
    stop an intruder. Lack of Segmentation: Without network
    segmentation, the threat actor can reach most systems over the
    network. Defenses: Only minimal barriers exist -- e.g., if some
    systems require separate logins or there's an isolated server, but
    that's uncommon. Likely lack of internal MFA means stolen creds work
    everywhere.
  \item
    Barracuda XDR indicates

    \begin{itemize}
    \item
      The data for 2024 shows that lateral movement is the clearest sign
      of ransomware activity. Just under half (44\%) of the ransomware
      attacks were spotted by the lateral movement detection engine.
    \end{itemize}
  \item
    Expert insights (2024) suggests \appr50\% of companies had
    XDR (though didn't specify whether these were SMEs)
  \item
    SMEs are less likely to implement robust EDR/XDR and associated
    well-staffed SOC resources, so there is a good chance that lateral
    movement is not detected.
  \item
    If the SME has an isolated backup server - then that may prevent
    lateral movement to all necessary machines.
  \item
    Kent Invicta citing Beaming research (2024): Just one in four UK
    businesses today consistently pursue good backup practices. 22\% of
    companies systematically back up data to a specialist offsite
    facility or provider, with full knowledge and control of backup
    procedures and where their data is held. 34\% maintain an air-gapped
    data backup physically isolated from the internet
  \item
    Analysis: Typically, lateral movement works assuming that privileges have
      already been escalated (which we've already accounted for).
      because once in, attackers can use admin shares, RDP, or deploy
      tools like PsExec essentially unopposed. But there will be
      occasions where it fails, for example most especially where an SME
      has adopted good backup practices. We select 65\% as a best
      estimate and 50\%, 80\% as a credible interval.
  \end{itemize}
\item
  Collection

  \begin{itemize}
  \item
    MITRE T1039 -- Data from Network Share

    \begin{itemize}
    \item
      No mitigations since based on abuse of normal system features
    \item
      Only defenses are detection of: command, file, network share,
      network traffic
    \end{itemize}
  \item
    MITRE T1560 -- Archive Collected Data

    \begin{itemize}
    \item
      Audit (detect unauthorized archival utilities)
    \item
      Detections similar to above
    \end{itemize}
  \item
    lllustrative defenses: Little to none. Sensitive data on file
    servers or shares is typically not encrypted at rest in SMEs.
    There's usually no Data Loss Prevention internally. Audit logging on
    file access might exist but is not actively monitored. Attackers can
    freely gather files; compressing data (using zip or RAR) likely
    won't be blocked (no application control on such tools).
  \item
    To maximize extortion, the affiliate will quietly collect
    confidential files before encryption. In SMEs, if the attacker has
    achieved admin privileges (through privilege escalation that is
    already accounted for), the attacker can read most or all data. No
    Internal DLP: SMEs generally lack DLP or alerts for large file
    access. If the attacker creates archives of data (T1560) or stages
    data on an internal server (T1074), it's unlikely to be detected.
    There are essentially no mitigations to prevent data gathering -- it
    abuses normal file access channels.
  \item
    According to StrongDM: Only 17\% of small businesses encrypt data.
  \item
    Clearly if data is encrypted,then extortion will be harder for the
    attacker to achieve.
  \item
    Analysis:

    \begin{itemize}
    \item
      In most cases the attacker will be able to collect data. It seems
      there is some chance that some enterprises may have encrypted data
      (and the attacker cannot access the decryption keys). So we select
      a best estimate of 90\%, with credible intervals of 75\%--95\%.
    \end{itemize}
  \end{itemize}
\item
  Command-and-Control

  \begin{itemize}
  \item
    Potentially many techniques are applicable. Here, we list
    mitigations for a common technique.
  \item
    MITRE T1071 -- Application Layer Protocol

    \begin{itemize}
    \item
      Filter network traffic (firewalls in network and on endpoints)
    \item
      Network intrusion prevention (looking for network traffic
      signatures)
    \end{itemize}
  \item
    Illustrative defenses:

    \begin{itemize}
    \item
      Basic network firewall allows most outbound traffic (common in
      SMEs). Few SMEs whitelist egress by domain/IP -- typically only
      high-security sectors do. SMBs might not have proxy inspection or
      advanced network monitoring. Some endpoint AV may flag known C2
      beacons, but attackers often use HTTPS or DNS tunneling to blend
      in.
    \end{itemize}
  \item
    After initial compromise, the affiliate's malware needs to
    communicate out to give instructions. Outbound web traffic is
    usually unrestricted (aside from perhaps basic web filtering).
    Attackers commonly use HTTPS, web protocols or even legitimate cloud
    services for C2, which are unlikely to be blocked. Without dedicated
    egress filtering or an intrusion prevention system or pre-configured
    detections, SMEs will generally not notice the C2 traffic. The
    adversary might also use encryption (TLS) or proxy through benign
    hosts, and SMEs lack the means to detect this subtle traffic.
  \item
    Analysis:

    \begin{itemize}
    \item
      High probability of success, best estimate, 90\%
    \item
      Credible interval is 80\% to 100\%
    \end{itemize}
  \end{itemize}
\item
  Exfiltration

  \begin{itemize}
  \item
    Potentially many techniques are applicable.
  \item
    MITRE T1567 -- Exfiltration Over Web Service

    \begin{itemize}
    \item
      Data Loss Prevention (detect and prevent sensitive data being
      uploaded via web browsers)
    \item
      Restricted web-based content (web proxies prevent use of
      unauthorized external services)
    \end{itemize}
  \item
    Illustrative defenses: Very few. SMEs rarely employ data encryption
    in transit or strict outbound bandwidth controls. No formal Data
    Loss Prevention solutions (which might detect sensitive data
    leaving). Possibly rate limits on network or basic alerts if an
    outbound connection is extremely large, but often not. Most rely on
    ISP and cloud security defaults
  \item
    The affiliate will attempt to send the stolen data out to their own
    servers before launching ransomware. If the SME does not have DLP
    (adn most are assumed not to have it), then there may be little to
    prevent exfiltration, though it may be noticed by large usage/bills
    for data traffic out of the network or more basic threshold type
    rules monitoring the network. Attackers might upload data via HTTPS
    to cloud storage or use their C2 channels. SMEs typically won't
    detect an HTTPS upload to an unknown server. The exfiltration might
    be limited by bandwidth or noticed if it crashes a network link.
  \end{itemize}
\item
  Impact

  \begin{itemize}
  \item
    MITRE T1486 -- Data Encrypted for Impact

    \begin{itemize}
    \item
      Behaviour prevention on endpoint
    \item
      Data backup (ideally stored off system)
    \end{itemize}
  \item
    MITRE T1490 -- Inhibit System Recovery

    \begin{itemize}
    \item
      Data backup
    \item
      Execution prevention (e.g block utilities like diskshadow.exe)
    \item
      OS configuration (prevent disabling of certain services or
      deletion of certain files)
    \item
      User account management (limit access to backups only as
      necessary)
    \end{itemize}
  \item
    MITRE T1489 -- Service Stop

    \begin{itemize}
    \item
      Network segmentation (operate SOC etc on separate system to
      production system)
    \item
      Out of band comms channels (for use during security incident)
    \item
      Restrict file and directory permissions
    \item
      Restrict registry permissions
    \item
      User account management (so only admins can change services)
    \end{itemize}
  \item
    illustrative defenses: Regular data backups (varies by SME -- many
    have at least nightly backups, but often connected to network or
    cloud). Quality of backups differ: some SMEs keep offline backups,
    but many have NAS or cloud shares that could be deleted. Few SMEs
    have immutable backups. Ransomware prevention tools (like
    anti-encryption behavioral blockers) are not common, aside from what
    baseline AV provides. Some critical files may be backed up to cloud
    services (OneDrive, etc.) providing partial resilience.
  \item
    In the final stage, the affiliate deploys the ransomware payload to
    encrypt data and disrupt operations, and they try to destroy backups
    to prevent recovery.
  \item
    Whilst most of our SMEs do maintain backups, often these are online
    or accessible from the network. We have already accounted for the
    ``poor backup hygiene'' aspect to some extent in the lateral movement
    step so we avoid double counting it here.
  \item
    According to Invenio IT, 58\% of backups fail during recovery due to factors like outdated technology, inadequate testing, or malware infection.
  \item
    Hence, even if the threat actor did not get to the backup servers,
    there's a good chance that the backups will prove insufficient to
    recover from whatever the attacker has encrypted (with issues, for
    example, in terms of when the backup was last run, and whether it
    was configured properly).
  \item
    Sophos: 57\% of backup compromise attempts were successful
  \item
    Some endpoint security might detect mass encryption (Microsoft
    Defender's Controlled Folder Access, etc.), but SMEs will not always
    have enabled or configured these. A lot of EDRs will detect the
    creation of ransom notes and changes to file extensions.
  \item
    By the time the attacker has got this far in the attack, they will
    have established significant privileges and control and it can be
    expected that there is a good chance that this last impact step will
    succeed
  \item
    Analysis:
    \begin{itemize}
        \item
          The best estimate probability of success should be quite high, and
          we go with 80\%
        \item
          A lower (5\%) bound of 60\% (given 58\% of backups fail)
        \item
          Upper bound of 90\% (given \appr60\% backups fail and the
          threat actor likely already has high privileges, and prevalence of
          anti-encryption protections are likely low)
        \item
          MITRE T1657
        \item
          Only relevant mitigation listed is ``User training''
        \item
          An SME likely lacks in-house training or capability in ransom
          negotiation, but could buy it in.
        \item
          But the best effort number is used in the calculation of the ransom
          payout in the Impact section below.
        \item
          Chainanalysis: ``According to our data, around 30\% of negotiations
          actually lead to payments or the victims deciding to pay the
          ransoms''.
        \item
          Coveware also support this \appr30\% payout figure
        \item
          For the 5\% bound: Given reports that companies are becoming more
          reluctant to pay ransoms, we'll elect a lower bound of 15\%
        \item
          For the 95\% bound: Since definitely there appears to be a trend in
          the literature indicating more reluctance to pay ransoms these days,
          we will select an upper bound of 50\%
    \end{itemize}
  \end{itemize}
\end{itemize}

\textit{Prob of success (excluding the last T1657 Financial Theft term) = 6.4\%}

\subsubsection*{Impact}\label{impact}
There are a wide number of factors that can determine the actual cost of
a specific attack, these include at least:

\begin{itemize}[leftmargin=*]
\item
  Whether the attacker completes the full double extortion attack (both
  data exfiltration and data encryption)
\item
  Whether the attacker is partially successful and exfiltrates data (but
  does not achieve data encryption)
\item
  Whether the attacker is partially successful and encrypts systems (but
  does not complete data exfiltration)
\item
  Whether the ransom is paid or not

  \begin{itemize}
  \item
    Which besides the cost of the ransom itself, may also impact
    recovery costs (e.g.~whether decryption keys are then provided,
    making recovery easier)
  \end{itemize}
\item
  Dependency on the value of any data that is exfiltrated
\item
  Dependency on the quantity and importance of files which have been
  encrypted
\item
  Whether backups exist
\item
  Whether the attacker is successful in corrupting the backups
\end{itemize}

\cref{tab:baseline_estimates_impact} summarizes our estimates for the baseline values of both the ransom payment and the recovery cost: 
{\renewcommand{\arraystretch}{1.5}
\begin{table}[!h]
\centering
\begin{tabular}{llcl}
\toprule
Cost to defenders per attack & 5\% & Most likely value & 95\% \\
\midrule
Ransom payment & \$50k & \$165k$^\dagger$ & \$400k \\
Recovery cost & \$0.3m & \$0.65m & \$1.5m \\
\bottomrule
\end{tabular}
\captionsetup{width=0.6\linewidth}
\caption{Impact estimates ($^\dagger$assumes 30\% pay ransom at \$550k mean, and 70\% pay no ransom).}
\label{tab:baseline_estimates_impact}
\end{table}
}

\underline{Rationale}

\textbf{Ransom Payment}

\begin{itemize}[leftmargin=*]
\item
  \citep{sophos:state_of_ransomware_2023} reports:
  \begin{itemize}
\item
  In SME's with revenue \$10m to \$50m bracket in 2023:
\item
  Mean demand was \$1.7m
\item
  Median demand was \$0.33m
\item
  Elsewhere in the report, it states that proportion of ransom demand
  paid in the \$10m--\$50m revenue SME market is 93\%
\item
  Note: it seems that the questioning to respondents in the Sophos
  survey related to the respondent's most significant ransomware attack.
  So, it's possible that the figures provided are biased upwards in
  cost.
\item
  Nevertheless, we assume 0.93$\times$\$1.7m=\$1.6m for mean ransomware
  payout.
\end{itemize}
\item
  \citep{sophos:state_of_ransomware_2025} indicates:
  \begin{itemize}
    \item 
    for SMEs in the \$10m to \$50m revenue bracket, the average payout is \$106k
\item
  If the ratio of mean/median was similar as 2024 then this might
  suggest a mean payout of \$106k $\times$ (1.7/0.33) = \$550k
  \end{itemize}
\item
  Coveware indicate:
  \begin{itemize}
\item
  A mean ransomware payout of \appr\$550k in Q1 2025, and an
  average payout of \appr\$200k. They indicate that the average
  size of companies impacted by ransomware is 228 employees.
\item
  This Coveware figure covers all sizes of companies. The comparable
  mean ransom payout figure for all companies as listed by~\citet{sophos:state_of_ransomware_2023}is \$3.9m.
\end{itemize}
\item
  Crowdstrike claim: The average ransom payment in 2024 is \$2.73 million, up from \$1.82 million in 2023.
\item
  BlackFog reports that the average ransom demand in Q1, 2025, was
  \$663,582, based on 93 known ransom demands
\item
  There seems to be a great deal of volatility in payments and
  demands:
\item
  \citep{sophos:state_of_ransomware_2023}: \ldots{} the mean ransom payment has increased x2.6 in the last year. Organizations that paid the ransom
  reported an average payment of \$3.9 million, up from \$1.5m in
  2023.
\item
  \citep{sophos:state_of_ransomware_2025} reports: ``The \ldots average ransom payment fell by 50\% in the last year, down from \$2 million in 2024 to \$1 million in 2025''
\item
  If the ratio of mean/median is assumed to be the same as in 2024
  (i.e.~\$3.9m/\$2m = \appr2) then mean, as measured across
  all enterprise sizes would be \appr\$2m
\item
  IBM report: The average ransom demand in 2024 also saw a significant
  increase, rising to 2.73 million USD, nearly 1 million USD more than
  in 2023.
\item
  Analysis:
    \begin{itemize}
\item
  The Sophos (2025) ``all enterprise sizes'' figure for mean ransom
  payment appears to be \appr75\% of Crowdstrike's latest
  (2024) number, approximately 400\% of Coveware's equivalent number,
  and 300\% of BlackFrog's number.
\item
  At least some of the questions in the Sophos survey did ask for info
  about each company's most impactful ransomware attacks, which might
  bias their figures upwards somewhat.
\item
  Nevertheless, Sophos is understood to be a reliable source, so we take
  the figure that we computed based on the ``average'' data provided in
  their 2025 report (and the mean/median ratio computed based on their
  2024 report): i.e.~\$550k
\item
  We then need to weight this assumed mean payout figure (\$550k)
  according to whether a successfully completed attack results in a
  ransom payout (where success here is defined in terms of successful
  data exfiltration and encryption):
\item
  Chainanalysis: ``According to our data, around 30\% of negotiations
  actually lead to payments or the victims deciding to pay the
  ransoms''.
\item
  Coveware also support this \appr30\% payout figure
\item
  Sophos --  56\% paid ransom and got data back (across all
  enterprise sizes), which presumably means that an even larger
  percentage paid a ransom (since some will have paid ransom and NOT got
  data back). But this figure is very different to those provided by
  Chainanalysis and Coveware. One possible explanation is that Sophos'
  questioning in its survey was referring to victims' worst incidents.
\item
  Since the Sophos information is a bit unclear regards definition of
  terms, and a number of sources cite declining prevalence in paying of
  ransoms, let's assume 30\% payout.
\end{itemize}
\item
  Hence expected best estimate ransom payout for an attack that is
  successfully executed from a technical perspective is 0.3 $\times$ \$0.55
  \appr\$165k
\item
  5\% - Given the great deal of uncertainty regards what the actual mean
  ransomware payout is, and also the great volatility in payouts year on
  year (up to 500\% reported by Sophos) we'll assume a lower credible
  interval value of \$50k
\item
  95\% - If we take the Sophos figure \$0.55m (which we think is likely
  to be on the high side, compared to other estimates), and then also
  assume 0.75 actually pay the ransom, this would give us \$400k
\end{itemize}

\textbf{Recovery cost (including downtime, people time, device cost,
network cost, lost opportunity)}

\begin{itemize}[leftmargin=*]
\item
  Sophos reports, for 2024:
\item
  For organisations with revenue $<$\$10m
\item
  Recovery costs were \$1.2m (up from \$0.16m in 2023)
\item
  Assumed to be mean, not median, since Sophos report: ``Globally, median
  recovery costs doubled from \$375,000 to \$750,000 over the last
  year'', and this will cover also large corporations\ldots{} in 2024,
  organizations reported a mean cost to recover from a ransomware attack
  of \$2.73M
\item
  For organisations with revenue of \$10m to \$50m
\item
  \$1.5m up from \$1.1m in 2023)
\item
  Sophos' question to CISOs: ``What was the approximate cost to your
  organization to rectify the impacts of the most significant ransomware
  attack (considering downtime, people time, device cost, network cost,
  lost opportunity etc.)''
\item
  ``The report is based on the findings of an independent,
  vendor-agnostic survey commissioned by Sophos of 5,000
  IT/cybersecurity leaders across 14 countries in the Americas, EMEA,
  and Asia Pacific. All respondents represent organizations with between
  100 and 5,000 employees. The survey was conducted by research
  specialist Vanson Bourne between January and February 2024, and
  participants were asked to respond based on their experiences over the
  previous year''
\item
  \citet{sophos:state_of_ransomware_2025} reports, for the equivalent question in 2025:
\item
  For companies of 100--250 employees recovery costs were \$638k
\item
  As averaged across all company sizes, recovery cost decreased from
  \$2.83m in 2024 to \$1.53m in 2025 (with the 2023 figure being
  \$1.82m). Hence, the 2025 figure is 54\% of the 2024 figure.
\item
  Analysis:
  \begin{itemize}
\item
  Actual recovery costs for any specific attack, will vary according to
  a long list of factors identified in the list above this table.
\item
  The above figures provided by Sophos presumably also cover a spectrum
  of possible scenarios, including a) double extortion vs single
  extortion, b) backups compromised vs backups not compromised, c)
  ransoms paid (and decryption keys provided, data sometimes returned)
  vs ransoms not paid vs both ransom paid and backups used etc. The
  statistics may implicitly also cover a range of fully successful
  attacks and partially successful attacks (from the attacker
  perspective).
\item
  One could potentially try and account for a different recovery cost
  dependent on all the factors listed above the table, however, since we
  are looking for a mean, and Sophos have already provided us with one,
  then we will go with the Sophos figures.
  \end{itemize}
\item
  We assume for our SME targets the recovery cost is
  \appr\$650k based on~\citet{sophos:state_of_ransomware_2025}
\item
  For 5\% assume \$300k (allowing for some possible deflation in
  ransomware payments over the coming year, and noting large year on
  year on swings)
\item
  For 95\%, assume \$1.5m (allowing for inflation over the coming year,
  and noting that the 2024 figure was twice as high as the 2025 figure)
\end{itemize}

\textit{Sum of costs = \$165k+\$650k = \$815k}

\subsubsection*{Total Risk} \label{total-risk}
Total risk = Estimated number of Actors (10) x Mean number of attacks per year per actor (200) x Probability of successful attack (0.064) x Impact per successful attack (\$0.8m) = \$102m / year (note that this is calculated as a simple multiplication, not through fitting distributions and running Monte Carlo simulations)

\underline{Sanity check calculations}

\begin{itemize}[leftmargin=*]
\item
  Consider incomes of individuals in the threat actor group

  \begin{itemize}
  \item
    Ransom payout is \$2.1m per OC3 group (200 $\times$ 0.064 $\times$ \$165k). With
    10 individuals per OC3 group, this suggests a revenue (before costs)
    of \$210k per person in the gang (a proportion of this income will
    go to the RaaS operator, which is the main variable cost of the
    threat actors (\appr20\% of the ransom payment goes to the
    RaaS operator according to~\citet{theregister:2023_ransomware_affiliates_money}), so each individual would make a return on the order of
    perhaps \$150k, once other costs are taken into account.
  \item
    Assuming \$30k/year for salary of Russian cyber professional, the
    figures would seem plausible, though the returns would be
    unexceptional for a cybersecurity professional working in Western
    Europe/US. This income is also on the low side compared to the
    \$1.9m that Astimorov of LockBit was supposed to have extorted over
    a 3 year period~\citep{doj:2024_lockbit_affiliates_guilty}.
  \end{itemize}
\item
  Harm as proportion of global harm

  \begin{itemize}
  \item
    \citet{chainalysis_ransomware_2025} state global ransom payouts in 2024 were \$813m
  \item
    Based on this figure, the \$20m/year ransom payout figure seems like
    it could be plausible:

    \begin{itemize}
    \item
      Since we are considering 10 larger (10 person) affiliate-groups,
      amounting to 100 individuals, out of a potential pool of
      \appr3000 people

      \begin{itemize}
      \item
        We assume 700 affiliates in total (see ``number of actors''
        computations above) comprising 200, 10-man OC3-groups and 500, 1
        or 2 person OC2 groups
      \end{itemize}
    \item
      And since, if every affiliate-individual earns broadly the same
      amount then this would suggest that the aggregate ransom revenue
      (based on our computed figures for revenue per OC3 group) would be
      3000/100 $\times$ \$20m = \$600m.
    \item
      Which is similar to the ChainAnalysis figure of \$813m
    \end{itemize}
  \end{itemize}
\end{itemize}

\section{Model Uplift and LLM-Estimated Parameters}
\label{app:F}
In~\cref{tab:benchmark_mapping}, we present the allocation of our two chosen benchmarks, Cybench and BountyBench, to the risk indicators in the OC3 Ransomware SME risk model, alongside a rationale for each decision. \cref{tab:uplift_estimates_app} shows the LLM-estimated values for these factors at both the SOTA (current) and saturated level of capabilities.

\begin{longtable}{p{3cm}lp{8cm}}
\caption{Mapping of our two chosen benchmarks to risk indicators} \label{tab:benchmark_mapping} \\

\toprule
Risk Factor & Benchmark & Rationale \\
\midrule
\endfirsthead

\multicolumn{3}{l}{\textit{(Continued from previous page)}} \\
\toprule
Risk Factor & Benchmark & Rationale \\
\midrule
\endhead

\midrule
\multicolumn{3}{r}{\textit{(Continued on next page)}} \\
\endfoot
\bottomrule
\endlastfoot

Initial Access & BountyBench &  BountyBench contains tasks involving exploiting public-facing applications and services\\
Execution & BountyBench & Tasks involve realistic Kali Linux execution environment.\\
Privilege Escalation & BountyBench & Tasks target privillege escalation specifically. \\
Lateral Movement & Cybench & Neither Cybench nor BountyBench has an ideal fit for this step. Cybench has a few multi-server tasks. \\
Impact & Cybench & Cybench contains cryptography-related tasks that better capture encryption steps in ransomware attacks. \\
Financial Theft / Extortion & Cybench & Neither BountyBench nor Cybench is ideal for ransom negotiations. Cybench is used here due to cryptography tasks providing some indicator of the quality of encryption for ransom, which may be a factor in ransom negotiations.\\
Number of Actors & BountyBench & BountyBench tasks are more relevant to early attack steps (Initial Access and Execution) which we expect to correlate with the number of actors attempting these attacks. \\
Number of Attempts Per Actor & BountyBench & BountyBench tasks are more relevant to early attack steps (Initial Access and Execution) which we expect to correlate with the number of attempts threat actors can make. \\
Damage Per Attack: Recovery & Cybench & Cybench is more relevant to cryptography components of attack which will define the cost of data recovery. \\
Damage Per Attack: Ransom & Cybench & We note that ransom payment size is likely not heavily correlated with either benchmark and may be extrinsic to the uplift model altogether. This said, the effectiveness of data encryption used will be a factor in ransom negotiations, which Cybench is a better indicator of. 

\end{longtable}

\begin{longtable}{llccc}
\caption{Uplift Estimates. Quantities are computed across 100,000 Monte Carlo samples from LLM simulated experts.} \label{tab:uplift_estimates_app} \\

\toprule
Capabilities & Factor & 5th Percentile & Mode (KDE Estimated) & 95th Percentile \\
\midrule
\endfirsthead

\multicolumn{5}{l}{\textit{(Continued from previous page)}} \\
\toprule
Capabilities & Factor & 5th Percentile & Mode (KDE Estimated) & 95th Percentile \\
\midrule
\endhead

\midrule
\multicolumn{5}{r}{\textit{(Continued on next page)}} \\
\endfoot

\bottomrule
\endlastfoot

SOTA & Reconnaissance & 1 & 1 & 1 \\
Saturated & Reconnaissance & 1 & 1 & 1 \\
SOTA & Resource Development & 1 & 1 & 1 \\
Saturated & Resource Development & 1 & 1 & 1 \\
SOTA & Initial Access & 0.1756 & 0.6623 & 0.9284 \\
Saturated & Initial Access & 0.2504 & 0.7653 & 0.9205 \\
SOTA & Execution & 0.0996 & 0.6116 & 0.9230 \\
Saturated & Execution & 0.2252 & 0.6008 & 0.8981 \\
SOTA & Privilege Escalation & 0.1812 & 0.7601 & 0.9472 \\
Saturated & Privilege Escalation & 0.2586 & 0.8225 & 0.9467 \\
SOTA & Discovery & 0.7133 & 0.8111 & 0.8796 \\
Saturated & Discovery & 0.7132 & 0.8106 & 0.8790 \\
SOTA & Lateral Movement & 0.4479 & 0.7021 & 0.8641 \\
Saturated & Lateral Movement & 0.4646 & 0.7117 & 0.8868 \\
SOTA & Collection & 0.7548 & 0.8976 & 0.9545 \\
Saturated & Collection & 0.7484 & 0.9040 & 0.9552 \\
SOTA & Command and Control & 0.8144 & 0.9281 & 0.9720 \\
Saturated & Command and Control & 0.8162 & 0.9337 & 0.9708 \\
SOTA & Exfiltration & 0.7274 & 0.9091 & 0.9665 \\
Saturated & Exfiltration & 0.7261 & 0.9090 & 0.9664 \\
SOTA & Impact & 0.6851 & 0.8248 & 0.9297 \\
Saturated & Impact & 0.6868 & 0.8410 & 0.9494 \\
SOTA & Financial Theft / Extortion & 0.1227 & 0.3369 & 0.6281 \\
Saturated & Financial Theft / Extortion & 0.1222 & 0.2978 & 0.6213 \\
SOTA & Damage per Attack: Recovery & 279149.9923 & 703448.7619 & 1838613.2362 \\
Saturated & Damage per Attack: Recovery & 381479.8551 & 761200.2752 & 1532674.0606 \\
SOTA & Damage per Attack: Ransom & 192373.5585 & 627288.6288 & 1138280.3394 \\
Saturated & Damage per Attack: Ransom & 223085.5330 & 625053.7436 & 1284713.9777 \\
SOTA & Number of Actors & 3.7162 & 11.9861 & 57.7922 \\
Saturated & Number of Actors & 5.1268 & 17.7117 & 63.9287 \\
SOTA & Number of Attempts per Actor & 70.6827 & 215.1253 & 587.9752 \\
Saturated & Number of Attempts per Actor & 71.5314 & 249.7416 & 712.1891 \\
\hline \textbf{Aggregated} \\
SOTA & Probability of Successful Attack & 0.0044 & 0.0163 & 0.1513 \\
Saturated & Probability of Successful Attack  & 0.0131 & 0.0444 & 0.1760 \\
SOTA & Risk (Ransom) & 1303544.0885 & 19558561.6269 & 373735748.5034 \\
Saturated & Risk (Ransom) & 4245705.7889 & 36926476.0946 & 673054052.0885 \\
SOTA & Risk (Recovery) & 6150346.2680 & 76550105.0840 & 1380727437.6985 \\
Saturated & Risk (Recovery) & 22280199.1123 & 155332176.6640 & 2285349245.8274 \\
SOTA & Total Risk & 8803585.3608 & 103297672.4354 & 1748881431.3670 \\
Saturated & Total Risk & 30307020.5801 & 195782280.9930 & 2909501181.1624 \\
\end{longtable}

\end{appendices}

\end{document}